\documentclass[a4paper,11pt]{article}

\usepackage{jheppub} 

\usepackage{graphicx}
\usepackage{multirow}
\usepackage{subfig}
\usepackage{float}

\title{Perturbative QCD description of jet data\\ from LHC Run-I and Tevatron Run-II}

\author[a,1]{Stefano Carrazza,\note{Corresponding author.}}
\author[b,c]{Jo\~{a}o Pires}

\affiliation[a]{Dipartimento di Fisica, Univerit\`a di Milano \& INFN,
  Sezione di Milano,\\Via Celoria 16, Italy}
\affiliation[b]{Dipartimento di Fisica, Univerit\`a di Genova
  and INFN Sezione di Genova,\\ Via Dodecaneso 33, I-16146
  Genova, Italy}
\affiliation[c]{Dipartimento di Fisica, Univerit\`a di Milano-Bicocca,\\ Piazza della Scienza 3, 20126 Milan, Italy}

\emailAdd{stefano.carrazza@mi.infn.it}
\emailAdd{joao.pires@mib.infn.it}

\abstract{We present a systematic comparison of jet predictions
  at the LHC and the Tevatron, with accuracy up to next-to-next-to-leading
  order (NNLO). The exact computation at NNLO is completed for
  the gluons-only channel, so we compare the exact predictions for
  this channel with an approximate prediction based on threshold resummation,
  in order to determine the regions where this approximation is
  reliable at NNLO. The kinematic regions used in this study are
  identical to the experimental setup used by recently published jet
  data from the ATLAS and CMS experiments at the LHC, and CDF and D0
  experiments at the Tevatron. We study the effect of choosing different renormalisation and
  factorisation scales for the NNLO exact prediction and as an exercise assess their impact on a PDF fit including these
  corrections. Finally we provide numerical values of the NNLO $k$-factors
  relevant for the LHC and Tevatron experiments.}

\begin{document} 

\maketitle

\flushbottom

\newpage

\section{Introduction}
\label{sec:intro}

Single jet inclusive and dijet observables are the most fundamental QCD processes measured at hadron colliders.
They probe the basic parton-parton scattering in QCD and thus allow for a determination of the parton distribution functions
in the proton and for a direct probe of the strong coupling constant up to the highest energy scales that can be attained in collider experiments. 
In particular, gluon scattering is a direct contribution to the production of high-$p_T$ jets. 
For this reason, jet data is included in PDF fits with the goal of assessing the gluon distribution in the proton at medium to large values
of the momentum fraction $x$. 

Improvements at the level of accuracy of the theoretical predictions for the single jet inclusive
cross section beyond next-to-leading order (NLO) in QCD perturbation theory have been achieved recently.
First, the exact next-to-next-to leading order (NNLO) prediction for the gluons-only channel has been published in~\cite{Currie:2013dwa,Ridder:2013mf}.
Second, an approximate NNLO prediction based on threshold resummation is presented in~\cite{deFlorian:2013qia}.

In this work we perform a systematic study comparing theoretical
predictions at leading (LO), next-to-leading (NLO) and next-to-next-to-leading
order (NNLO), to recent data from the LHC and Tevatron
experiments. The aim of this study is to understand and characterize the validity
of the NNLO threshold approximation~\cite{deFlorian:2013qia} by comparing it to the exact
computation in the gluon-gluon channel~\cite{Currie:2013dwa,Ridder:2013mf}. In
Ref.~\cite{deFlorian:2013qia} the threshold approximation is compared to the
exact calculation in the gluon-gluon channel showing, after integration
over rapidity, a good agreement at large $p_{T}$.
However, for small $p_{T}$ regions it tends to diverge from the exact
computation. Our objective is to determine the
experimental regions where this breakdown of the threshold
approximation occurs. A rejection criteria to exclude approximate predictions will be suggested based on the gluon-gluon
channel which is dominant in the small $p_{T}$ region. In this region the full NNLO 
computation is dominated by the gluon-gluon channel and therefore the predictions from the
exact NNLO calculation in this channel are reliable to determine the kinematic regions for which the threshold
terms become accurate. Moreover, and contrary to the study made in Ref.~\cite{deFlorian:2013qia}, we will
compare both predictions using the same factorisation and renormalisation scales in both calculations and discuss
the effects of making different scale choices in the theory predictions.

In order to obtain the exact NNLO predictions, the calculation in~\cite{Currie:2013dwa,Ridder:2013mf} used
the antenna subtraction scheme~\cite{GehrmannDeRidder:2005cm} to perform the cancellation of IR singularities between real
and virtual corrections at NNLO~\cite{Glover:2010im,GehrmannDeRidder:2011aa,Ridder:2012dg,Currie:2013vh}. For hadron collider observables this includes
contributions due to radiative corrections from partons in the initial state~\cite{Daleo:2006xa,Daleo:2009yj,Boughezal:2010mc,Gehrmann:2011wi,GehrmannDeRidder:2012ja}.  
The cancellation of IR singularities is achieved analytically in all intermediate steps of the calculation thereby producing a strong check on the correctness of the calculation.
In this calculation the exact two-loop~\cite{Glover:2001af,Glover:2001rd,Bern:2002tk}, one-loop~\cite{Bern:1993mq} and tree-level~\cite{Mangano:1990by} QCD matrix elements
at NNLO are included in a parton-level generator NNLOJET, which integrates them over the exact full phase space to compute any infrared
safe two-jet observable to NNLO accuracy. For the purposes of the present study we compute the single jet inclusive cross section $pp\to j +X$
where we require to observe at least one jet in the final state and integrate inclusively any additional radiation.

In Ref.~\cite{deFlorian:2013qia} approximate NNLO results for the same observable were derived using the formalism of threshold resummation for single jet production
in hadron-hadron collisions. This formalism was first developed in~\cite{Kidonakis:1998bk,Kidonakis:1998nf,Laenen:1998qw} and predictions
in a scheme where jets are assumed to be massless at the partonic threshold were produced in~\cite{Kidonakis:2000gi}.
In a study performed in~\cite{Kumar:2013hia} it was shown that the NLO terms in this scheme~\cite{Kidonakis:2000gi} 
fail to match a full NLO calculation even in a regime where threshold logarithms should dominate. On the other hand in Ref.~\cite{deFlorian:2013qia} the structure of the threshold logarithms allows jets to have a non-vanishing
invariant mass at threshold. We will compare our predictions at NLO and NNLO with the predictions obtained in the latter scheme~\cite{deFlorian:2013qia}. 
In this framework the threshold limit is defined by the vanishing of the invariant mass of the system that recoils against the observed jet $s_4=P_X^2\to0$.
In this limit the phase space available for additional soft radiation is restricted such that the higher $k$th order coefficient functions are dominated by large
logarithmic corrections,

\begin{equation}
\alpha_s^k w_{ab}^{(k)}\to\alpha_s^k \left(\frac{\log^m(z)}{z}\right)_{+},\qquad m\le2k-1,\qquad z=\frac{s_4}{s}.
\end{equation}

The next-to-leading logarithmic (NLL) threshold calculation~\cite{deFlorian:2013qia} 
then performs a systematic resummation of these logarithmic enhanced contributions for all partonic channels
to all orders in the strong coupling $\alpha_s$, by determining the
three leading logarithmic contributions $\propto$ $(\log^3(z)/z)_+$,
$(\log^2(z)/z)_+$, $(\log(z)/z)_+$ and keeping full dependence of the
cross section on the jet rapidity~\cite{deFlorian:2013qia}. The soft
contribution $\delta(z)$ as well as non-enhanced regular terms in $z$
of NNLO accuracy are not computed in this approach.  For this reason
it has been in shown in~\cite{deFlorian:2013qia} that different
approximate NNLO predictions can be derived from the threshold
formalism if the variables used in the computation differ away from
the threshold $z=0$ limit (but are otherwise identical at $z=0$). This
effect can lead to a significant change in the shape of the
approximate NNLO threshold prediction~\cite{deFlorian:2013qia} and
increases its uncertainty.

Together with the comparison between the predictions at NNLO obtained
in the threshold formalism and in the exact fixed-order calculation,
we also provide the NNLO/NLO $k$-factors relevant for the Tevatron and
the LHC experiments.  The paper is organized as follows. In
Sect.~\ref{sec:bench} we present the jet data selected for the
comparison, and the setup of the computational tools used for the
generation of the theoretical predictions. In Sect.~\ref{sec:lhc} and
Sect.~\ref{sec:tev} we show the results for the LHC and the Tevatron
experiments respectively. In Sect.~\ref{sec:PDFfit} we present as an
exercise an aNNLO PDF fit, where ``a'' stands for approximate, i.e. we
use only approximate NNLO $k$-factors computed in Sect.~\ref{sec:lhc}
and Sect.~\ref{sec:tev} and not the full NNLO predictions, since the
exact full prediction is not available yet. In this paper the aNNLO
notation is used in order to reserve the terminology NNLO PDF fit for
when the full NNLO predictions for jet production are available. In
Sect.~\ref{sec:conclusion} we present our conclusions and directions
for future work. An appendix is enclosed which provides tables with
$k$-factors in the gluon-gluon channel at the LHC and the Tevatron.

\section{Benchmark predictions for jet production}
\label{sec:bench}

\subsection{LHC and Tevatron jet data}

In order to provide realistic comparisons, based on real data which is
already included in the extractions of parton distribution functions
(PDFs)~\cite{Ball:2012cx}, we have selected recent data sets obtained
during the LHC Run-I and the Tevatron Run-II.  Using data from both
colliders provides the possibility of investigating differences and
similarities between datasets for different collision energies and
kinematic coverage. A summary of the experimental data included in our
analysis is presented in Table~\ref{tab:experiments}. As we will show
in Sect.~\ref{sec:lhc} and Sect.~\ref{sec:tev}, the region of validity
of the threshold approximation is very dependent on the experimental
details.

From the LHC experiments we have included the CMS measurements of
the double differential jet cross sections at $\sqrt{s}=7$
TeV~\cite{Chatrchyan:2012bja}, where jets are reconstructed up to
$|\eta|<2.5$. We have also included the ATLAS measurements of
inclusive jet cross sections at $\sqrt{s}=7$ TeV~\cite{Aad:2010ad}
and $\sqrt{s}=2.76$ TeV~\cite{Aad:2013lpa}, where the rapidity
coverage reaches $|\eta|<4.4$. For both LHC experiments jets are
reconstructed with the anti-$k_{t}$ algorithm. The main differences
between CMS and ATLAS data is the choice of jet resolution parameter R,
which is R=0.7 for CMS and R=0.4 for ATLAS, and the $p_{T}$
coverage which for CMS covers the very high $p_{T}$ region, reaching
2~TeV, while ATLAS measures very low $p_{T}$ jets starting
from 20~GeV.

Concerning the Tevatron data, we have included the
most recent CDF Run-II $k_{t}$ jets~\cite{Abulencia:2007ez} and the D0
Run-II cone data~\cite{Abazov:2011vi}. In contrast to LHC data, the
center of mass energy of both sets is $\sqrt{s}=1.96$ TeV and their
coverage in rapidity and $p_{T}$ is smaller than ATLAS and CMS
experiments. It is important to highlight that CDF uses the $k_{t}$
algorithm to do the jet reconstruction, while D0 presents 
data reconstructed with the MidPoint cone algorithm which
is infrared unsafe at NNLO.

\begin{table}
  \begin{centering}
    \begin{tabular}{cc|c|c|c|c|c|c}
      \hline \multicolumn{2}{c|}{Data Set} & Points & $\sqrt{s}$ (TeV) & $R$ & $\eta$ coverage
      & $p_{T}$ coverage & Ref.\tabularnewline 
      \hline \hline
      \multirow{3}{*}{LHC} & CMS 2011 & 133 & 7 & 0.7 & $|\eta|<2.5$ & $[114,2116]$ GeV & ~\cite{Chatrchyan:2012bja} \tabularnewline
      \cline{2-8} & ATLAS 2010 & 90 & 7 & 0.4 & $|\eta|<4.4$ & $[20,1500]$ GeV & ~\cite{Aad:2010ad} \tabularnewline 
      \cline{2-8} & ATLAS 2011 & 59 & 2.76 & 0.4 & $|\eta|<4.4$ & $[20,500]$ GeV & ~\cite{Aad:2013lpa} \tabularnewline 
      \hline \hline
      \multirow{2}{*}{Tevatron} & CDF $k_{t}$& 76 & 1.96 & 0.7 & $|\eta|<2.1$ &$[54,700]$ GeV & ~\cite{Abulencia:2007ez} \tabularnewline
      \cline{2-8} & D0 cone & 110 & 1.96 & 0.7 & $|\eta|<2.4$ & $[50,665]$ GeV & ~\cite{Abazov:2011vi} \tabularnewline 
      \hline
    \end{tabular}
    \par\end{centering}
    \caption{\label{tab:experiments}Jet data included in the current
      analysis with the respective kinematic information.}
\end{table}

\subsection{Theoretical predictions}
\label{sec:benchTH}

Theoretical predictions presented in this work are computed
exclusively with the central value of the {\tt
  NNPDF23\_nnlo\_as\_0118} set, presented by the NNPDF collaboration
in Ref.~\cite{Ball:2012cx}. This set is used for predictions at all
perturbative orders. However, we are interested in comparing
predictions at the same order and thus the choice of the input PDF is
only marginally relevant.

At LO and NLO full exact predictions are available and these have been computed with the {\tt
  FastNLO}~\cite{Kluge:2006xs} interface for CMS, CDF and D0 and with
the {\tt APPLgrid}~\cite{Carli:2010rw} tables for the ATLAS
predictions. Tables used with both interfaces have been computed with
{\tt NLOjet++} program~\cite{Catani:1996vz,Nagy:2003tz}. Predictions at NNLO are computed using the exact fixed-order results in the gluon-gluon channel
and with the threshold approximation code.

To obtain the exact predictions at NNLO we use the parton level Monte Carlo NNLOJET code recently presented in
Ref.~\cite{Currie:2013dwa,Ridder:2013mf,Currie:2014upa} interfaced with {\tt libHFILL}\footnote{Available at: \url{http://libhfill.hepforge.org/}}, a histogram library
developed for this work and compatible with all MCs programs which
allows the automatic construction of jet $p_{T}$ distributions
from event weights, using the binning and kinematic regions presented
in Table~\ref{tab:experiments}. The Monte Carlo uncertainties
presented for the exact predictions are below the percent level. 
In this code, the $gg\to gg+X$ at full colour and the $q\bar{q}\to gg+X$~\cite{Currie:2013vh,Currie:2014upa} contributions at leading colour
are available at NNLO and the current limitations are the missing
partonic contributions for $qg$ and $qq$ scattering. 

To obtain the approximate NNLO predictions based on threshold
resummation we use the threshold approximation
code~\cite{deFlorian:2013qia} which implements predictions for all
channels.  We have also used the narrow-jet approximation code (NJA)
presented in Ref.~\cite{Jager:2004jh,Mukherjee:2012uz}, which computes
in analytic form the single jet inclusive cross section at LO and NLO
in the narrow-jet limit where both the matrix elements and phase space
are expanded around the narrow-jet limit. The original version of the
NJA and threshold codes have been improved through comparisons with
exact calculations and updated in order to use the {\tt
  LHAPDF}~\cite{Whalley:2005nh} interface to PDFs and by including the
bottom and the anti-quark PDFs contributions to the total
luminosity. After these modifications both codes show full agreement
at LO with the exact calculations and can be used for comparisons with
experimental data.

For all predictions the value of $\alpha_s$ is provided by the PDF set through the {\tt LHAPDF}~\cite{Whalley:2005nh} interface.
For the exact calculation we generate predictions using two different dynamical renormalisation and factorisation scales. One choice evaluates the
fixed-order single jet inclusive cross section using $\mu_R=\mu_F=\mu=p_{T1}$ where for each event the renormalisation and factorisation
scales are set equal to each other and equal to the $p_T$ of the leading jet in the event. The $p_T$ of the leading jet is obtained after clustering all parton final-state momenta
into jet momenta using the appropriate jet algorithm employed by each experimental setup. Whenever the jet algorithm determines a partonic clustering the 4D recombination scheme is applied, i.e.,
the 4-momenta of the jet's constituents is added, producing a list of final-state jets that are ordered in $p_T$ at the end of the clustering procedure.
As a second choice we computed the fixed-order single jet inclusive cross section using $\mu_R=\mu_F=\mu=p_{T}$ where in this case each jet in every event
is binned with the weight evaluated at the scale $p_T$ of the jet. While at LO the two final state partons generate two jets with equal transverse momentum
$p_{T1}=p_{T2}=p_{T}$ and the two scale choices coincide, radiative corrections can generate subleading jets and the effects of the different scale choice in the theory prediction
become apparent at NLO and NNLO.

The approximate threshold and the NJA predictions use
$\mu_R=\mu_F=\mu=p_{T}$, where the $p_T$ of the observed jet is
generated by the MC integration and the general structure of the
resummed cross section applies specifically to the anti-$k_T$
algorithm~\cite{deFlorian:2013qia}.

In the next section we present the theoretical predictions for all the experimental setups in Table~\ref{tab:experiments} at each order in perturbation
theory up to NNLO and perform a benchmark comparison of the various approximations. 

\begin{figure}[t]
  \begin{centering}
    \includegraphics[scale=0.35]{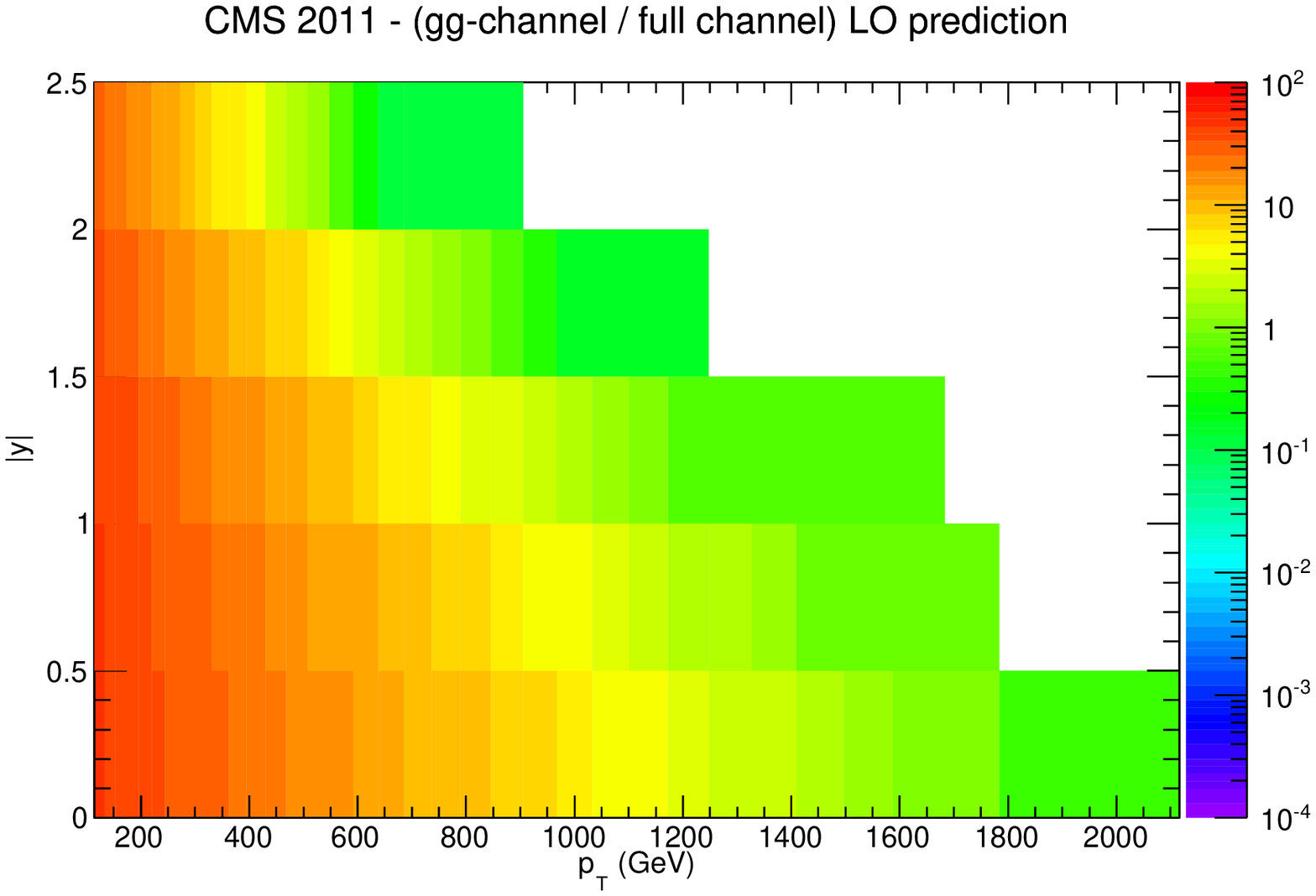}\includegraphics[scale=0.35]{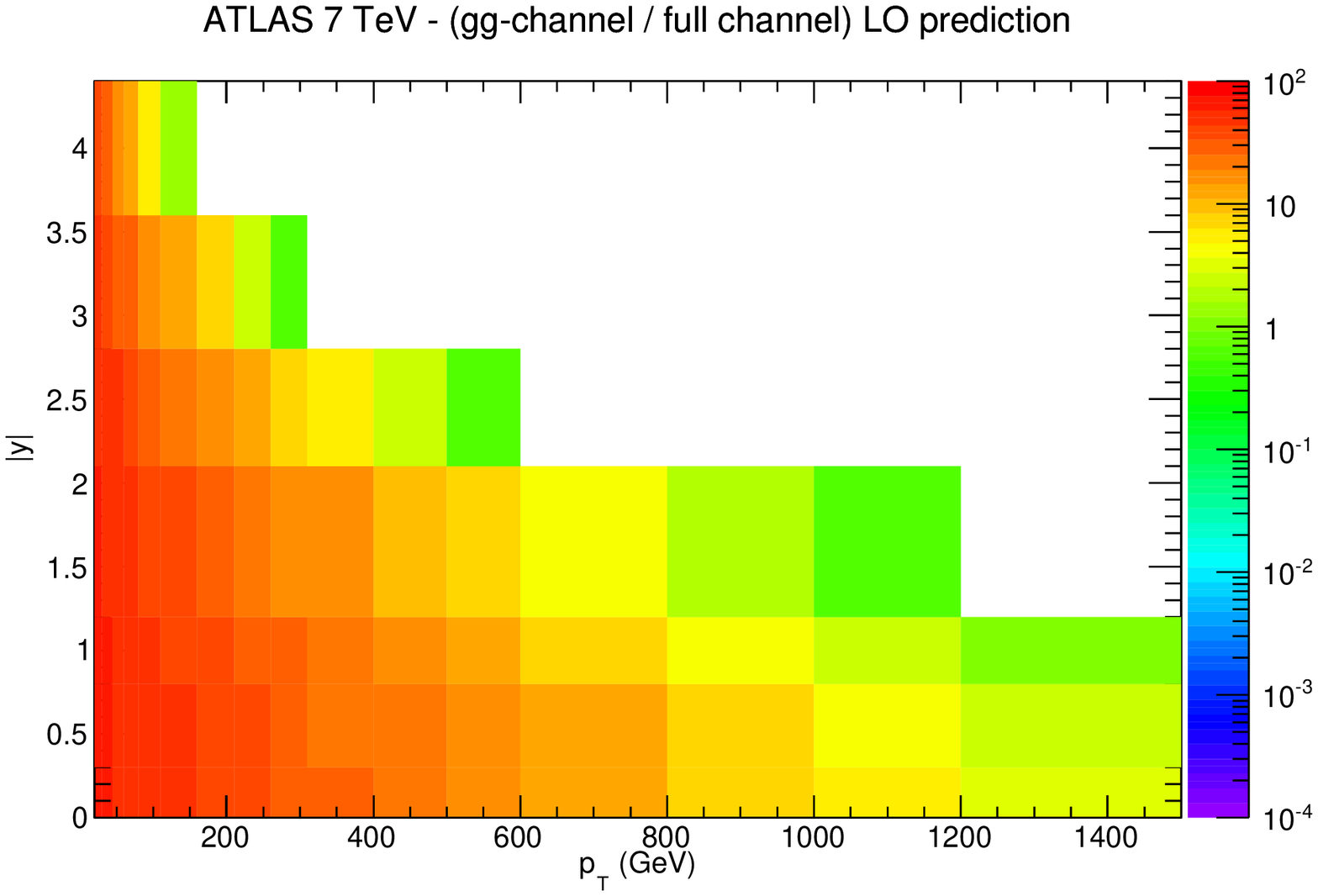}
    \par\end{centering}
  \begin{centering}
    \includegraphics[scale=0.35]{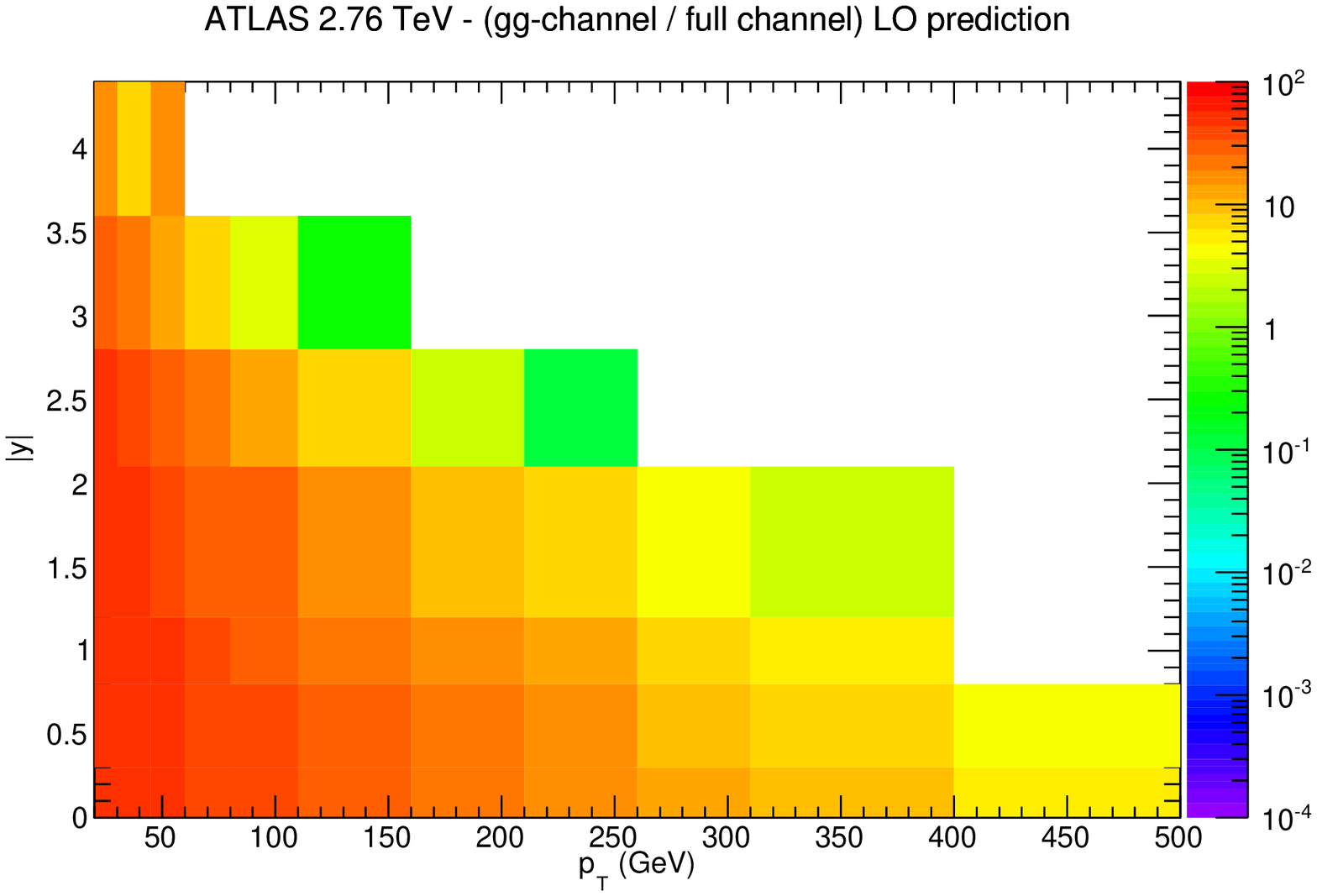}\includegraphics[scale=0.35]{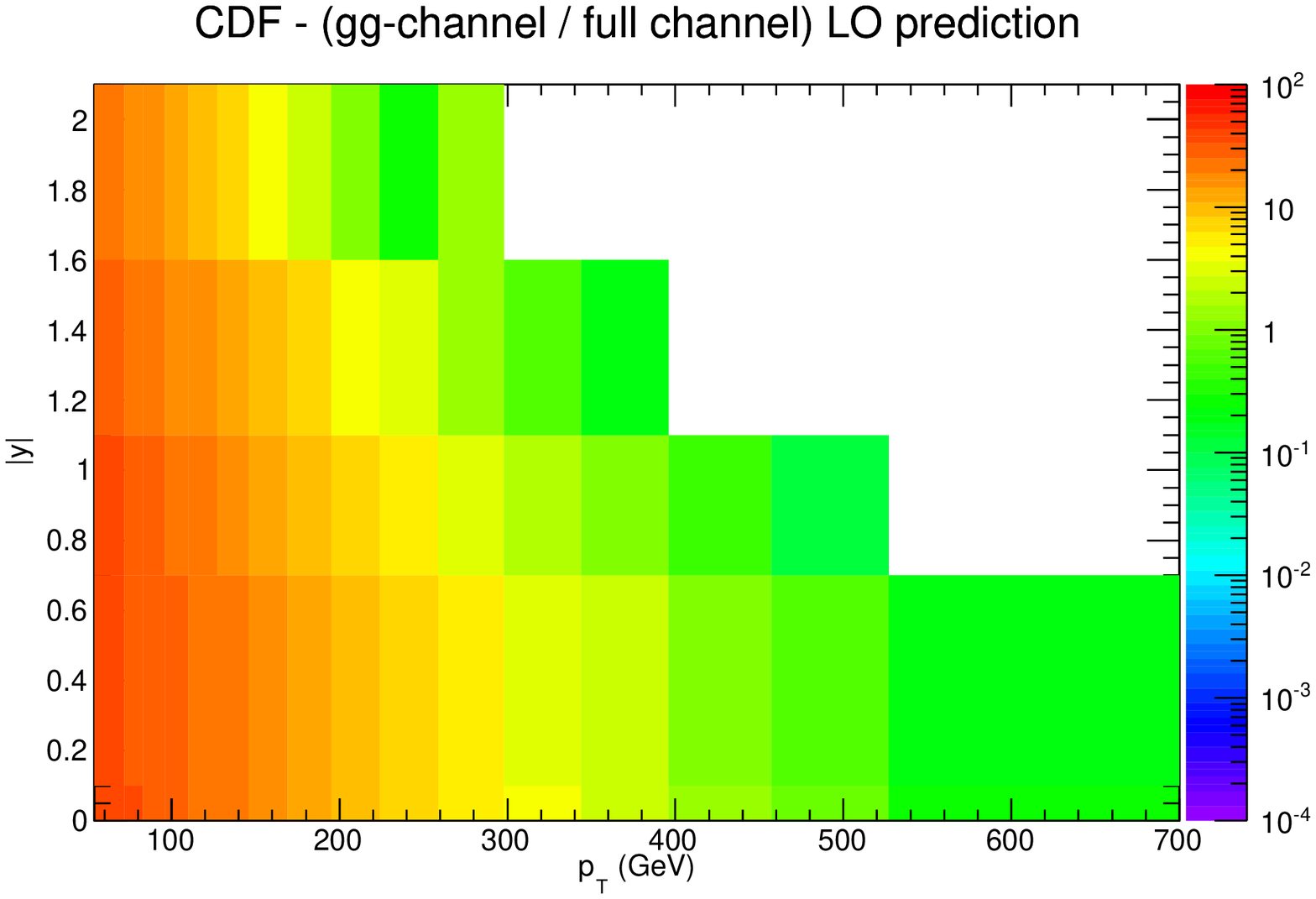}
  \par\end{centering}
\caption{
  \label{fig:ggfraction} Relative percentage-wise contribution of the $gg$-channel to the full all-channel hadronic jet production 
  as a function of $p_{T}$ and $|y|$ for CMS, ATLAS 7 TeV and 2.76 TeV and CDF bins.}
\end{figure}

When assessing the validity of the threshold prediction
at NNLO we will suggest a rejection criteria which is to exclude approximate predictions which are more than 10\%
off the exact prediction. We will apply this criteria in the gluons-only channel to
help determine exactly the regions in the experimental setup where the threshold approximation is applicable. In Fig.~\ref{fig:ggfraction} it is shown,
using the full LO prediction, the relative contribution of the $gg$-channel
in the various ($p_T$,$|\eta|$) regions of each experimental setup to the full result. We observe that in the high-$p_T$ region of each experiment the 
gluon-gluon scattering contribution is highly suppressed. The high-$p_{T}$ region corresponds precisely to the threshold region where the phase space
available for additional radiation is limited and for this reason the dominance of the gluon-gluon channel cannot be used as an argument to assess
the validity of the threshold approximation. However, by looking at the relative error of NLO-threshold vs. NLO exact for different partonic channels we
observe a convergence of the threshold approximation to the exact result at high-$p_T$ in each channel. Since the same comparison can be performed
reliably at NNLO in the $gg$-channel only, the criterium mentioned above to exclude approximate predictions is not a recommendation 
and its purpose is to fix a level of accuracy when using approximate predictions, while the full calculation is not avaliable.
In Sect.~\ref{sec:PDFfit} we will discuss the effects of being more restrictive or flexible with this choice.

\section{LHC jet data}
\label{sec:lhc}
\begin{figure}
  \begin{centering}
    \includegraphics[scale=0.8]{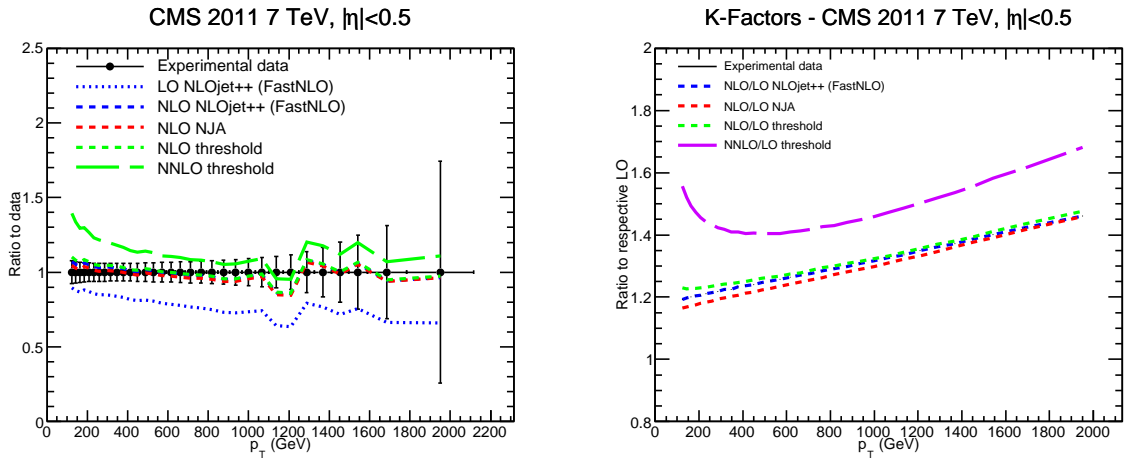}
    \par\end{centering}
    \caption{\label{fig:cmsfull} Comparison between CMS
      data and theoretical predictions computed with the exact and
      approximate codes for the first bin in $\eta$. On the left
      plot the theoretical predictions are normalized to the CMS measurement. On the
      right plot, $k$-factors for higher order cross sections over the leading order (LO) are presented. }
\end{figure}

For each experiment in Table~\ref{tab:experiments} we have generated all channel full LO, NLO and approximate NNLO predictions
and compared them directly against the experimental data. In
Figure~\ref{fig:cmsfull} we show an example of this analysis for the
first rapidity bin of the CMS jets 2011 dataset. On the left plot of
Fig.~\ref{fig:cmsfull}, the full channel theoretical predictions are
normalized to the CMS data, corrected by non-perturbative corrections,
where uncertainties are estimated from the diagonal of the covariance
matrix, which is extracted by considering systematic uncertainties
additively. 

From this plot we observe that the data is well described by the NLO predictions. We note that the NNPDF2.3 set used in this comparison is obtained from a NNLO fit that
includes jet data from the Tevatron and the LHC for which the corresponding theory predictions are known presently only to NLO accuracy. For this reason,
higher order theory effects beyond NLO are not taken into account in the jet prediction used in the fit. As a result, we observe that the approximate NNLO prediction based on threshold resummation
predicts a cross section above the data indicating the need to consistently include NNLO jet predictions in NNLO PDF fits of jet data. Part of this excess could be due to the inherent
approximated nature of this prediction and for this reason we aim to disentangle in the next sections the regions which correspond to a breakdown of the threshold approximation.

On the right plot of Fig.~\ref{fig:cmsfull} we quantify the size of
the higher order corrections by computing ratios of higher order cross
sections over the leading order one. These $k$-factors show that NLO
corrections vary between 20\% and 45\% with respect to the LO
prediction with the approximate NNLO threshold corrections varying
between 40\% and 70\%.  We also observe that the NLO/LO $k$-factors
using the threshold (in green) and NJA (in red) codes are in good
agreement with the exact computation (in blue).

\begin{figure}
  \begin{centering}
    \includegraphics[scale=0.4]{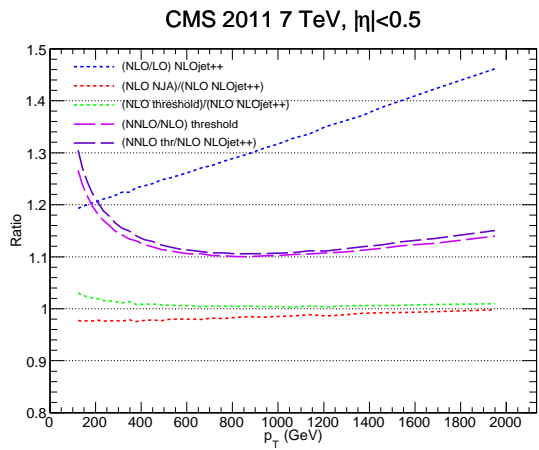}
    \par\end{centering}
    \caption{\label{fig:cmsfullratio} Ratios between exact and approximate
      predictions at the same order in perturbation theory. The level of agreement at NLO takes into account all partonic channels in the predictions.
      In the same plot we present the 
      exact NLO/LO $k$-factor (all partonic channels included) and the NNLO/NLO $k$-factors produced
      by the threshold approximation code (all partonic channels included).}
\end{figure}

A more detailed comparison between the all channels exact and approximate predictions
is presented in Figure~\ref{fig:cmsfullratio}. We conclude that
as expected the NLO threshold (in green) and the NJA (in red) predictions converge to the exact computation
for high values of $p_{T}$. The high-$p_T$ region corresponds precisely to the threshold region $s_4=0$ where
the phase space available for additional radiation is limited. In the low-$p_T$ region we observe instead
an instability in the approximate predictions.

In the same figure we present the exact NLO/LO $k$-factor (in blue) and with 
the approximate NNLO prediction we have constructed NNLO/NLO $k$-factors using two different choices for the
denominator: the approximate NLO threshold (in light magenta) and the NLO exact
(in dark magenta). As we can see
the effects due to this choice are at the few percent level, negligible in
comparison to the size of the approximate NNLO threshold correction.

We have performed this exercise for all experiments, however, in the
next subsections we limit the analysis to the gluon-gluon channel because we are
interested in determining the regions where the NNLO threshold
$k$-factors are in agreement with the exact computation, which is
available for that channel. The full report with all plots and tables
of $k$-factors is available on-line at:
\begin{center}
{\bf \url{http://libhfill.hepforge.org/JetStudy2014}~}
\end{center}
and a short summary of the main results which identify the experimental regions where approximate predictions
should be discarded is provided in Sect.~\ref{sec:PDFfit}.

\subsection{CMS jets}

\begin{figure}[!h]
  \begin{centering}
    \includegraphics[scale=0.4]{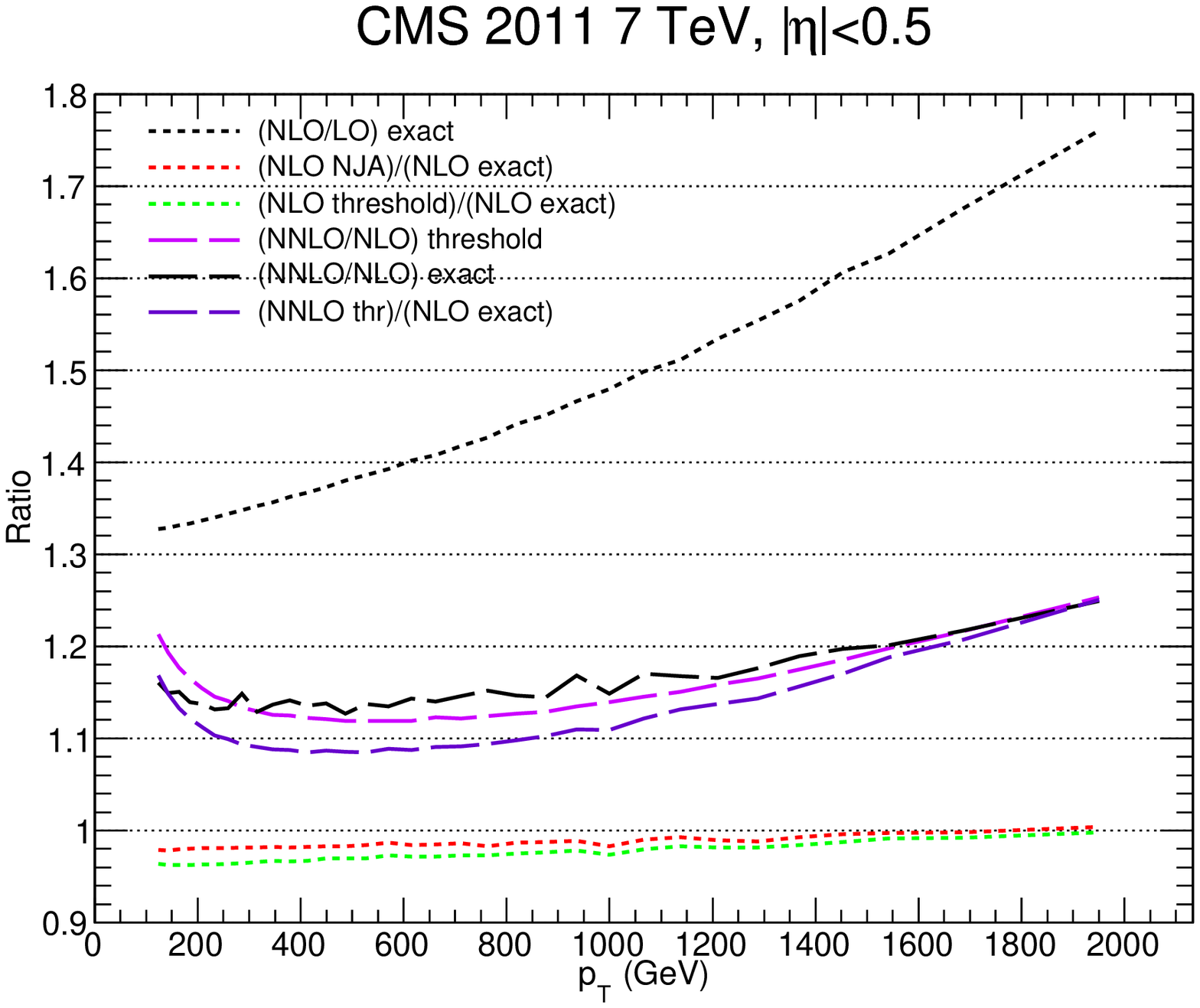}\includegraphics[scale=0.4]{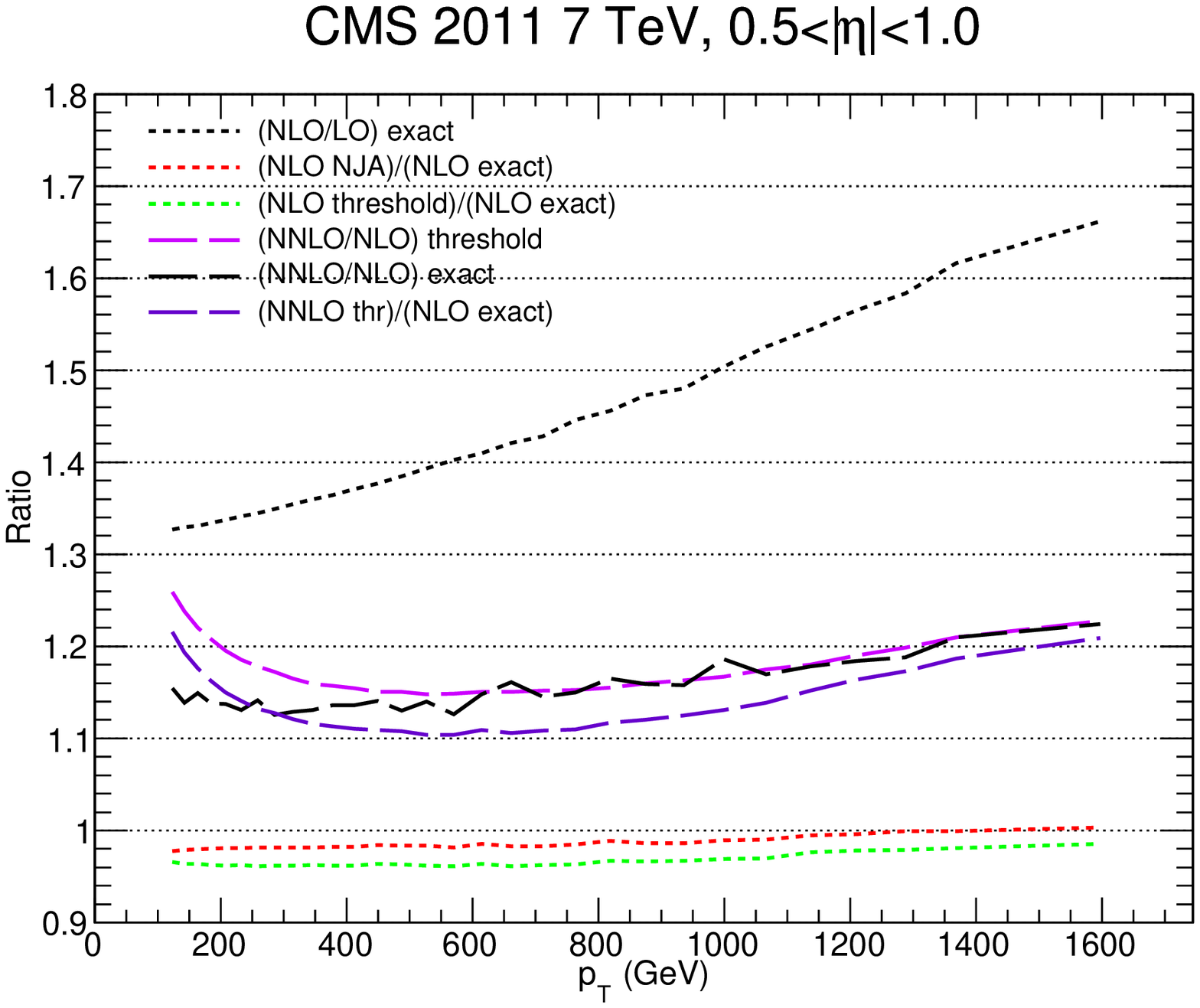}
    \par\end{centering}
  \begin{centering}
    \includegraphics[scale=0.4]{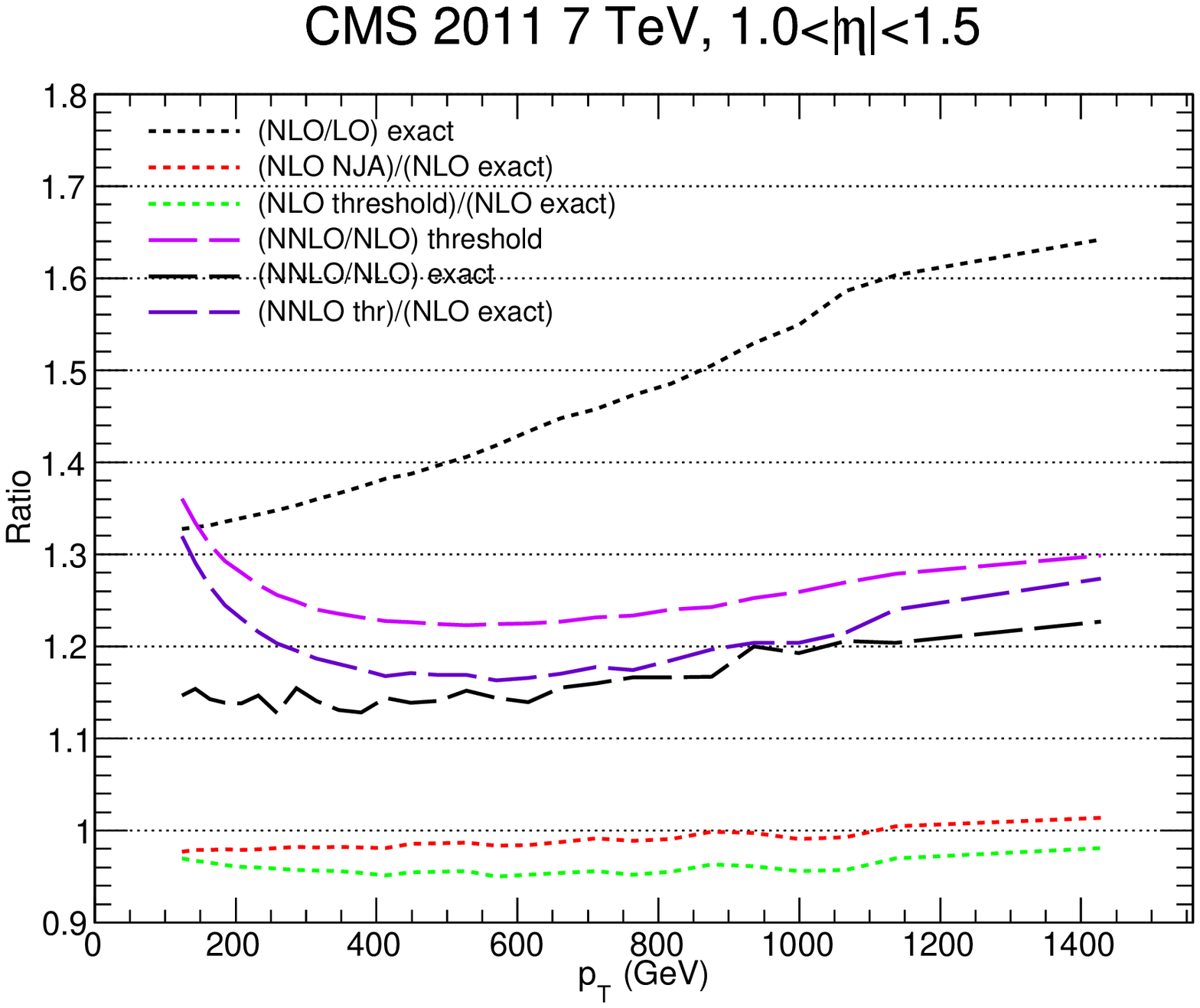}\includegraphics[scale=0.4]{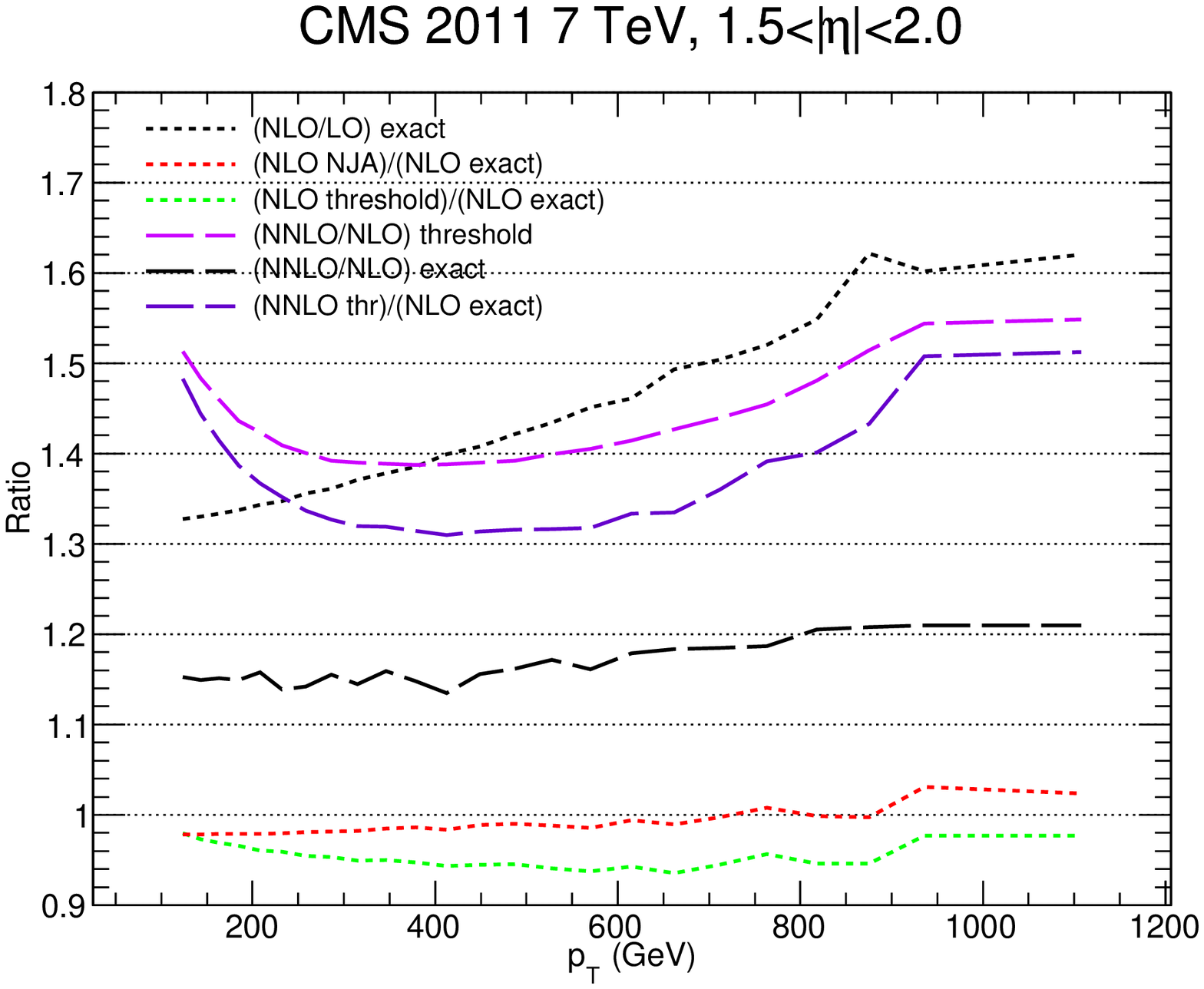}
  \par\end{centering}
  \begin{centering}
    \includegraphics[scale=0.4]{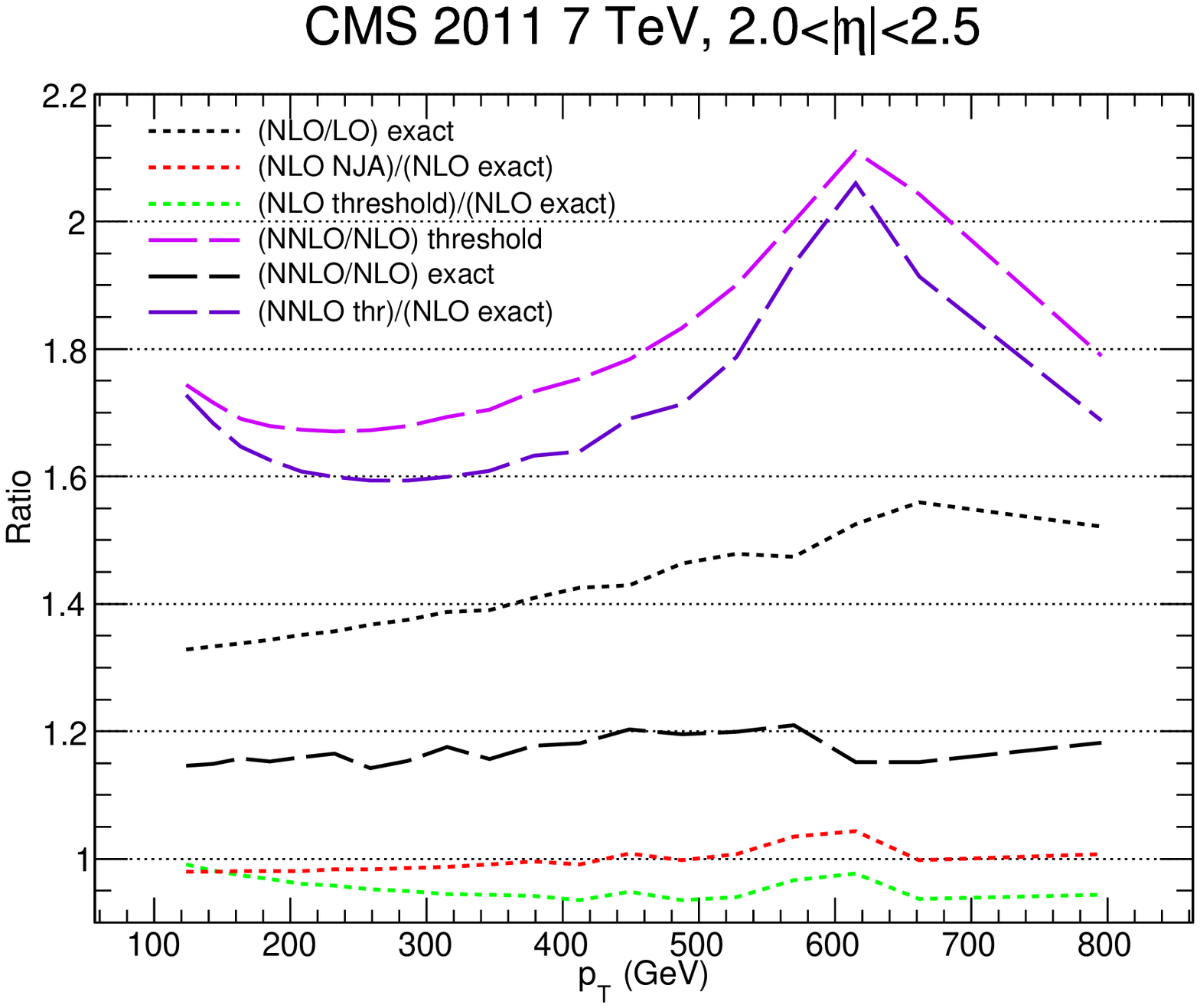}
  \par\end{centering}      
  \caption{\label{fig:cmsratio} Ratios between exact and approximate
      predictions at the same order (LO and NLO) in perturbation theory in the gluons-only channel.
      In the same plot we present the 
      exact NLO/LO and NNLO/NLO $k$-factors (gluons-only channel) and the NNLO/NLO $k$-factors produced
      by the threshold approximation code (gluons-only channel) for the CMS 2011 jet binning.}
\end{figure}

In Figure~\ref{fig:cmsratio} we show the ratios between predictions in the gluons-only channel
computed with the codes presented in Sec.~\ref{sec:bench} using the
rapidity and $p_{T}$ bins of the CMS jets 2011 dataset.
The LO, NLO, NNLO predictions labelled exact are obtained from the Monte Carlo
NNLOJET presented in~\cite{Currie:2013dwa,Ridder:2013mf,Currie:2014upa} and are compared 
with the NJA code~\cite{Jager:2004jh,Mukherjee:2012uz}
at NLO and with the threshold code~\cite{deFlorian:2013qia} at NLO and NNLO.
For all predictions we set the renormalisation and factorisation scales equal to each other
and equal to the $p_T$ of each individual jet in every event ($\mu_R=\mu_F=p_T$)

For all plots we first check the agreement at NLO between all
codes and then the agreement of the NNLO/NLO $k$-factors obtained with the threshold
approximation code and the exact computation. As we did in the
previous section, we provide two definitions for the NNLO threshold
$k$-factor by dividing the approximate NNLO threshold predictions with
the approximate NLO threshold predictions (in light magenta long-dashed curves) and also
by dividing them with the exact
NLO prediction (in dark magenta long-dashed curves). The NNLO/NLO $k$-factor
using the exact computation at NLO and NNLO is plotted in long-dashed black curves. The distance
between the long-dashed black curve and the long-dashed magenta curves in Figure~\ref{fig:cmsratio} and subsequent Figures 
indicates the level of disagreement between the $k$-factors produced by the exact NNLO computation
and the approximate NNLO threshold computations. 

By looking at Figure~\ref{fig:cmsratio} we conclude that at NLO the NJA 
code shows percent level differences at small $p_{T}$. Similarly the NLO
threshold code shows percent level differences at small $p_T$ which rise to
5\% at central $p_{T}$ for the last bins in rapidity.

Concerning the NNLO predictions we looked at the NNLO $k$-factors and
relative differences bin by bin. The relative difference between the exact computation and the threshold computations 
is documented in Tables~\ref{tab:kcms1} to~\ref{tab:kcms5} where, for each rapidity slice
of the experiment, we show for each $p_T$ bin, the experimental cross section, the experimental
error, the gluons-only exact NNLO and threshold NNLO $k$-factors together with their relative
percentage wise difference and finally the percentage wise relative difference between the two possible NNLO gluons-only
threshold $k$-factors. 

We first notice that for the entire kinematic range of the experiment the choice of the
denominator for the NNLO threshold $k$-factor produces relative differences which are
much smaller than the difference to the exact $k$-factor.
When comparing either of these with the exact NNLO computation we find for all rapidity bins an instability
at low-$p_T$ in the approximate NNLO results. In these regions the approximate NNLO threshold $k$-factor starts to rise
generating large perturbative corrections. While for the first two bins in rapidity $|\eta|<1.0$, the relative differences with 
the exact calculation are below 10\% we observe strong deviations for $|\eta|>1.0$.

Using the rejection criteria suggested at the end of Section~\ref{sec:benchTH} we conclude that for CMS
the NNLO threshold prediction is not applicable for the rapidity slices $|\eta|>1.5$ as
relative differences with respect to the exact computation are larger than 20\% and can rise up to 60\%. 
Furthermore, due to the instability of the approximate prediction at low-$p_T$, 
for the rapidity slice $1.0<|\eta|<1.5$ the first seven $p_T$ bins should be excluded.  

\begin{figure}
  \begin{centering}
    \includegraphics[scale=0.4]{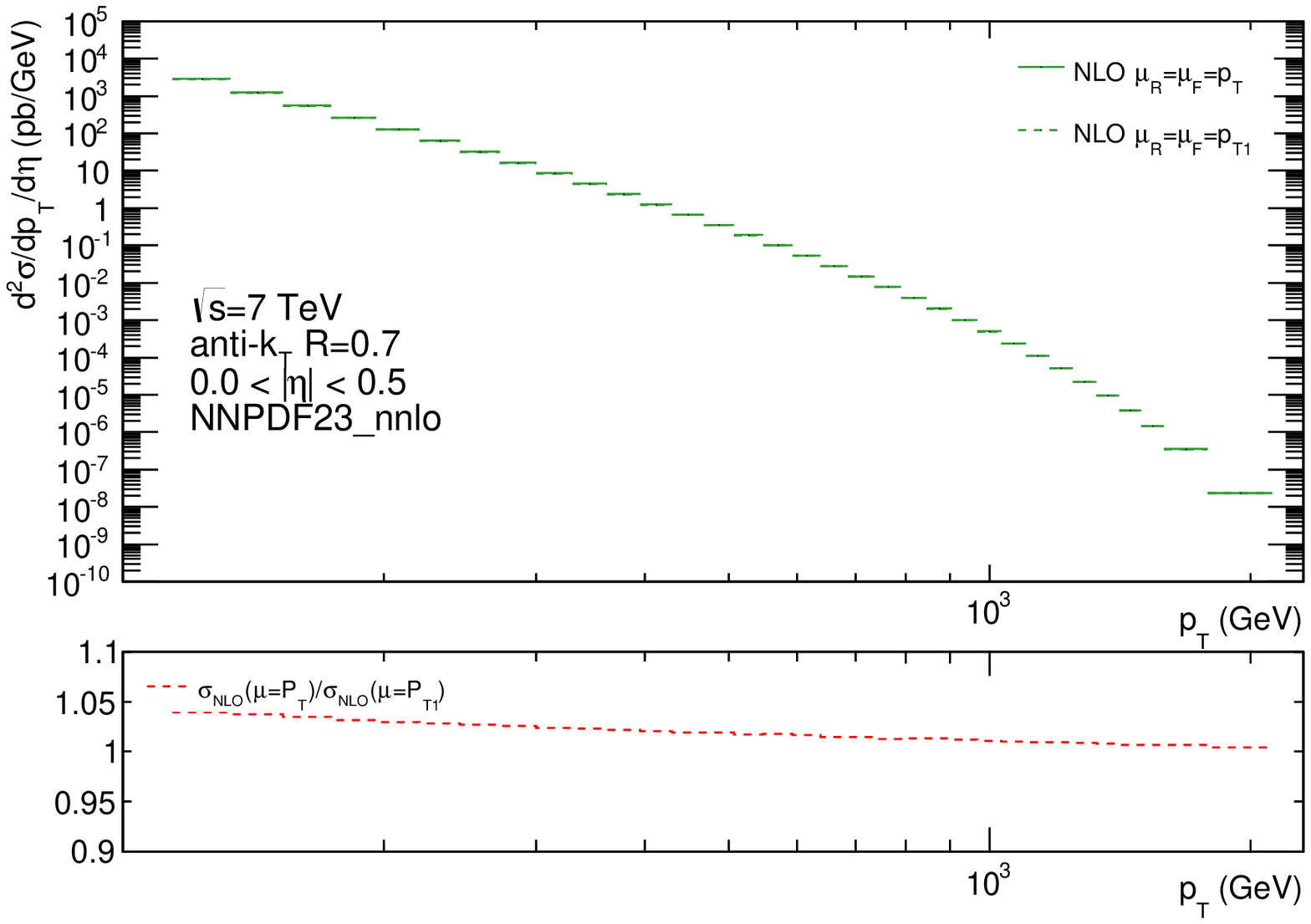}\includegraphics[scale=0.4]{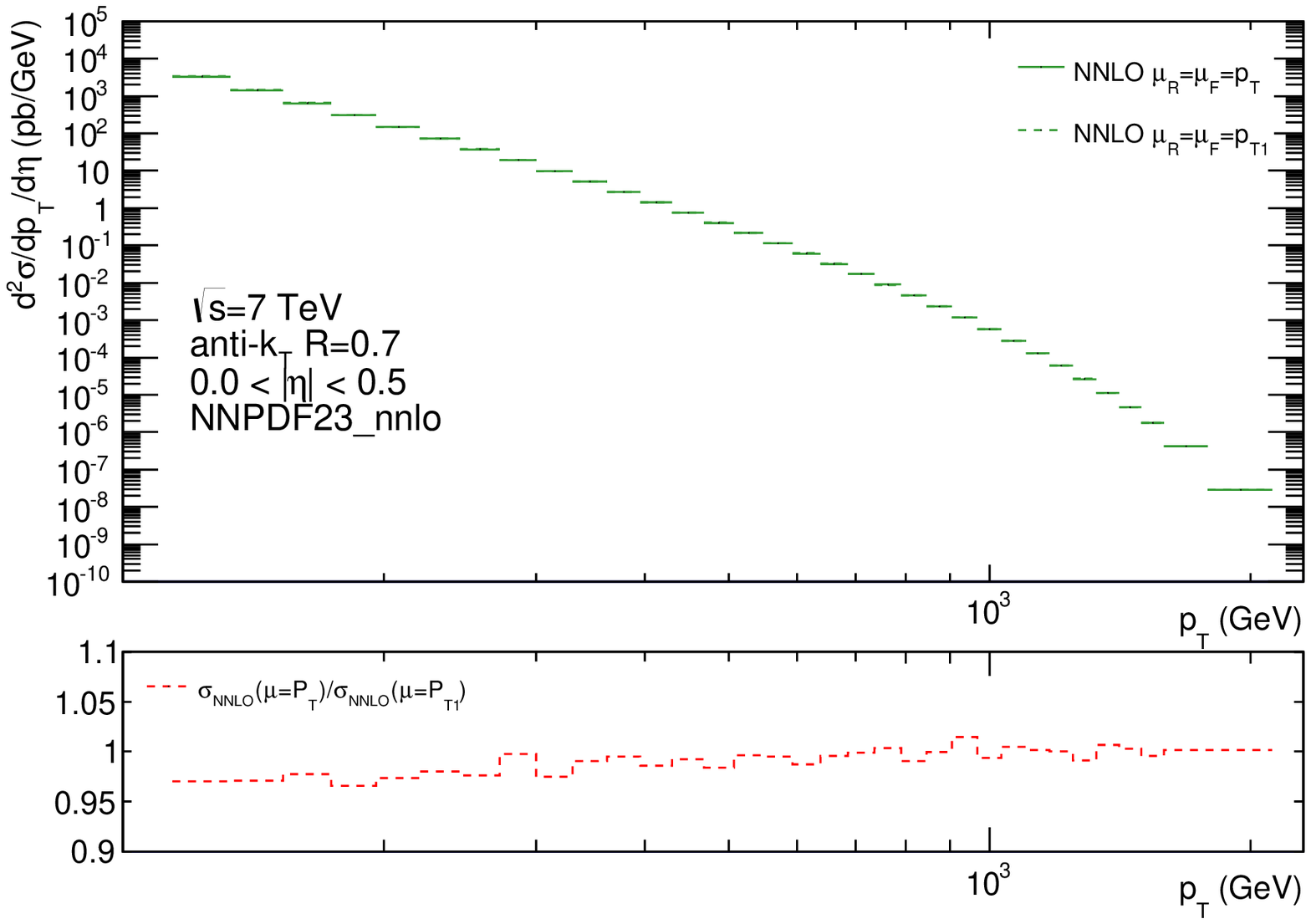}
    \par\end{centering}
  \caption{\label{fig:CMSPTvsPT1} NLO (left) and NNLO (right) exact $gg$-channel predictions for CMS evaluated with the renormalisation and factorisation scales
 $\mu_{R}=\mu_{F}=p_T$ and $\mu_{R}=\mu_{F}=p_{T1}$. In the lower pads we present the relative differences due to the different central scale choice.}
\end{figure}

As mentioned in the introduction, the comparison between the exact
fixed-order calculation and the threshold approximation performed
in~\cite{deFlorian:2013qia} used different central scale choices for
each prediction. The predictions from the threshold resummation
formalism were obtained using $\mu_R=\mu_F=p_T$, where $p_T$ is the
individual jet $p_T$ while the fixed-order calculation used
$\mu_R=\mu_F=p_{T1}$, where $p_{T1}$ is the transverse momentum of the
leading jet in each event. In order to eliminate this inconsistency,
and also to study the impact of the central scale choice in the
fixed-order predictions, we show in Figure~\ref{fig:CMSPTvsPT1} NLO
and NNLO gluons only cross sections evaluated at the two different
scales for the first rapidity slice of the CMS experiment. We observe
that at high-$p_T$ subleading jets tend to be soft and in this region
$p_{T1}\sim p_{T}$ and the predictions using either scale choice
coincide. In the low-$p_T$ region we observe low-$p_T$ jets
accompanying a high-$p_T$ object. In this case $p_{T} < p_{T1} $ and
we observe an increase in the NLO prediction of about 5\% when using
the scale $\mu_R=\mu_F=p_T$. This effect is due to the fact that the
virtual contribution which has Born kinematics remains identical with
either scale choice while the real radiation contribution is enhanced
when its weight is computed at a lower scale. At NNLO we observe
instead a reduction of the size of the fixed-order prediction at
low-$p_T$. As a consequence we conclude that the NNLO/NLO $k$-factor
is typically smaller in the low-$p_T$ region with the scale choice
$\mu_R=\mu_F=p_T$ as compared to the choice $\mu_R=\mu_F=p_{T1}$.  The
resulting reduced NNLO $k$-factor for the exact prediction then shows
that the disagreement between the exact calculation and the threshold
calculation is enhanced when both calculations are performed using the
same central scale choice. This observation is rapidity independent as
emissions in events with a high-$p_T$ central object can produce
low-$p_T$ jets entering the single jet inclusive $p_T$ distribution in
the forward regions.

\subsection{ATLAS jets at $\sqrt{s}=7$ TeV and $\sqrt{s}=2.76$ TeV}

The main difference between ATLAS and CMS jets is the different
kinematic coverage as shown in Table~\ref{tab:experiments}. ATLAS provide jet data at two distinct center of
mass energies and kinematic ranges.

\begin{figure}[H]
\begin{centering}
\includegraphics[scale=0.38]{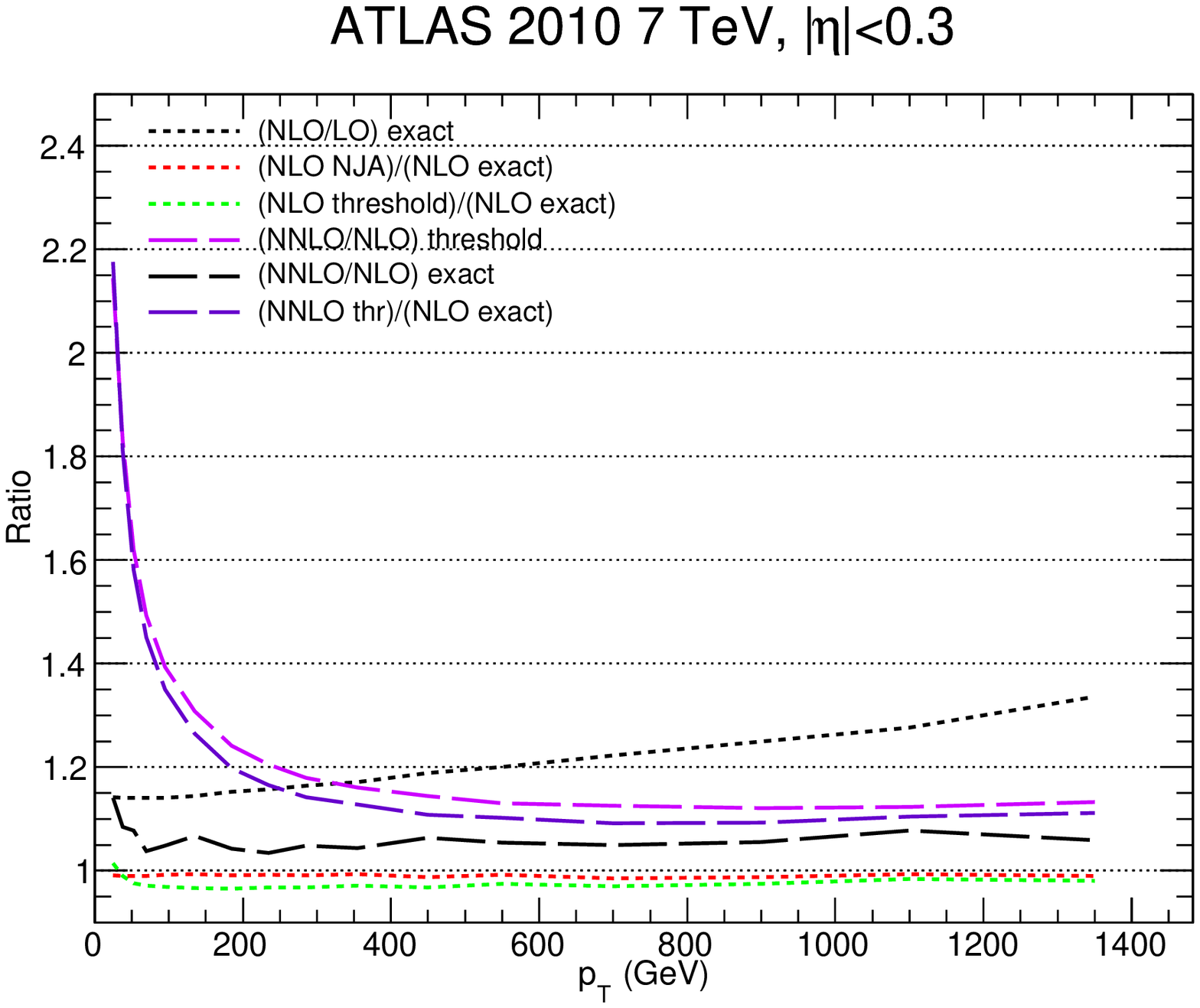}\includegraphics[scale=0.38]{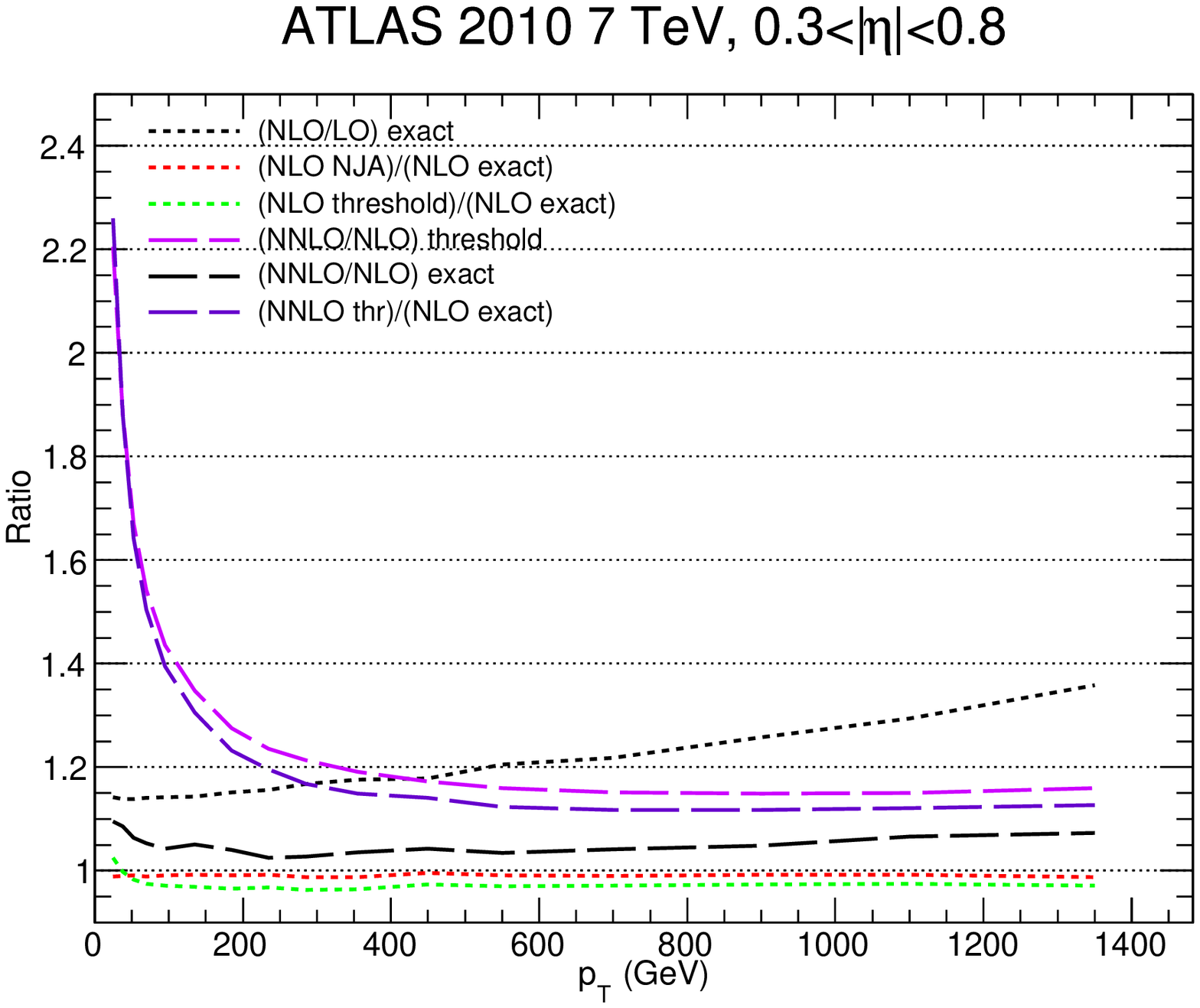}
\par\end{centering}

\begin{centering}
\includegraphics[scale=0.38]{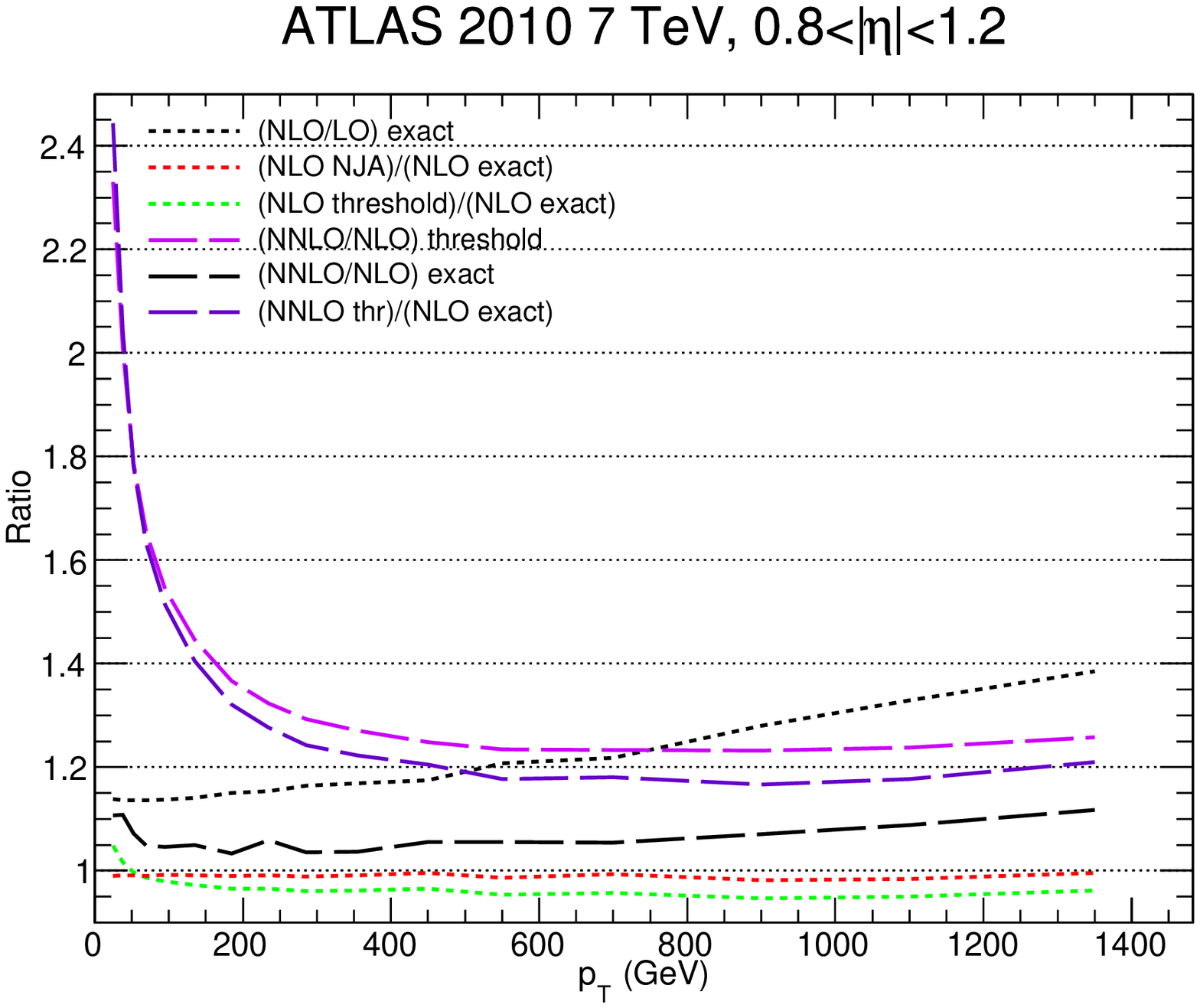}\includegraphics[scale=0.38]{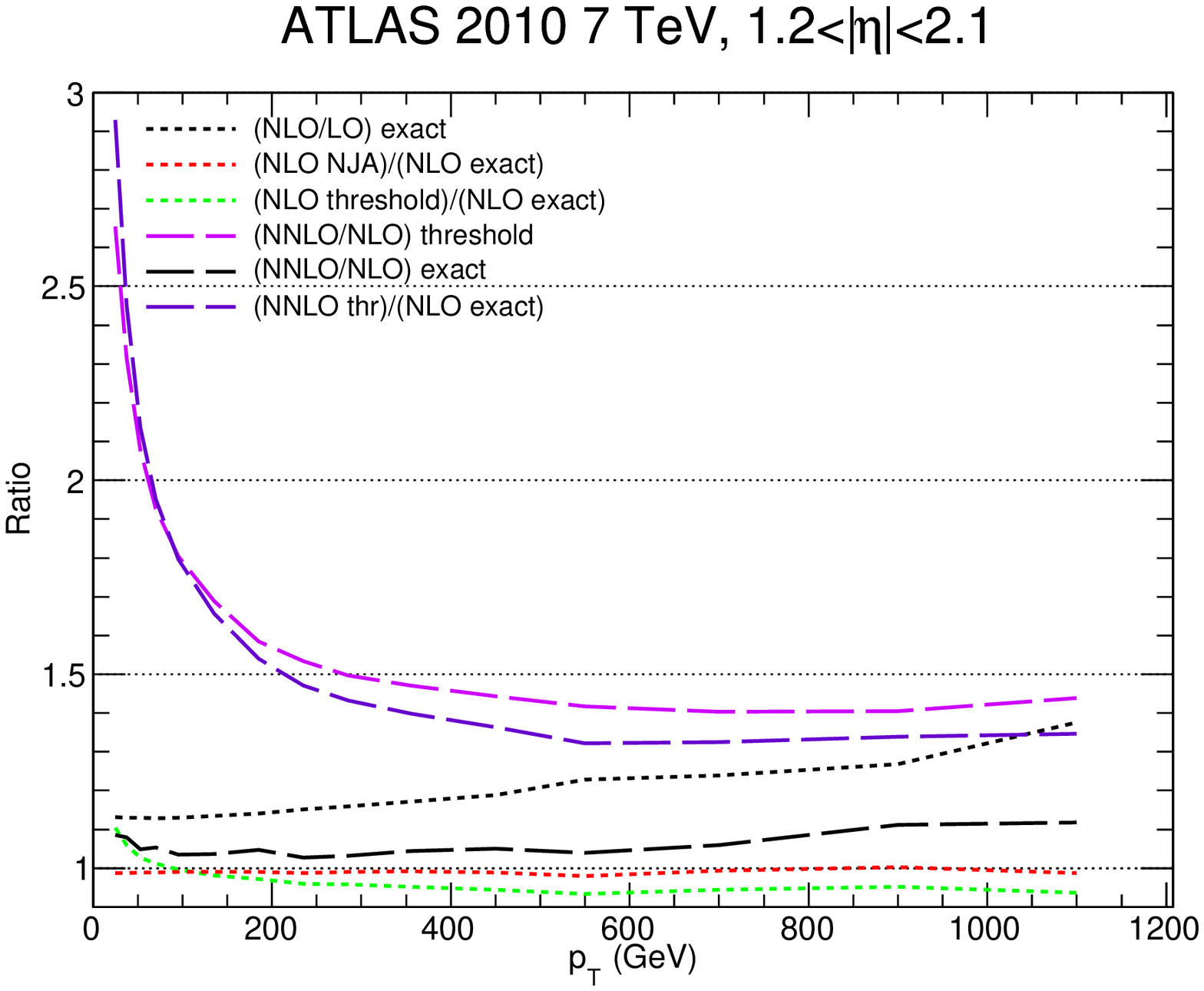}
\par\end{centering}

\begin{centering}
\includegraphics[scale=0.38]{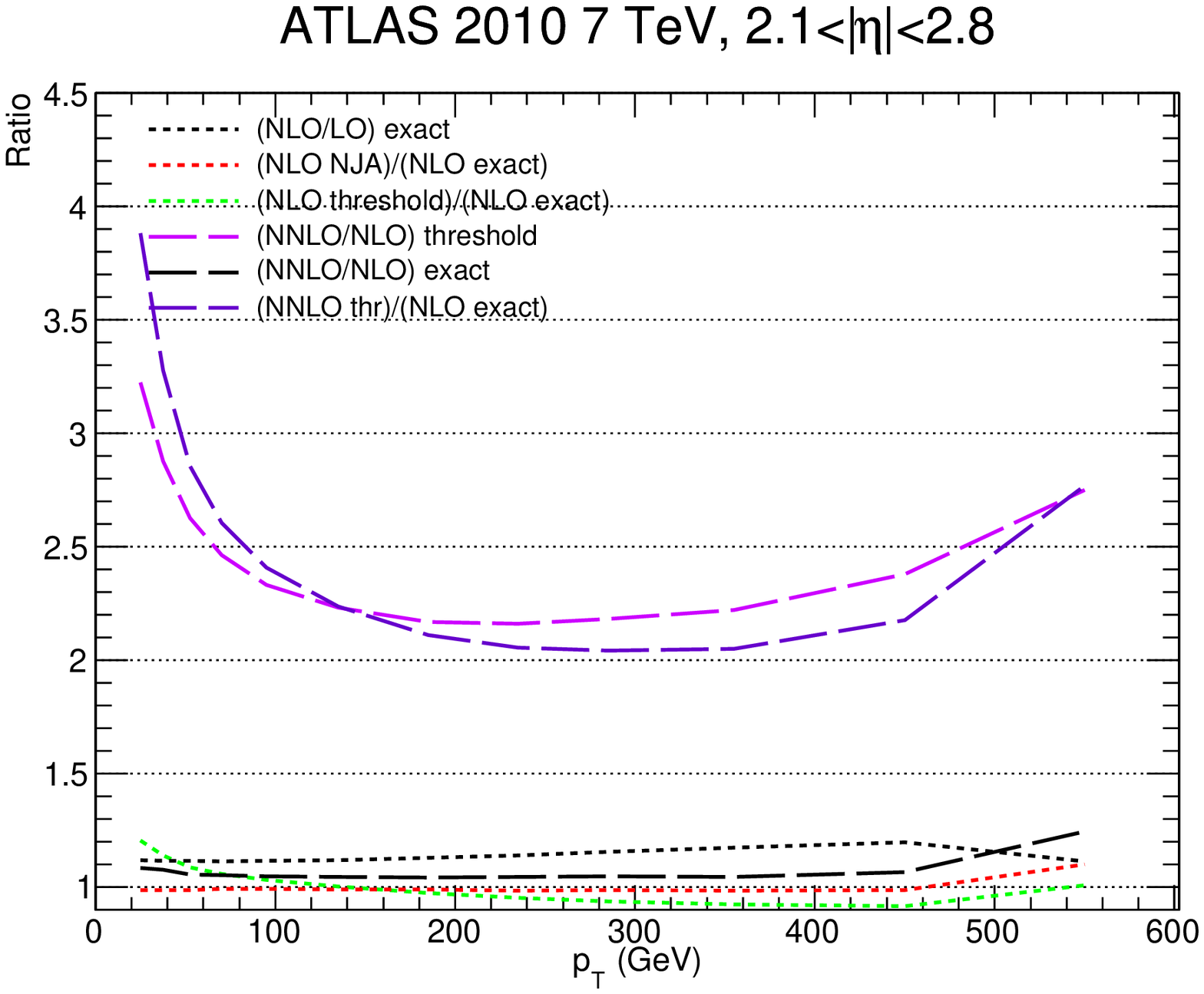}\includegraphics[scale=0.38]{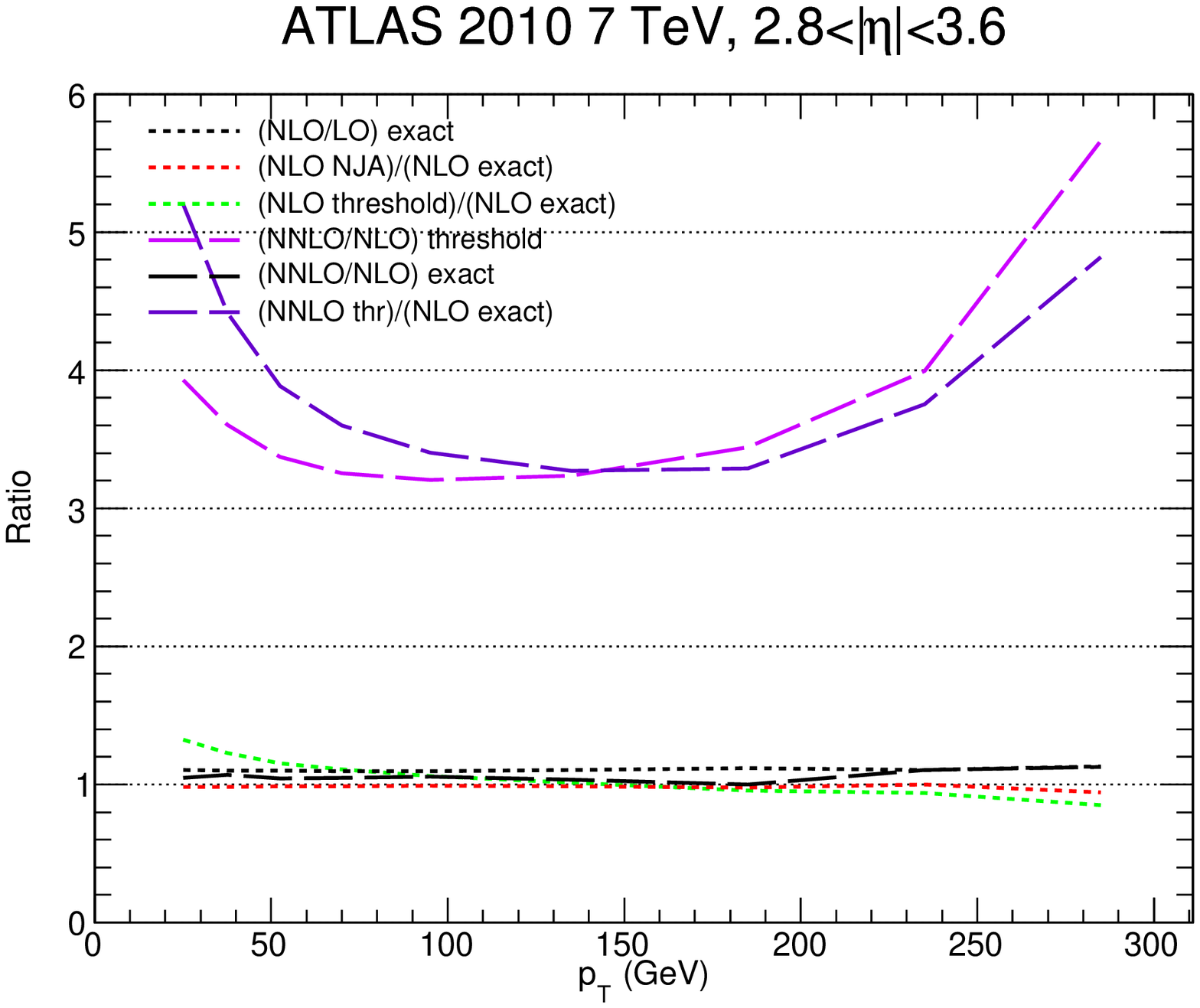}
\par\end{centering}

\begin{centering}
\includegraphics[scale=0.38]{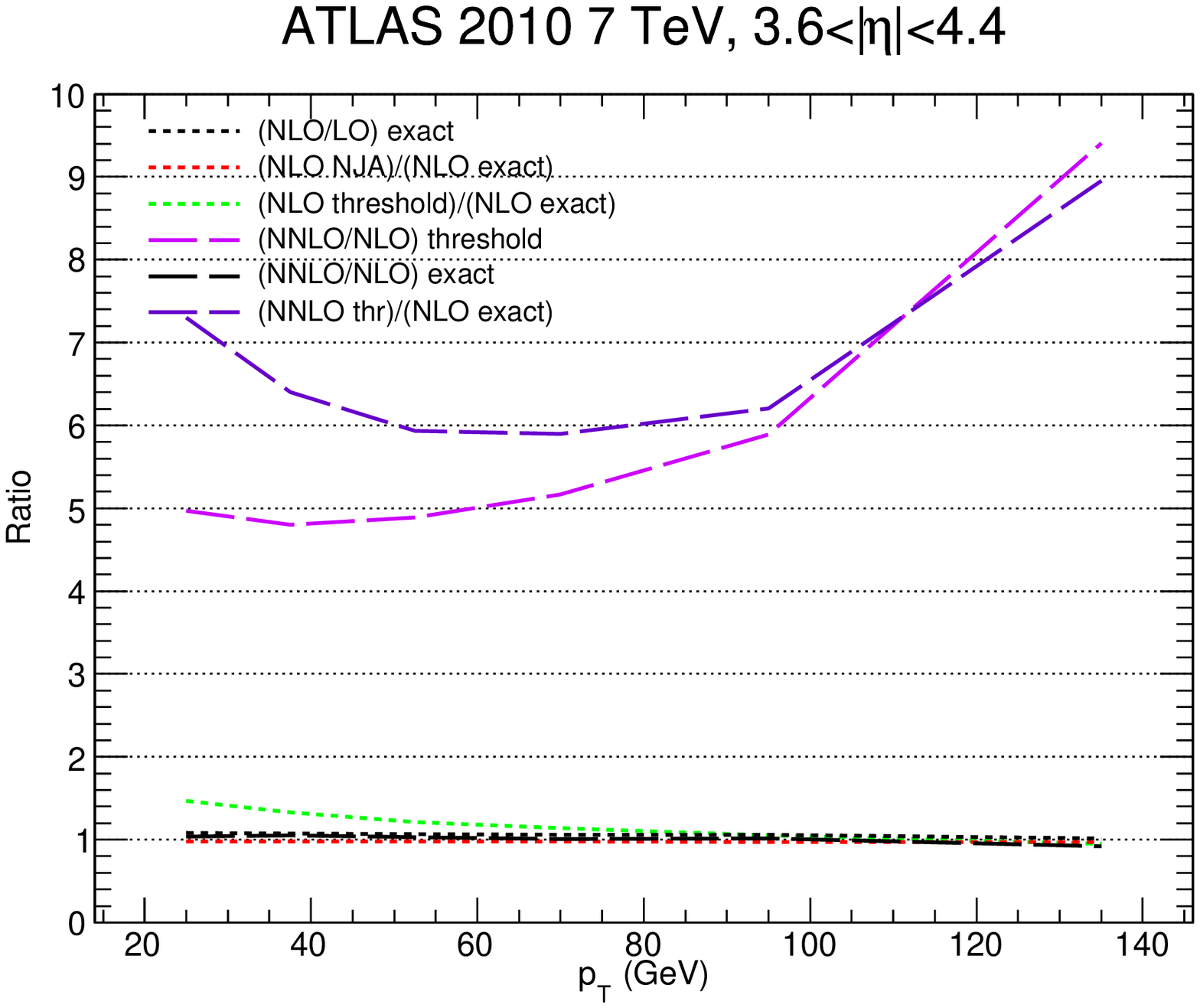}
\par\end{centering}

\caption{\label{fig:atlas7ratio}Ratio plots for $gg$-channel predictions together with NNLO $k$-factors for ATLAS jets at $\sqrt{s}=7$ TeV kinematics.}
\end{figure}

\begin{figure}[H]
\begin{centering}
\includegraphics[scale=0.38]{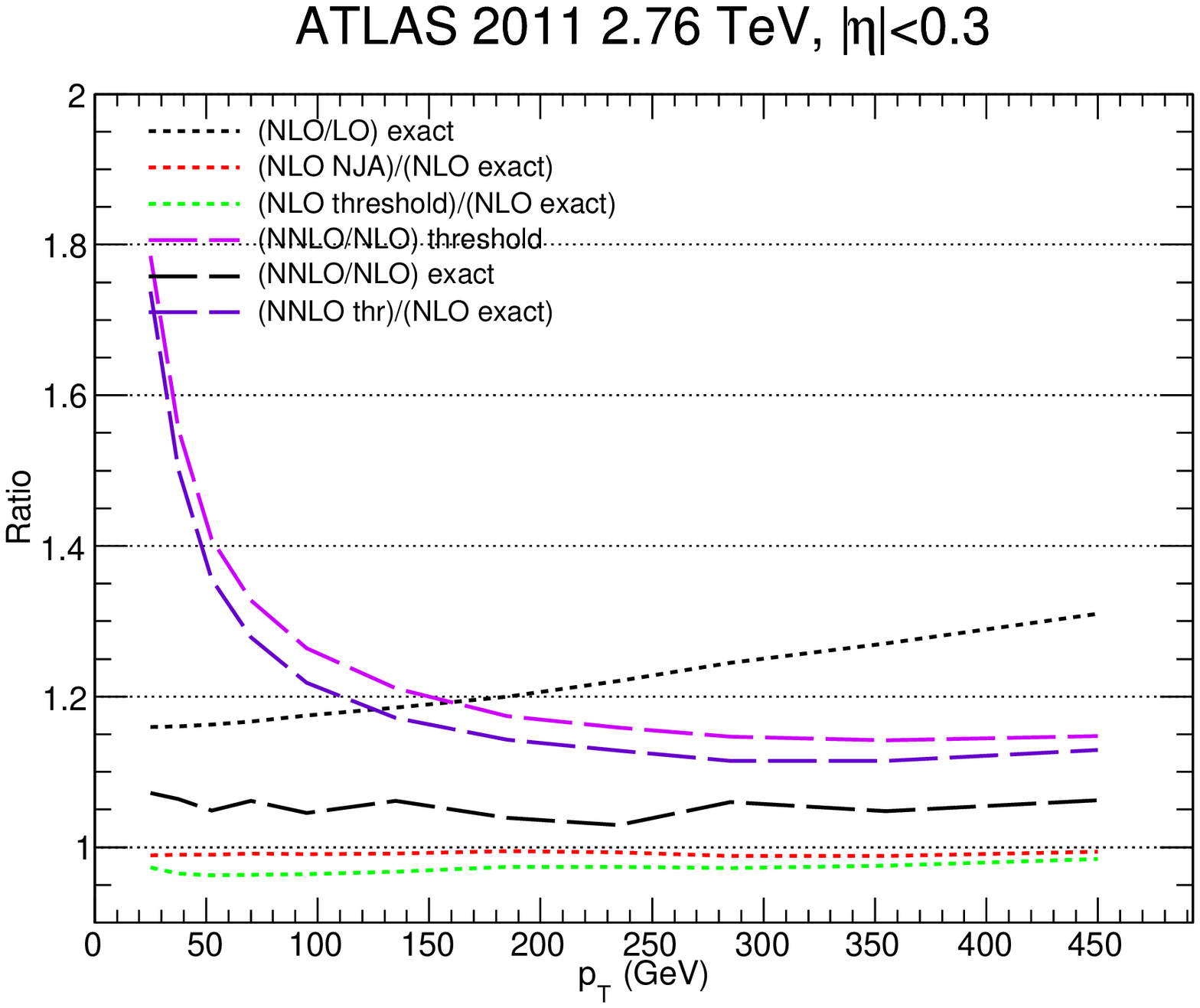}\includegraphics[scale=0.38]{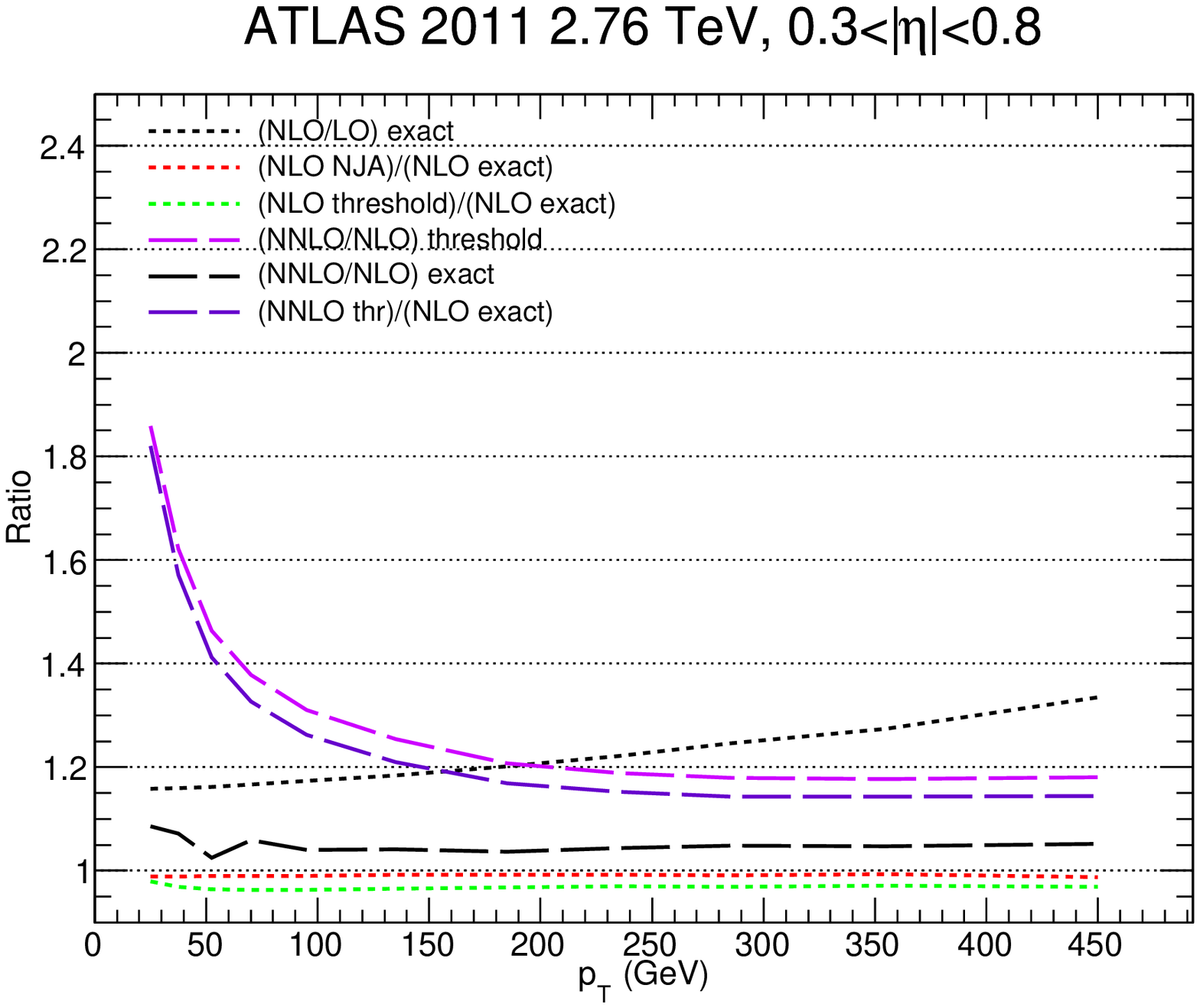}
\par\end{centering}

\begin{centering}
\includegraphics[scale=0.38]{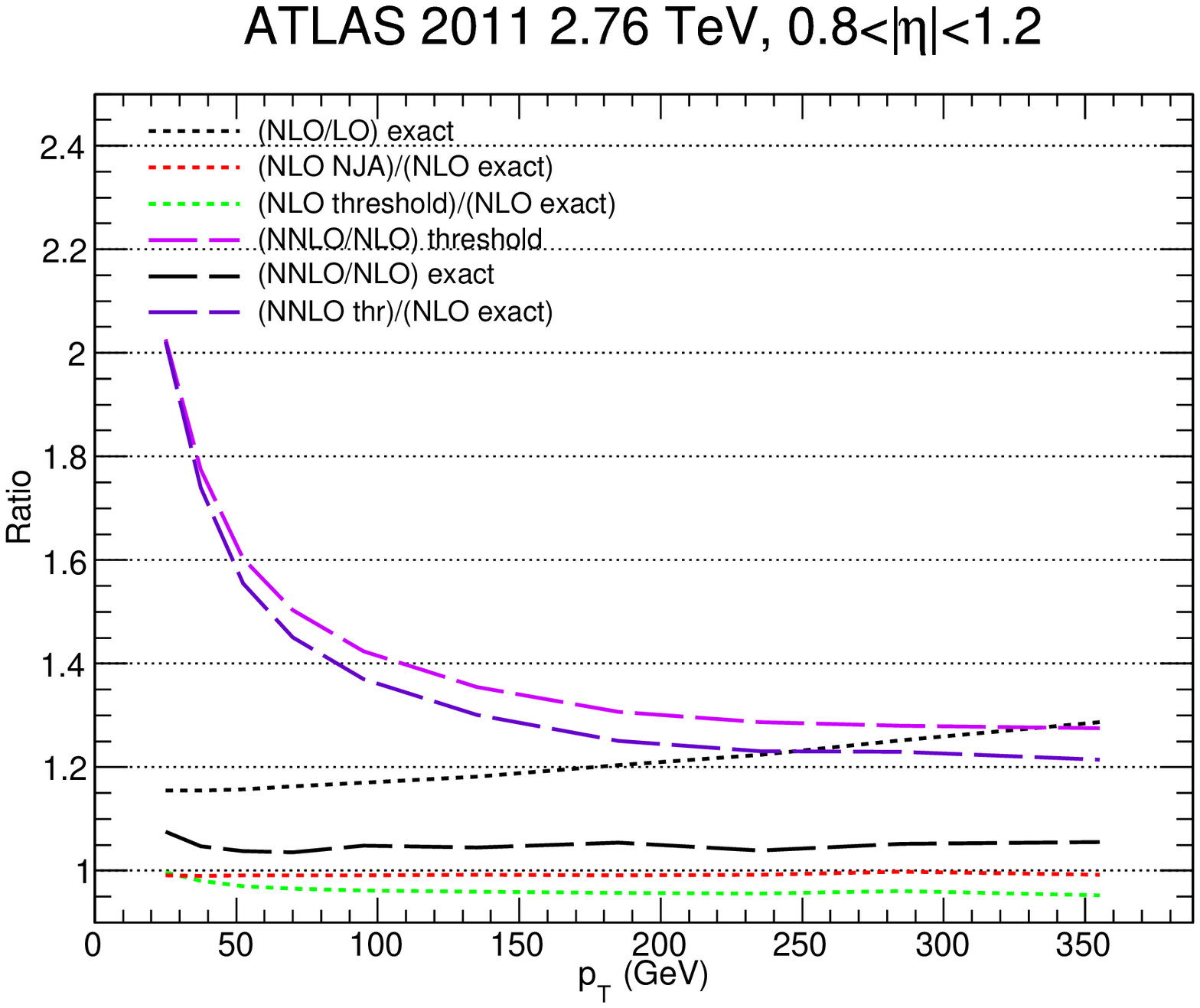}\includegraphics[scale=0.38]{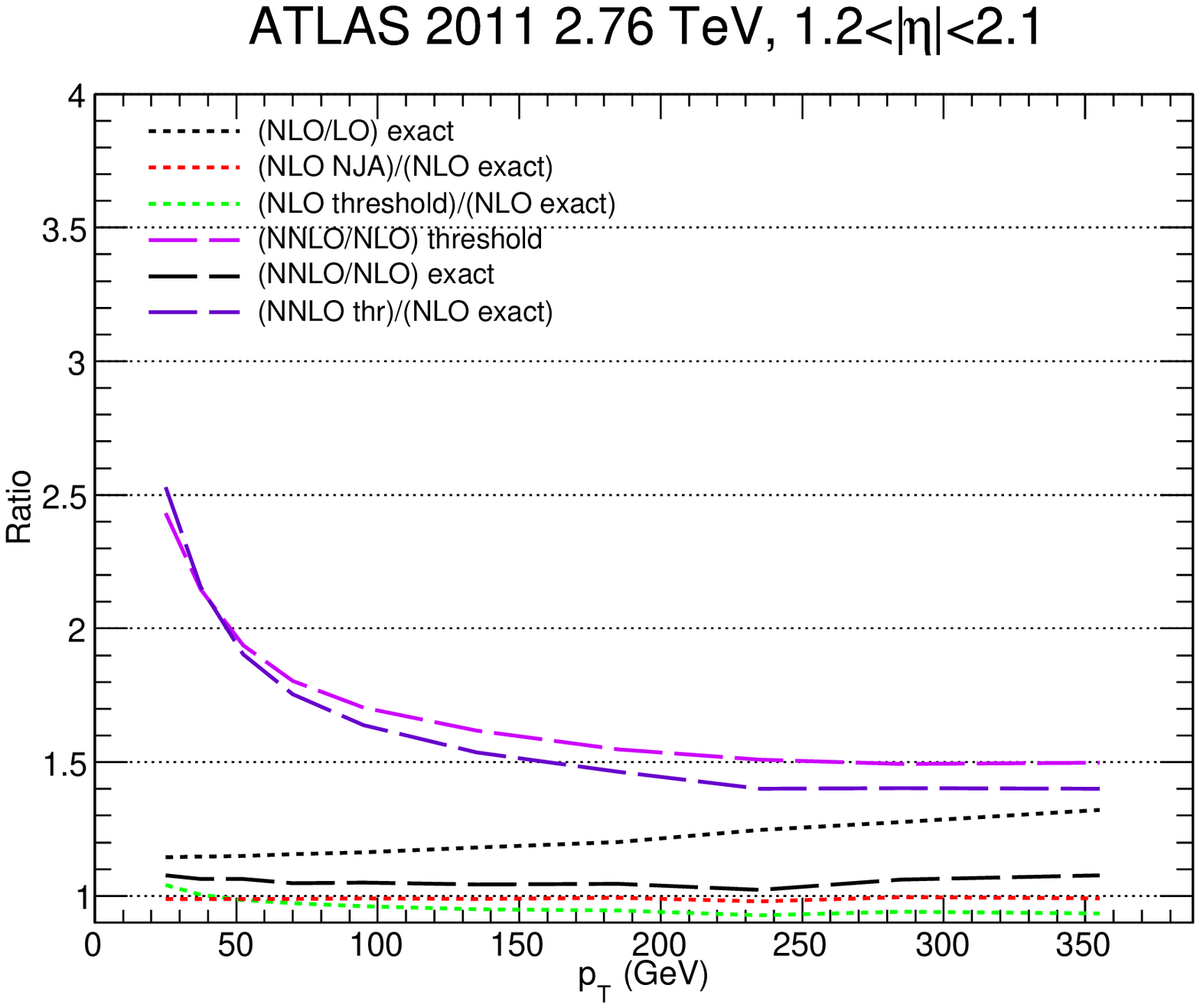}
\par\end{centering}

\begin{centering}
\includegraphics[scale=0.38]{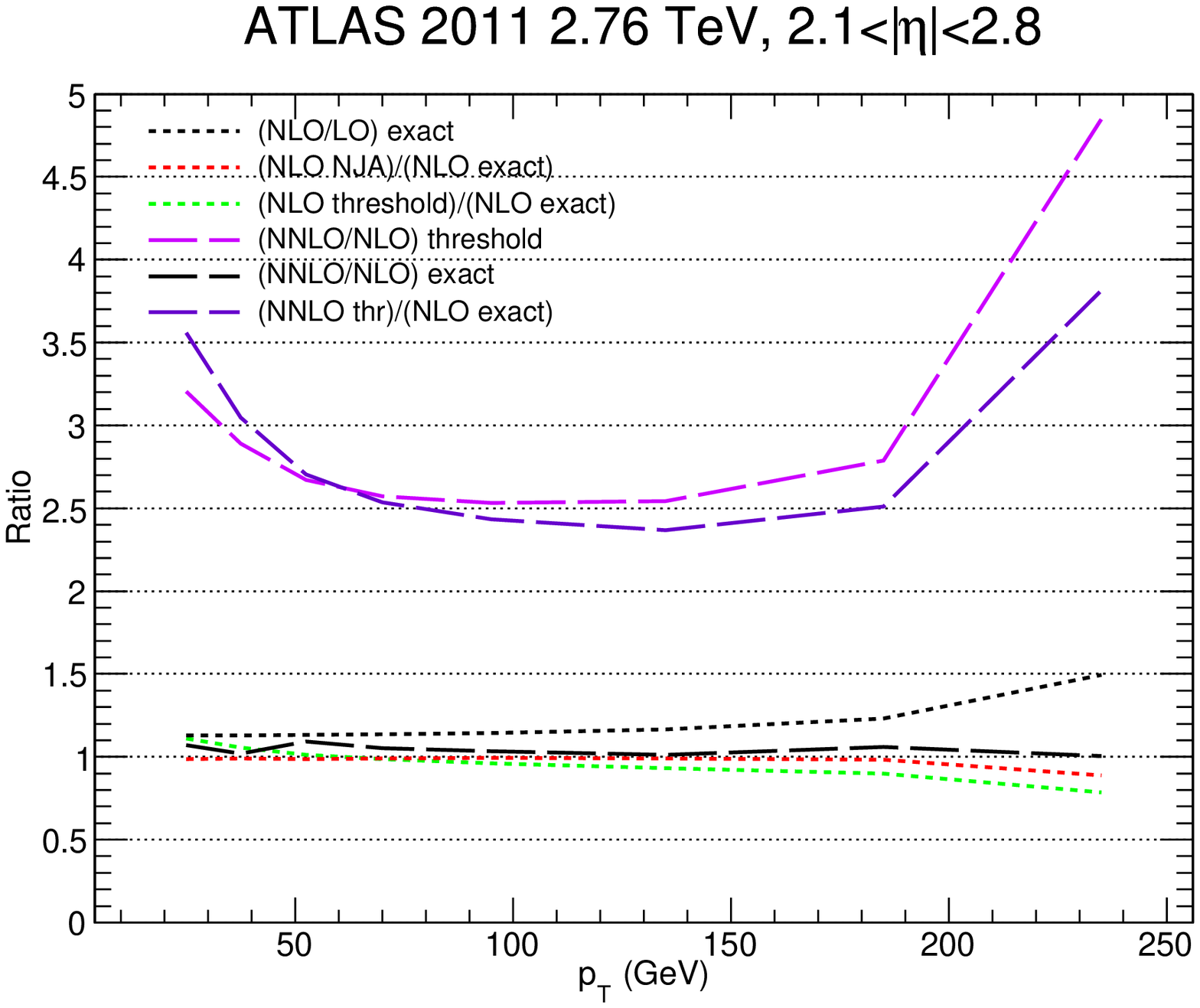}\includegraphics[scale=0.38]{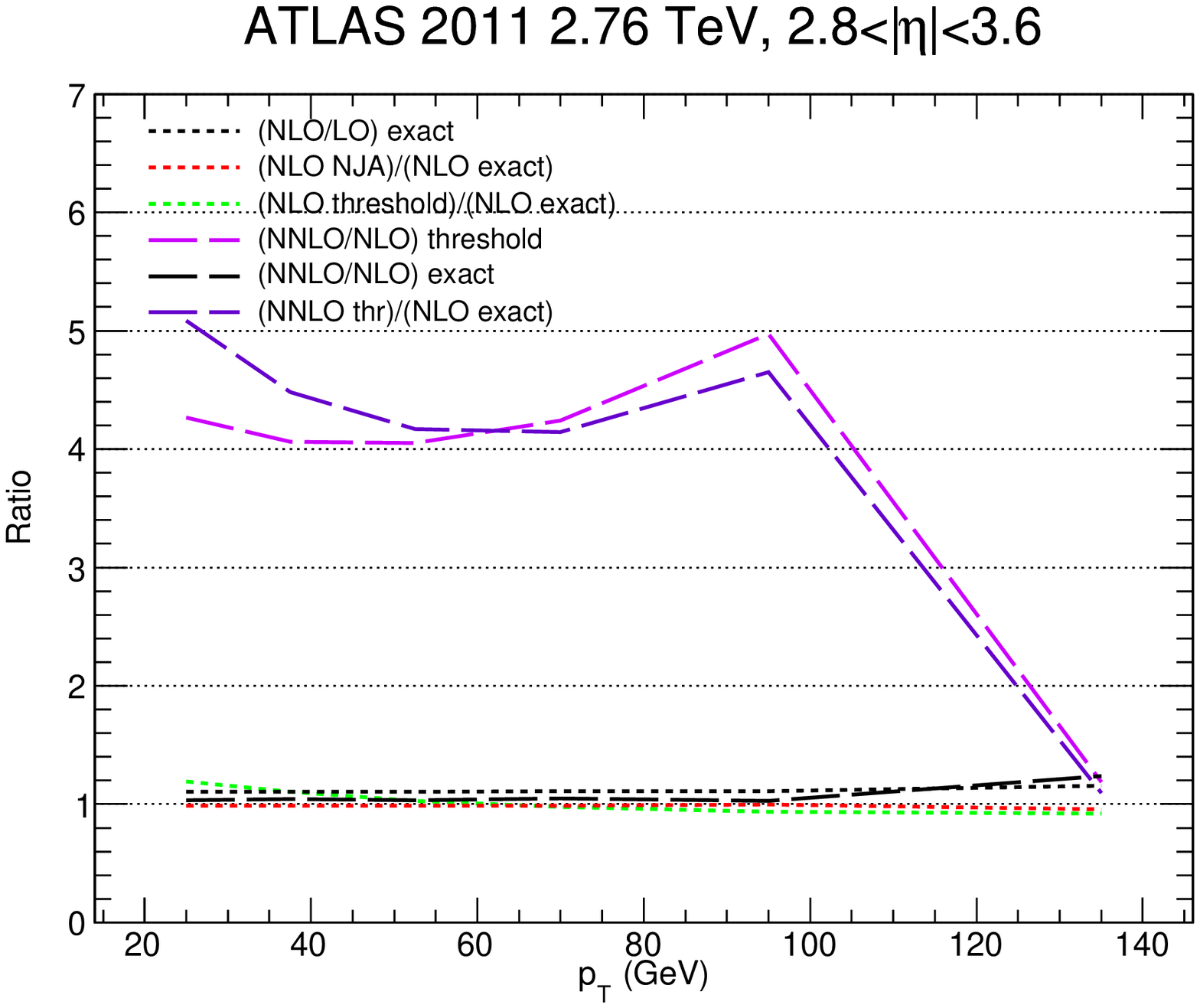}
\par\end{centering}

\begin{centering}
\includegraphics[scale=0.38]{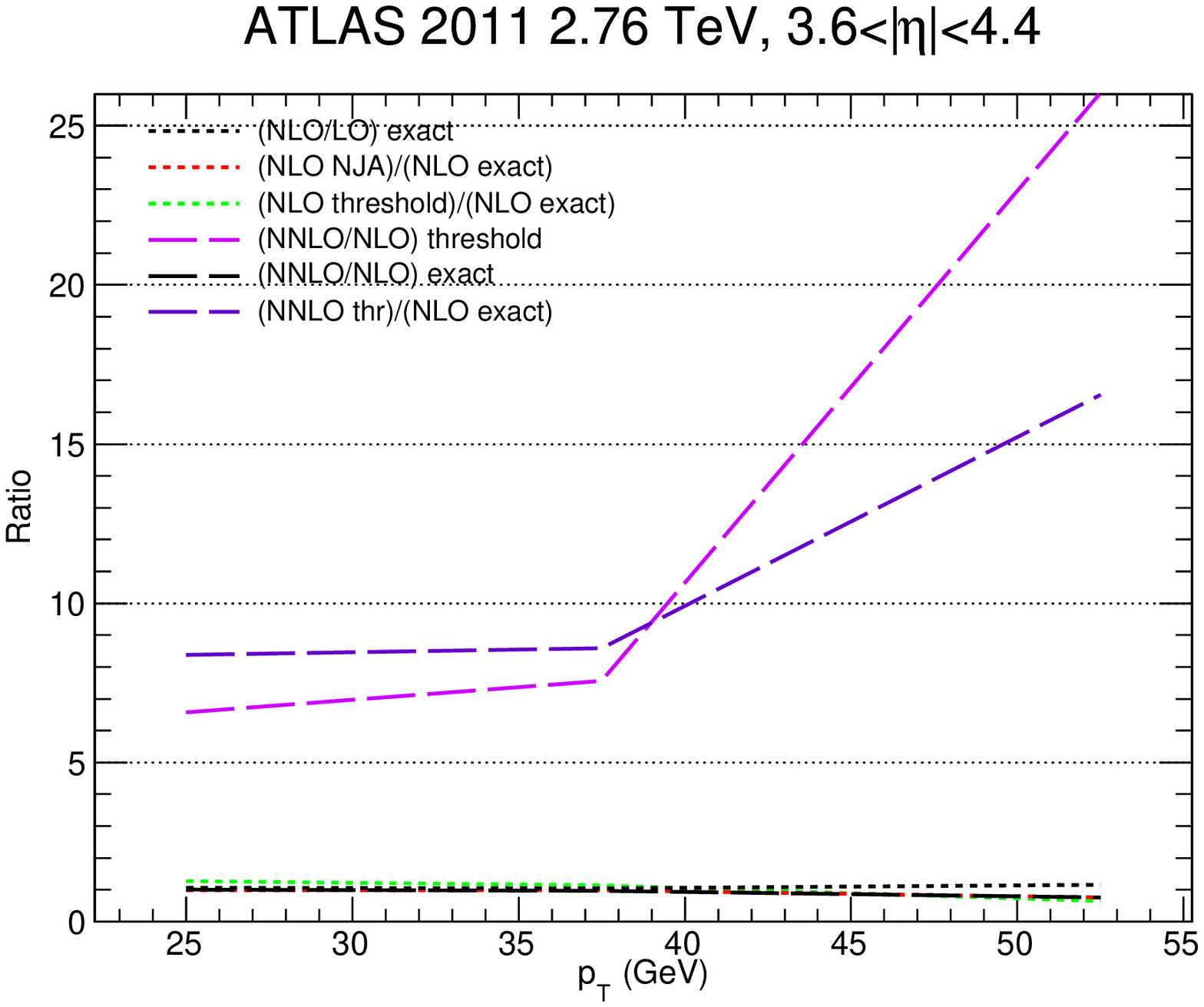}
\par\end{centering}

\caption{\label{fig:atlas2ratio}Ratio plots for $gg$-channel predictions together with NNLO $k$-factors for ATLAS jets at $\sqrt{s}=2.76$ TeV kinematics.}
\end{figure}

Figures~\ref{fig:atlas7ratio} and~\ref{fig:atlas2ratio} present the gluons-only theoretical predictions
for ATLAS jets at $\sqrt{s}=7$ TeV and $\sqrt{s}=2.76$ TeV
respectively. For all predictions we set the renormalisation and factorisation scales equal to each other
and equal to the $p_T$ of each individual jet in every event ($\mu_R=\mu_F=p_T$). At NLO the NJA (narrow-jet approximation) prediction shows
good agreement with the exact NLO result across the entire kinematic range. We note that the ATLAS
experiment employs R=0.4 for the anti-$k_t$ jet clustering procedure and therefore the agreement
between NJA approximation and the exact calculation is expected and observed to improve for smaller values of R.
The threshold prediction at NLO is also in good agreement with the exact calculation but shows an evident
instability at small $p_T$ where discrepancies vary between 10\% and 40\% increasing with rapidity.

The predictions at NNLO are in worse agreement when compared to the
results obtained at CMS. There is a constant gap between the threshold and the exact
predictions for ATLAS jets at $\sqrt{s}=7$ TeV and $\sqrt{s}=2.76$ TeV. The convergence 
of the threshold approximation code to the exact
computation is not evident as it is for CMS because the maximum $p_{T}$ values are
smaller for ATLAS.

On Tables~\ref{tab:katlas71} to~\ref{tab:katlas77} in the Appendix we document
the NNLO $k$-factor results. The general behaviour is similar to CMS:
differences between the exact NNLO computation and the approximate NNLO threshold computation
are large at small $p_{T}$ (between 20\% and 80\%) and increase with
rapidity. For the rapidity range $0.8<|\eta|<2.1$ and for all $p_T$ the disagreement between the two predictions is between $10\%-100\%$. 
For the rapidity regions $|\eta|>2.1$ the disagreement is larger than $100\%$ for all $p_T$. An application of the rejection criteria suggested before
excludes approximate predictions for the $p_{T}$ points for the first ten bins with
$|\eta|<0.3$ and the first thirteen points with $0.3<|\eta|<0.8$. In the regions of rapidity $|\eta|>0.8$ we should discard all points as differences
between the exact and approximate calculation are for all values of jet $p_T$ much larger than 10\%.

In particular, and contrary to statements made in Ref.~\cite{deFlorian:2013qia}, the large NNLO $k$-factors observed in the approximate threshold calculation (Figure~\ref{fig:atlas7ratio})
of the order 5 or so at $\eta\sim4$ are not present in the exact NNLO calculation. As can be seen in Figure~\ref{fig:atlas7ratio} while the exact NNLO $k$-factor decreases
with rapidity, the NNLO $k$-factor of the approximate NNLO calculation increases with rapidity.

The comparison between the NNLO $\sqrt{s}=7$ TeV ATLAS $k$-factors and the
NNLO $\sqrt{s}=2.76$ TeV  ATLAS $k$-factors show a moderate dependence on the center of mass
energy, with the predictions at $\sqrt{s}=2.76$ TeV  giving slightly smaller NNLO $k$-factors. 
With the results documented in Tables~\ref{tab:katlas21} to~\ref{tab:katlas27} and by 
extending the rejection criteria suggested to $\sqrt{s}=7$
TeV to ATLAS at $\sqrt{s}=2.76$ TeV we conclude that
the approximate calculation gives acceptable predictions only for the first rapidity slice $|\eta|<0.5$ after
removing the first eight $p_{T}$ bins.

\begin{figure}
\begin{centering}
\includegraphics[scale=0.38]{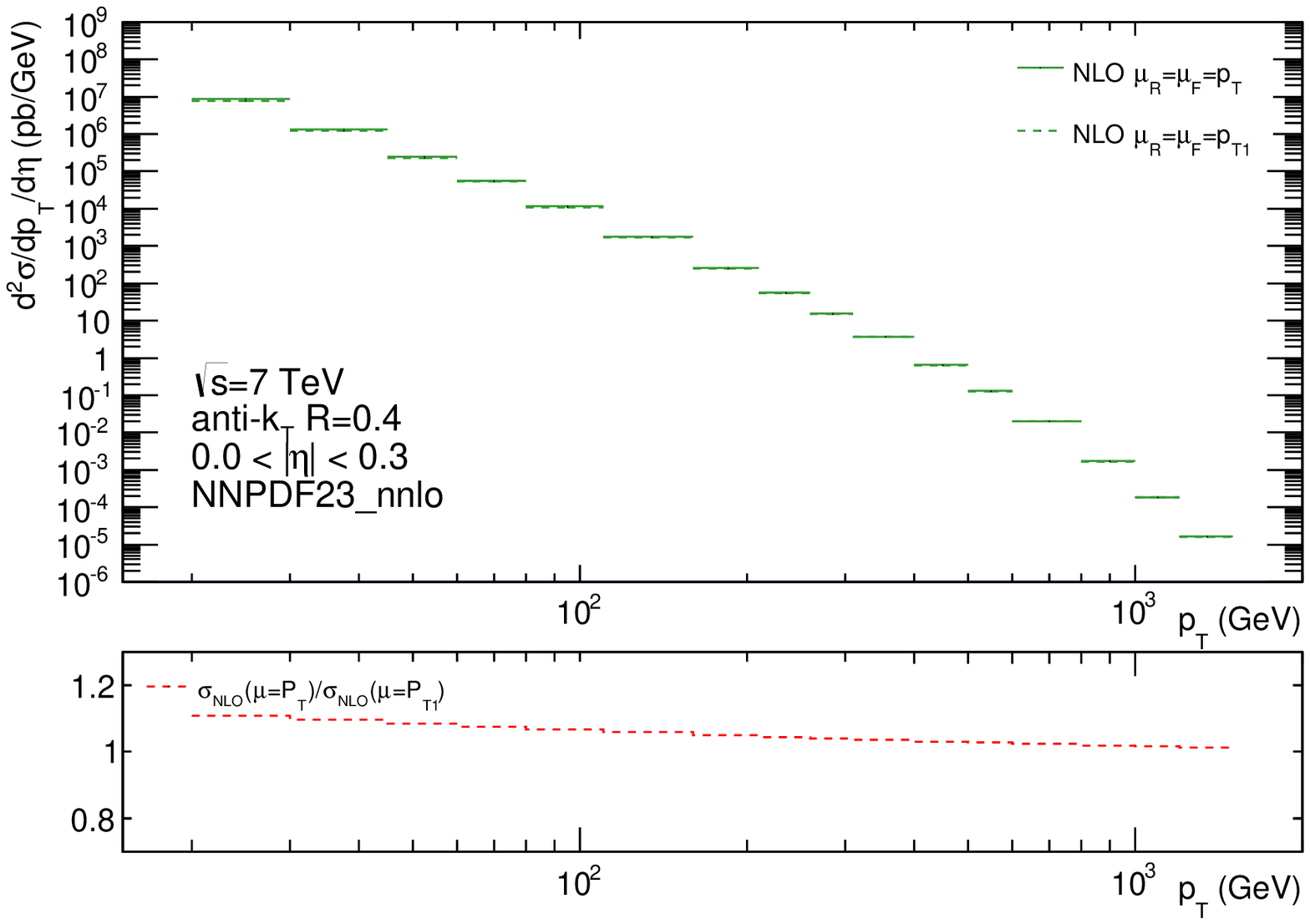}\includegraphics[scale=0.38]{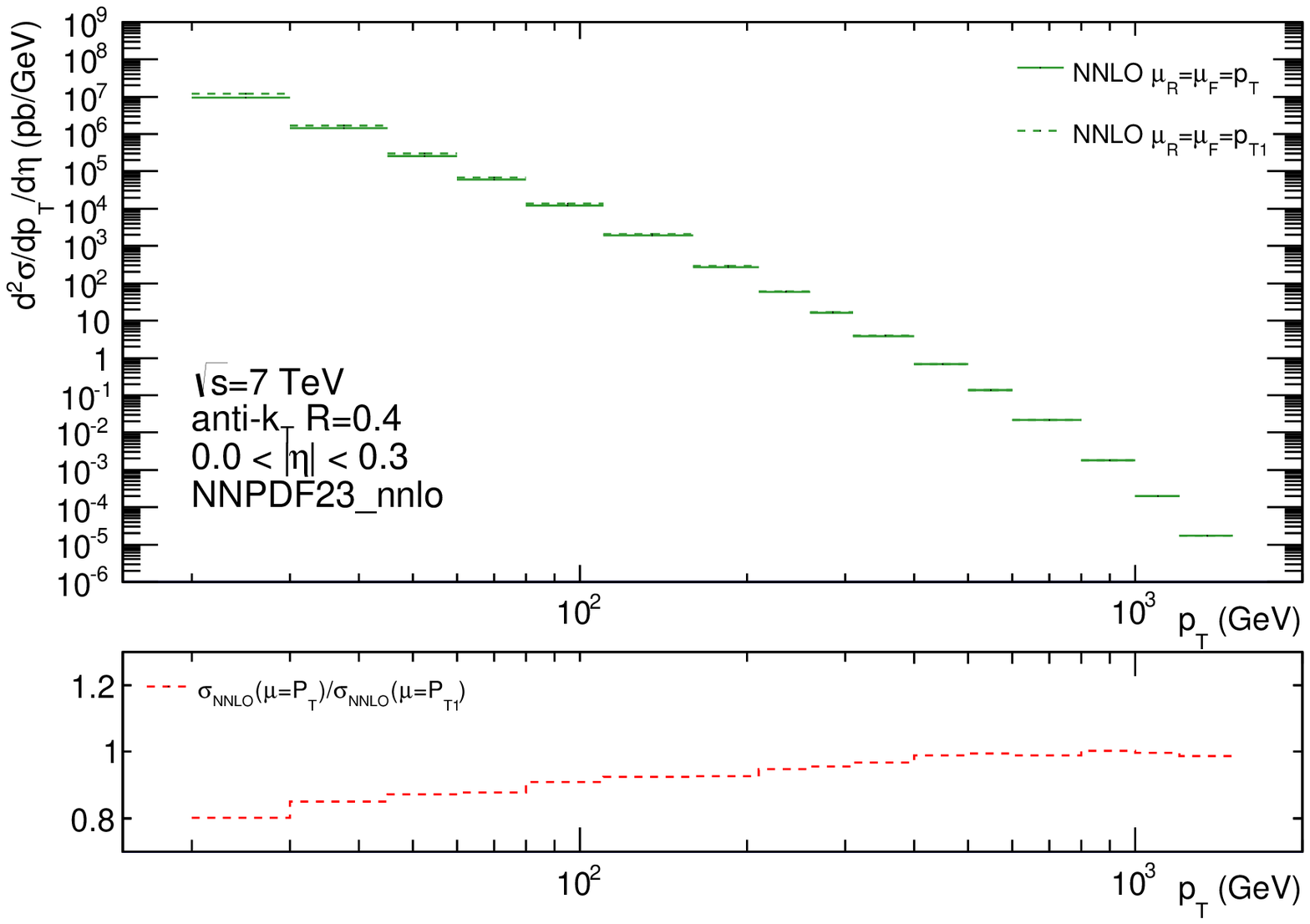}
\par\end{centering}
\caption{\label{fig:atlas7PTvsPT1} NLO (left) and NNLO (right) exact $gg$-channel predictions for ATLAS evaluated with the renormalisation and factorisation scales
 $\mu_{R}=\mu_{F}=p_T$ and $\mu_{R}=\mu_{F}=p_{T1}$. In the lower pads we present the relative differences due to the different central scale choice.}
\end{figure}

We conclude this section by comparing the exact fixed-order predictions in the $gg$-channel evaluated at the scales  $\mu_R=\mu_F=p_{T1}$ and  $\mu_R=\mu_F=p_{T}$.  
In Figure~\ref{fig:atlas7PTvsPT1} we show NLO and NNLO gluons only cross sections evaluated at the two different scales for the first rapidity slice of the ATLAS experiment. The
qualitative effect is similar to the one showed for CMS in the previous section. By choosing $\mu_R=\mu_F=p_T$ as a central scale we observe that at low-$p_T$ the NLO prediction
increases by about 10\% while the NNLO prediction is reduced by around 20\%, with respect to the results obtained using $\mu_R=\mu_F=p_{T1}$. 
Predictions at high-$p_T$ are as expected identical with either scale choice. We note that the due to the $p_T$ coverage by ATLAS being more extreme than at CMS 
(nearly two orders of magnitude in $p_T$ covered by ATLAS) we are at low-$p_T$ more often in the kinematical regions where 
$p_{T} \ll p_{T1}$ and therefore the effects that result from changing the central scale choice for the predictions
from $\mu=p_{T1}$ to $\mu=p_T$ are enhanced. 

\begin{figure}
\begin{centering}
\includegraphics[scale=0.38]{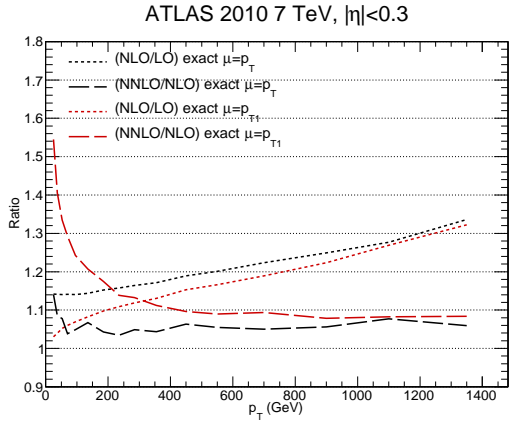}
\par\end{centering}
\caption{\label{fig:kfactPTvsPT1} NLO/LO and NNLO/NLO exact $k$-factors for the $gg$-channel evaluated with the renormalisation and factorisation scales
 $\mu_{R}=\mu_{F}=p_T$ and $\mu_{R}=\mu_{F}=p_{T1}$.}
\end{figure}

In Figure~\ref{fig:kfactPTvsPT1} we show the exact NLO/LO and NNLO/NLO
$k$-factors for each scale choice. Interestingly we observe that for
the first bin in $p_T$ the perturbative series behaves for
$\mu_{R}=\mu_{F}=p_T$ as 14\% NLO correction with respect to LO and
14\% NNLO corrections with respect to NLO. This compares with 3\% NLO
corrections with respect to LO and 54\% NNLO corrections with respect
to NLO when using $\mu_{R}=\mu_{F}=p_{T1}$. Therefore, the
convergence of the perturbative series in the fixed order calculation
is improved using $\mu_{R}=\mu_{F}=p_T$ as the NNLO/NLO $k$-factor is
smaller than the NLO/LO $k$-factor for all $p_{T}$ and rapidity.

\begin{figure}
\begin{centering}
\includegraphics[scale=0.38]{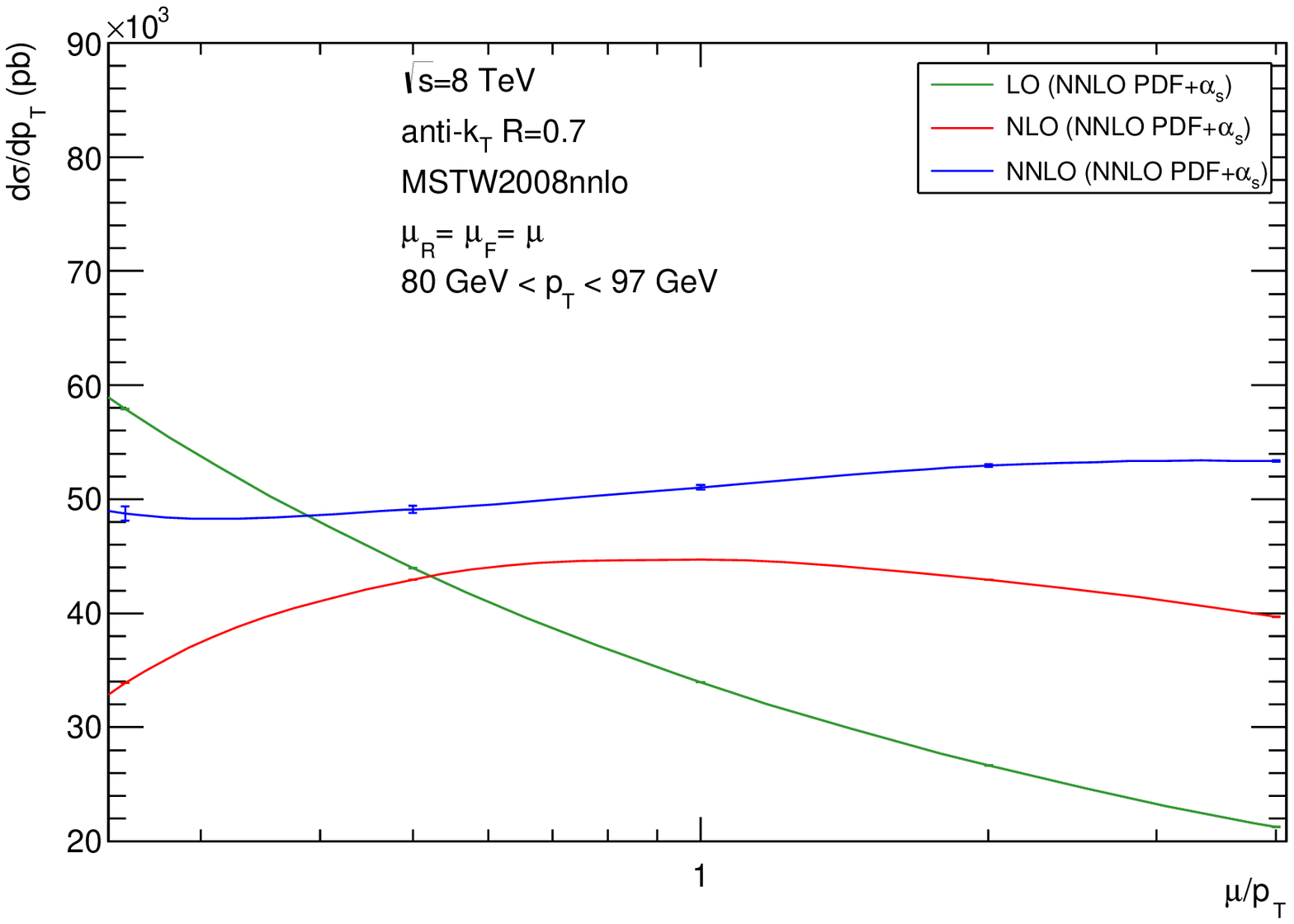}\includegraphics[scale=0.38]{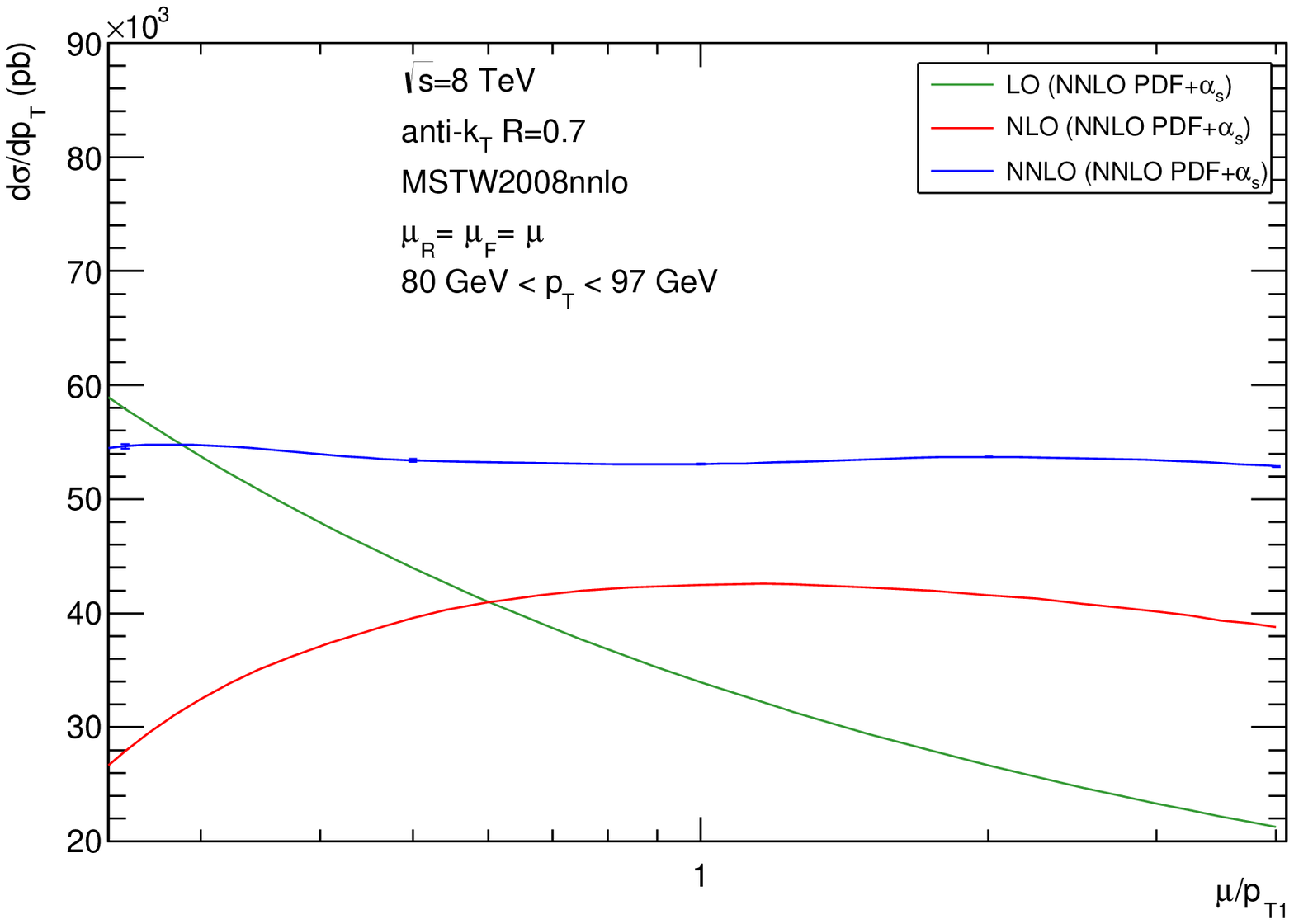}
\par\end{centering}
  \caption{Scale dependence of the inclusive jet cross section for $pp$ collisions at $\sqrt{s}=8~$TeV for 
  the anti-$k_T$ algorithm with $R=0.7$ at NNLO (blue), NLO (red) and LO (green) and with $|y|<4.4$ and 
 80 GeV$<p_{T}<$ 97 GeV for different central scale choices: 
  $\mu_R=\mu_F=p_{T}$ on the left and $\mu_R=\mu_F=p_{T1}$ on the right.}
  \label{fig:scale}
\end{figure}

In Figure~\ref{fig:scale} we study the scale dependence of the exact calculation
in the $gg$-channel at each order in perturbation theory for jets with $|y|<4.4$ and
80 GeV$<p_{T}<$ 97 GeV. For this study we have employed the same kinematical
setup used in~\cite{Ridder:2013mf} and changed only the renomalisation and factorization central scales
to generate the predictions to be $\mu_R=\mu_F=p_T$ instead of the leading jet $p_{T1}$. 
We observe again in this kinematical setup that changing the central scale choice to $\mu_R=\mu_F=p_T$
results in the NNLO/NLO $k$-factor becoming smaller as the blue curve (NNLO prediction)
is closer to the red curve (NLO prediction) with this scale choice.
We observe also a reduction of the scale dependence of the cross section at NNLO.

Similarly to the results presented for CMS in the previous section we
conclude that the disagreement between the exact calculation and the
threshold calculation is enhanced when both calculations are performed
using the same central scale choice. In particular, by cutting away
kinematical regions where the disagreement between both calculations
is larger than 10\% we keep only data points for which the NNLO
$k$-factors are typically smaller $\sim1.1-1.2$. As we will shown in
Sect.~\ref{sec:PDFfit} this has the consequence of improving the
$\chi^2$ of an aNNLO PDF fit including the approximate NNLO single jet
inclusive prediction.

\section{Tevatron jet data}
\label{sec:tev}

\subsection{CDF jets}

\begin{figure}
\begin{centering}
\includegraphics[scale=0.4]{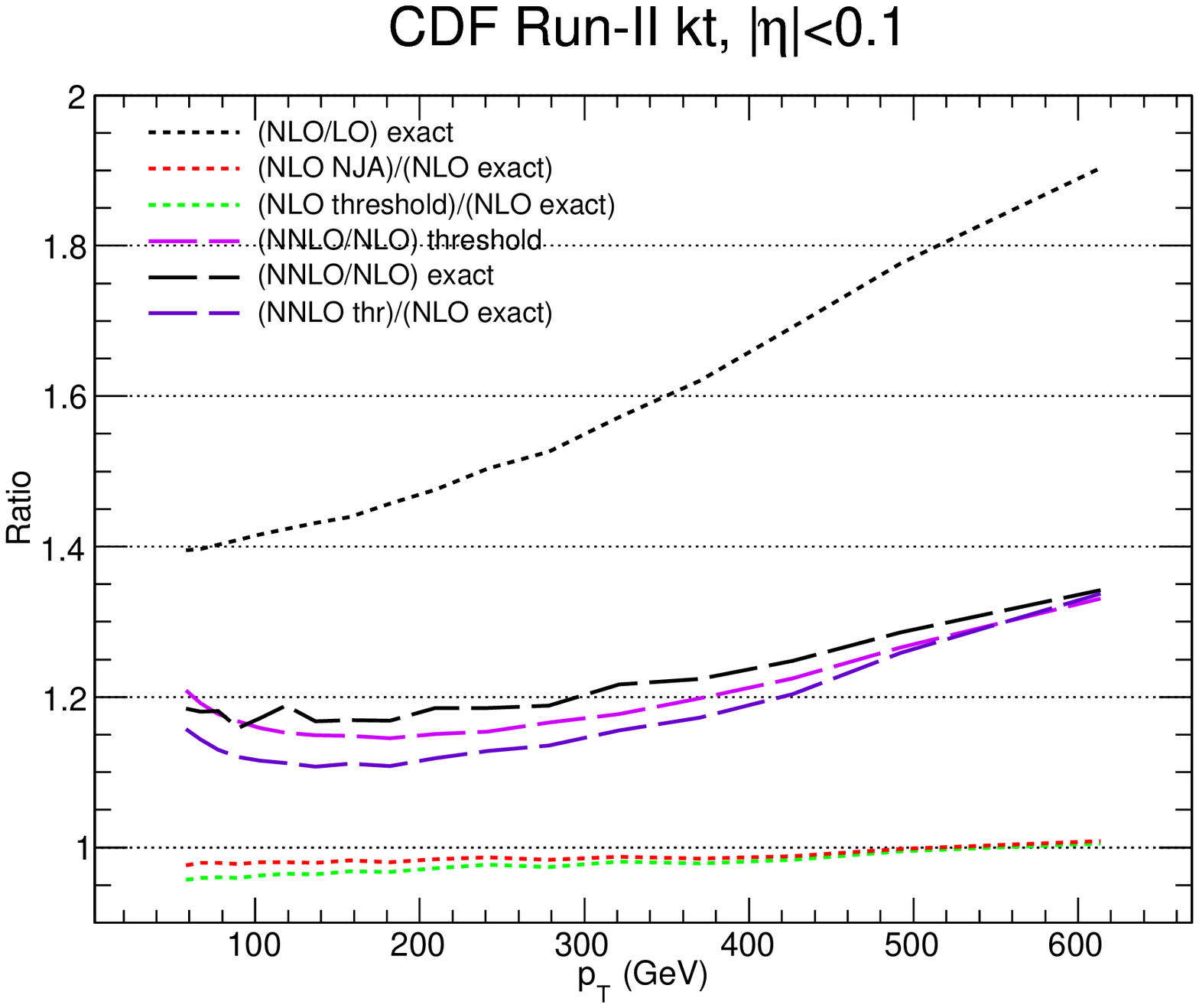}\includegraphics[scale=0.4]{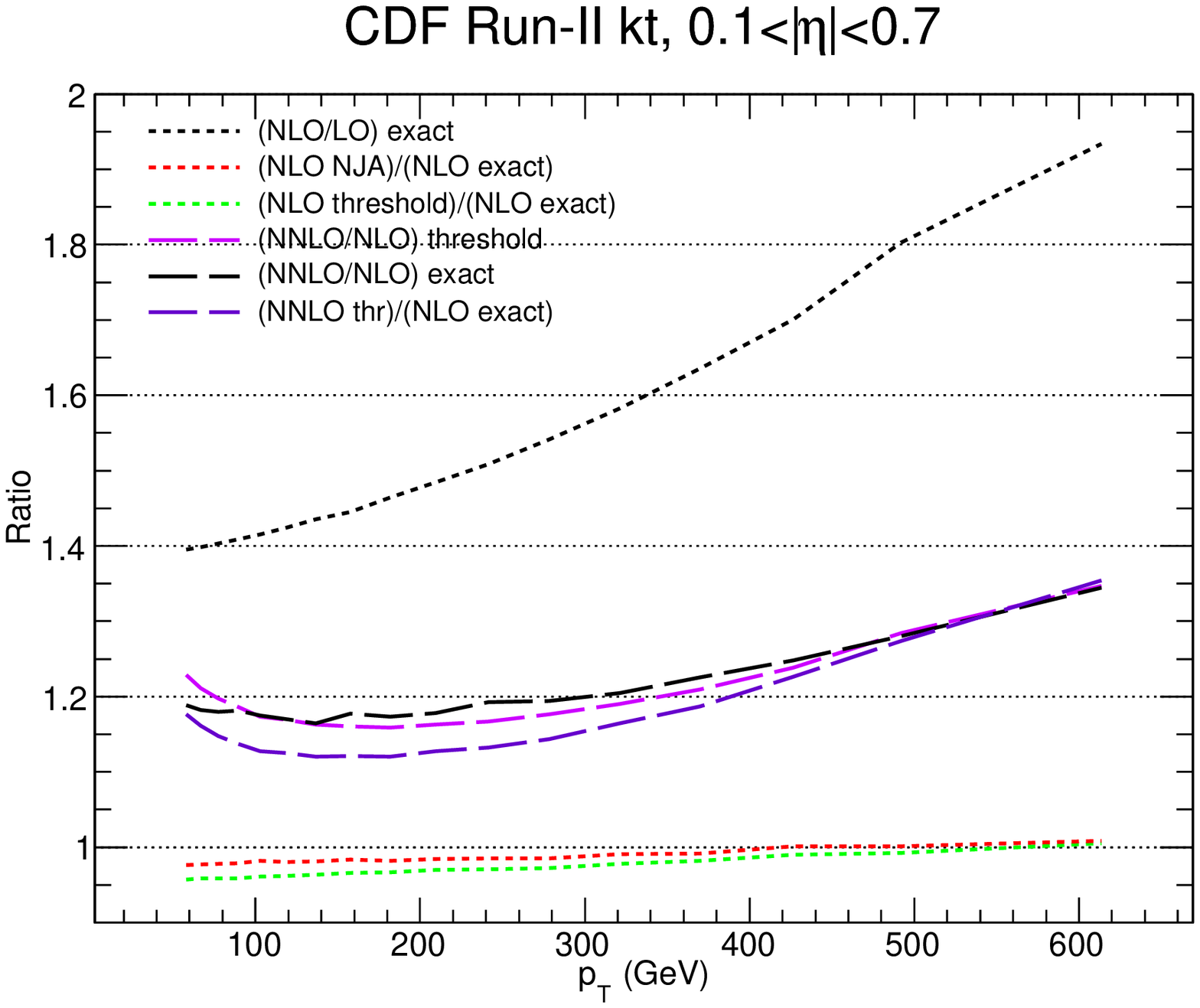}
\par\end{centering}

\begin{centering}
\includegraphics[scale=0.4]{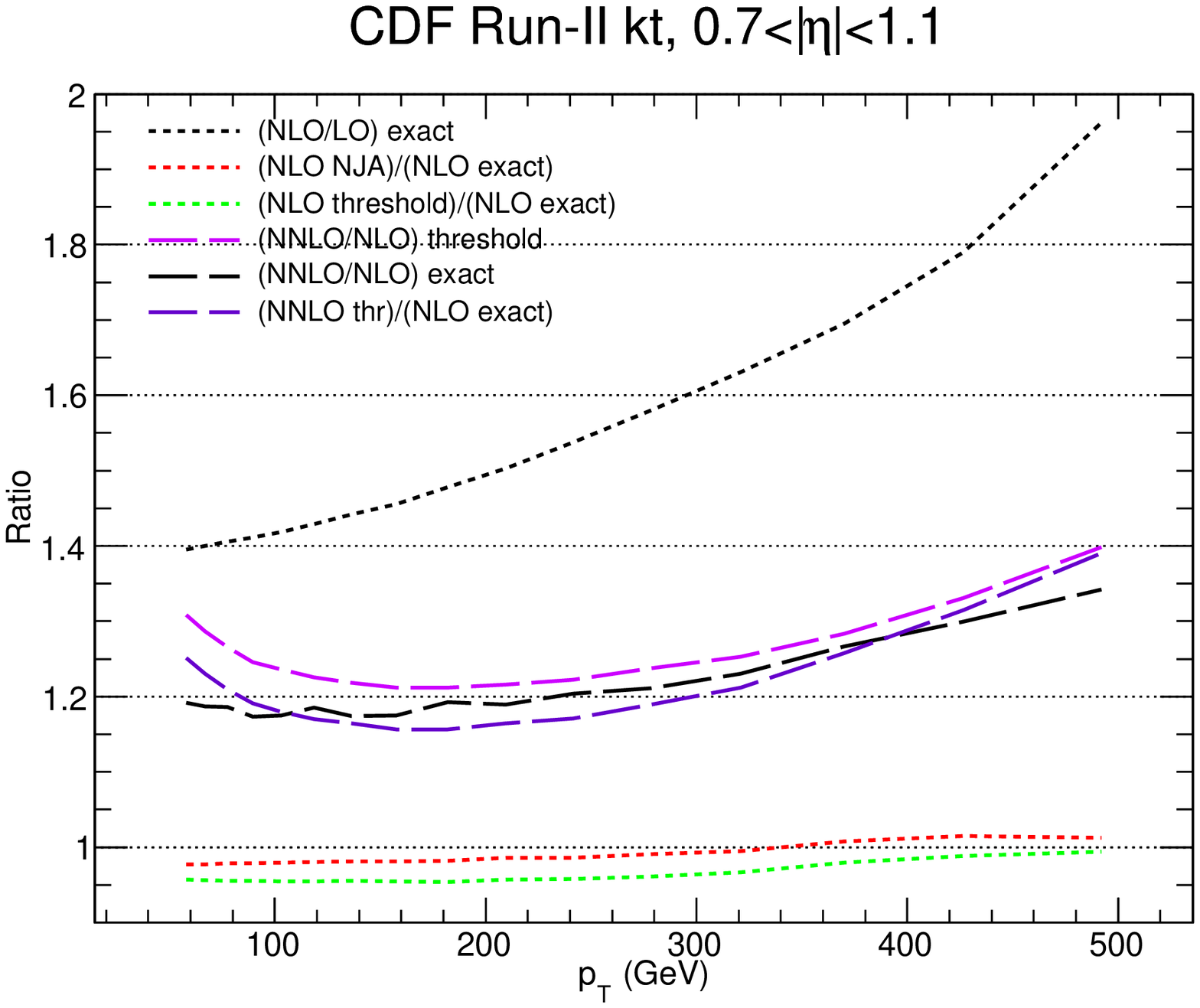}\includegraphics[scale=0.4]{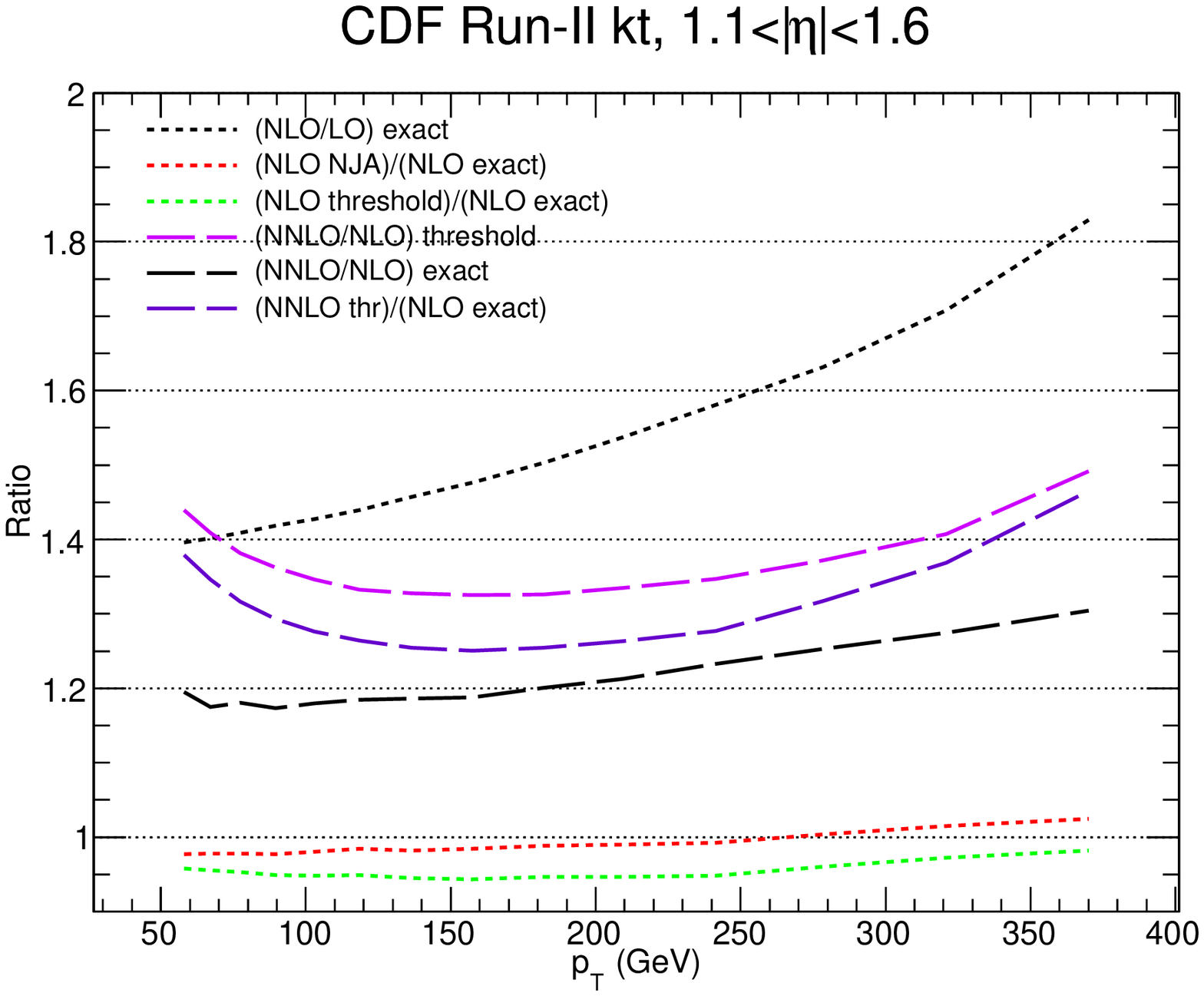}
\par\end{centering}

\begin{centering}
\includegraphics[scale=0.4]{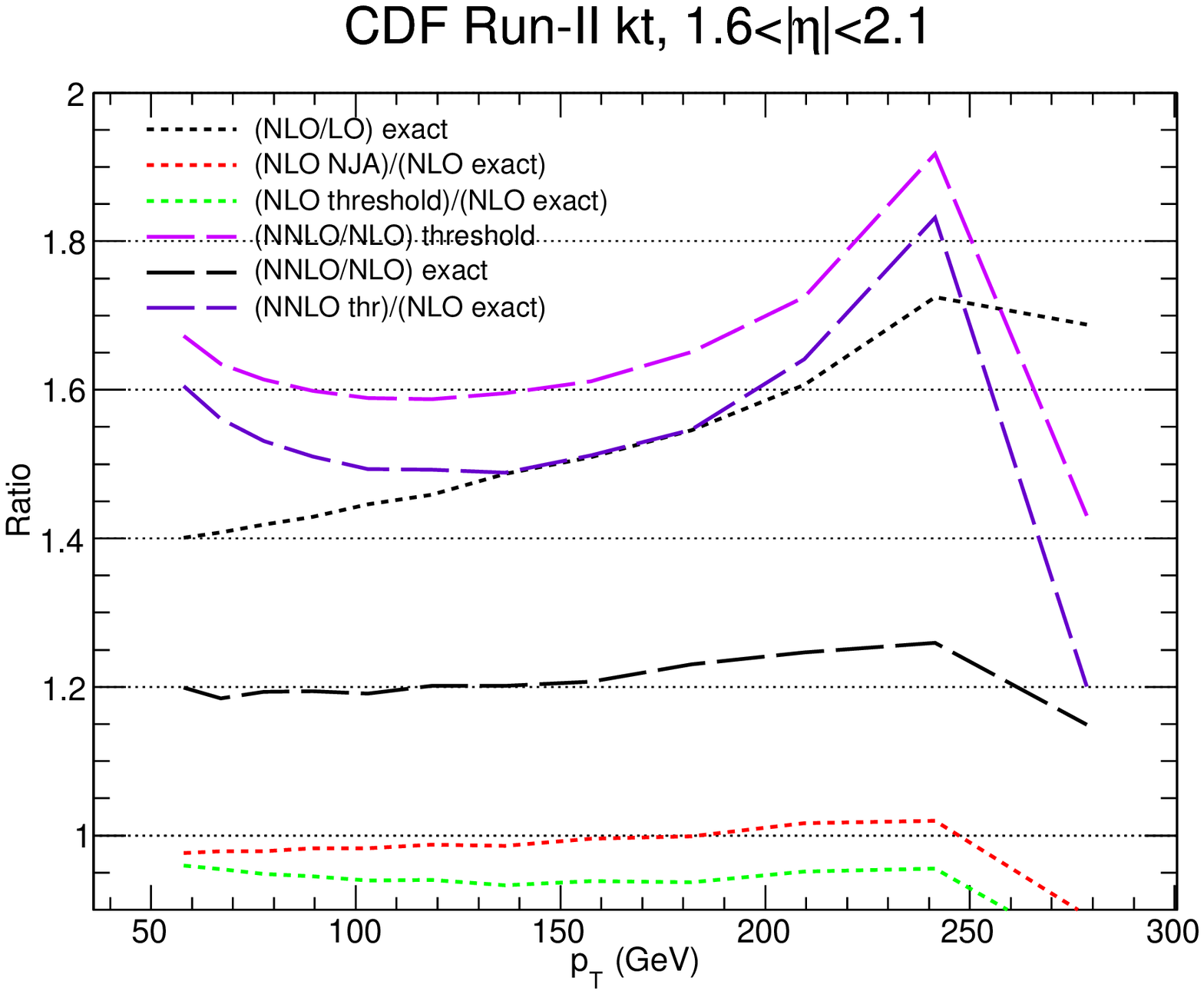}
\par\end{centering}

\caption{\label{fig:cdfratio}  Ratios between exact and approximate
      predictions at the same order (LO and NLO) in perturbation theory in the gluons-only channel.
      In the same plot we present the 
      exact NLO/LO and NNLO/NLO $k$-factors (gluons-only channel) and the NNLO/NLO $k$-factors produced
      by the threshold approximation code (gluons-only channel) for the CDF $\sqrt{s}=1.96$ TeV jet binning~\cite{Abulencia:2007ez}.}
\end{figure}

The gluons-only theory predictions for the CDF setup are presented in
Figure~\ref{fig:cdfratio}. We observe that the level of agreement at NLO is
better with respect to the LHC comparisons presented in the previous
section. As before, Tables~\ref{tab:kcdf1} to~\ref{tab:kcdf5} show the
$k$-factors for the $gg$-channel, where non-perturbative corrections have
been applied to the experimental data as performed for ATLAS and CMS.

At NNLO the situation has improved since differences between exact and
threshold approximation results are smaller than what we have observed
for the LHC experiments. With a rejection criteria of excluding points where the disagreement
is larger than 10\% we observe that the last two rapidity slices for $|\eta|>1.1$ should be excluded.

The main differences between the Tevatron and LHC setups are the different
center of mass energies, the projective particles and kinematic ranges
which are shorter at the Tevatron than at the LHC. For these reasons, the threshold approximation
code provides at the Tevatron predictions closer to the exact NNLO calculation.

\begin{figure}
\begin{centering}
\includegraphics[scale=0.4]{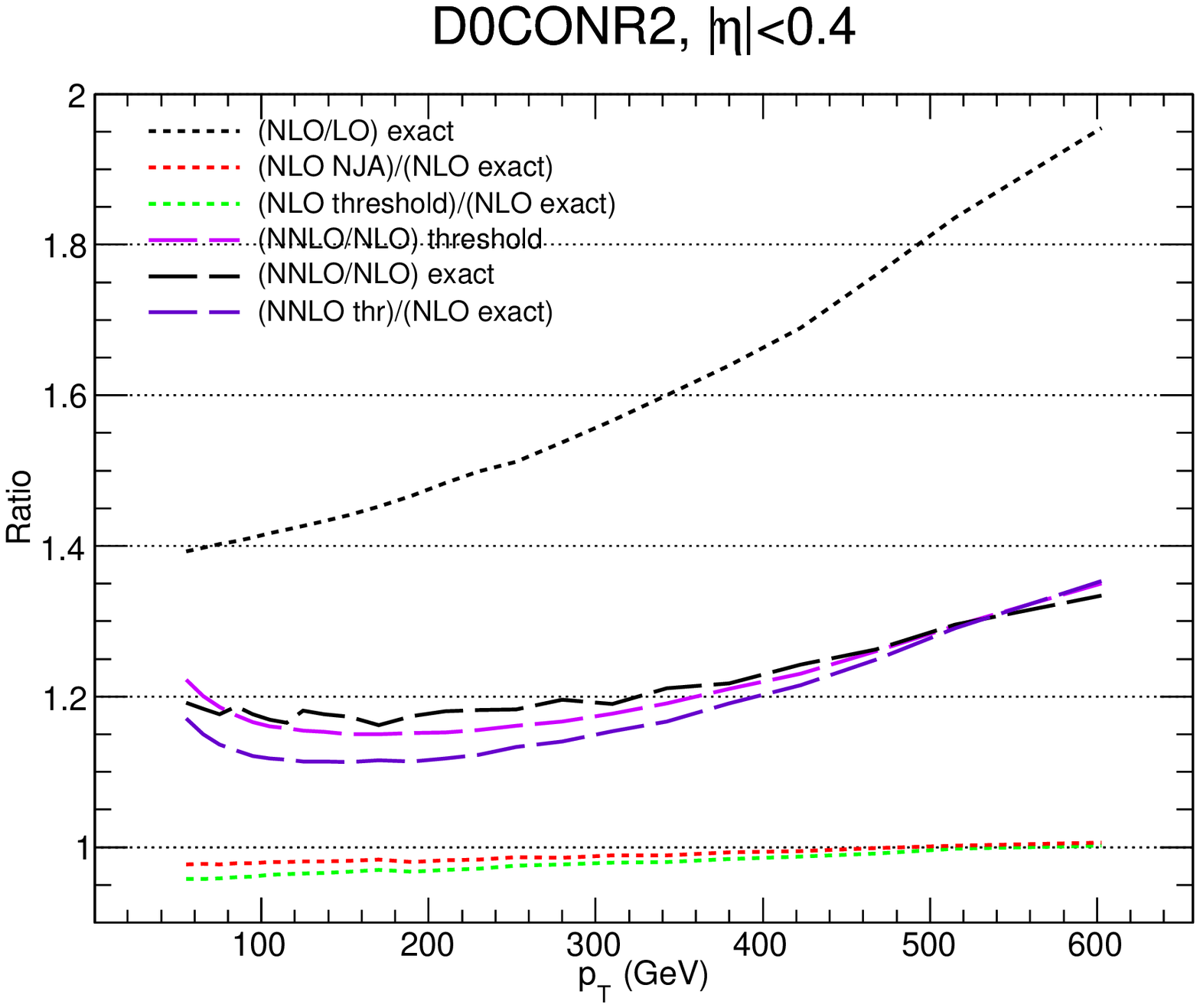}\includegraphics[scale=0.4]{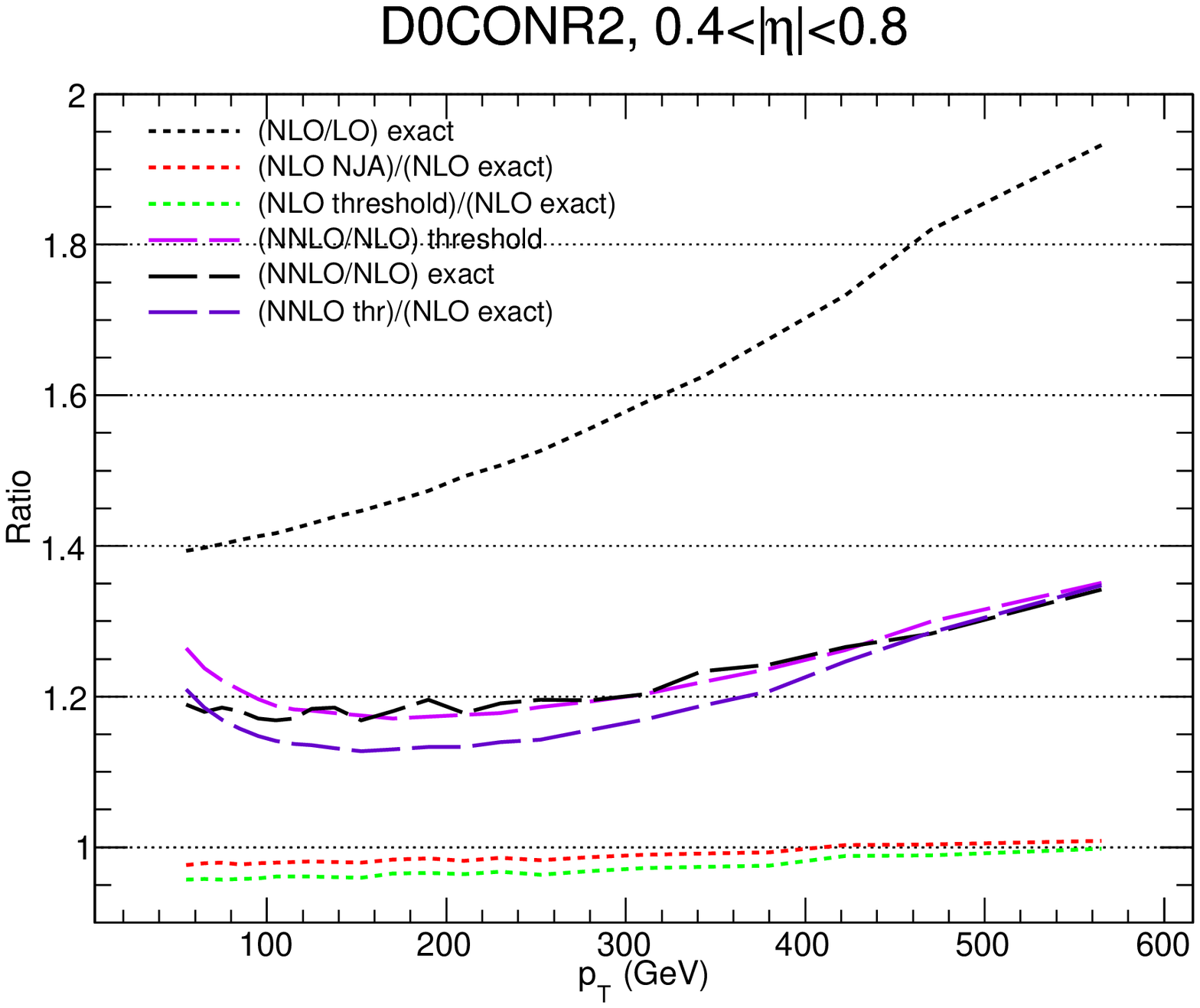}
\par\end{centering}

\begin{centering}
\includegraphics[scale=0.4]{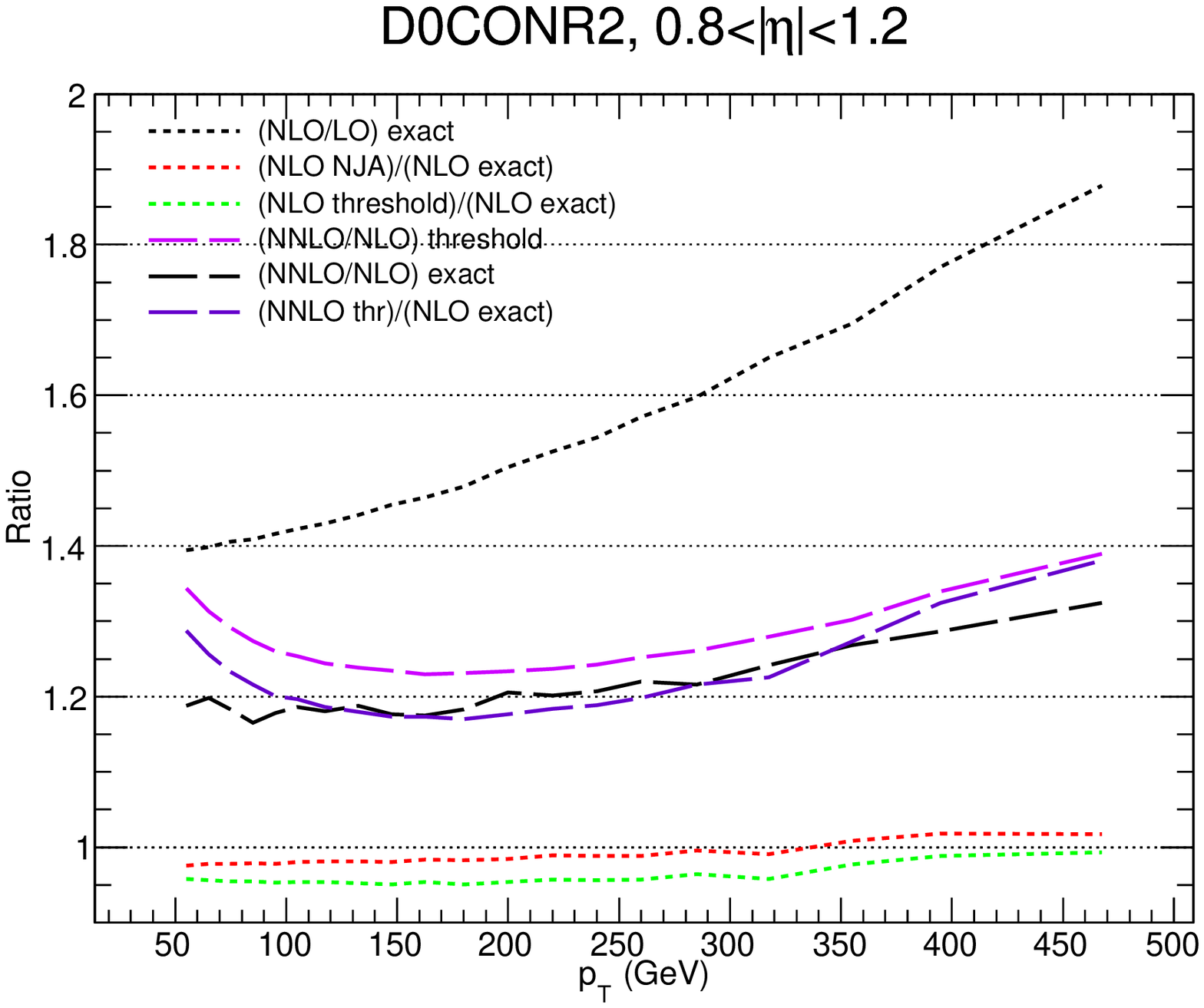}\includegraphics[scale=0.4]{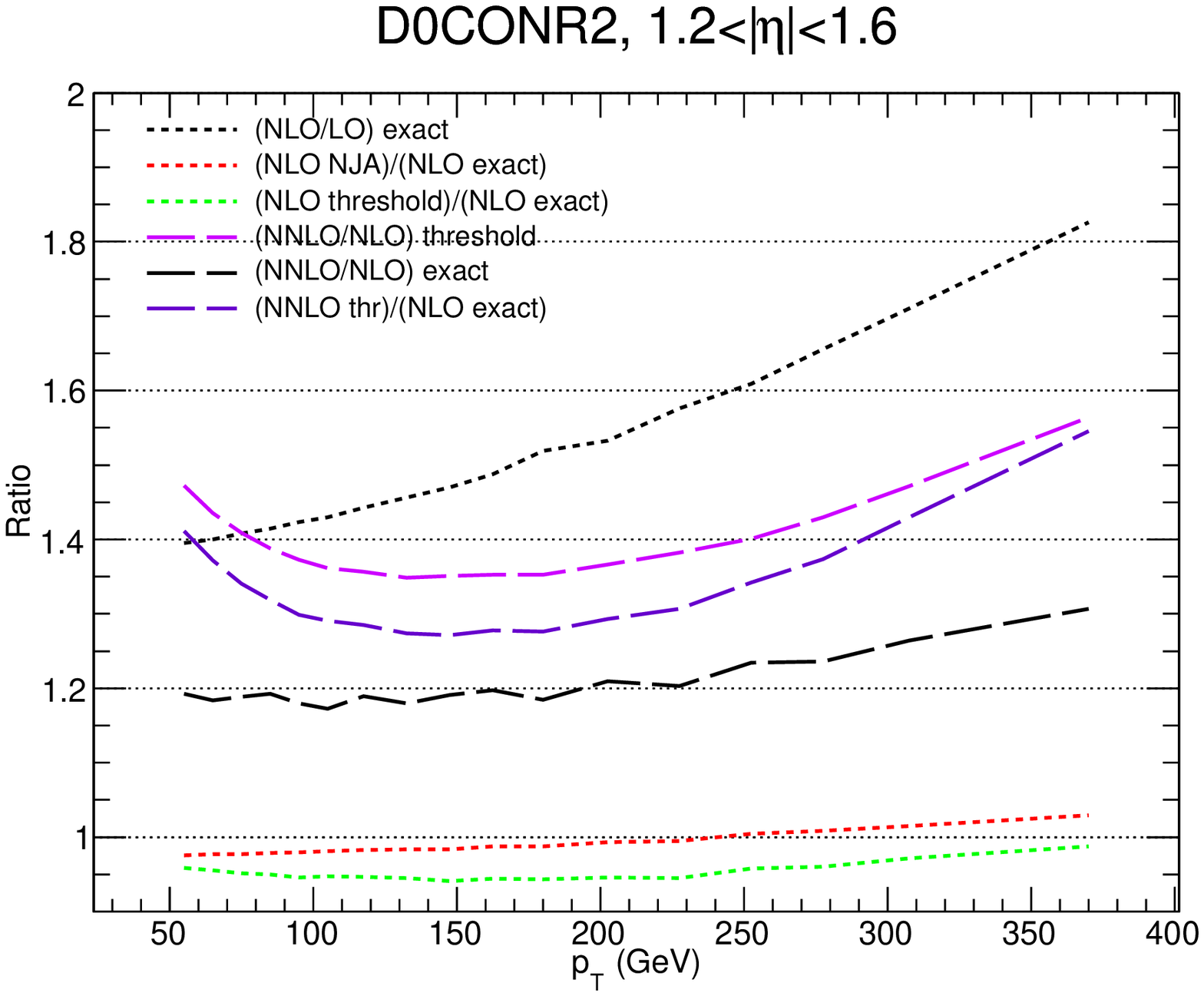}
\par\end{centering}

\begin{centering}
\includegraphics[scale=0.4]{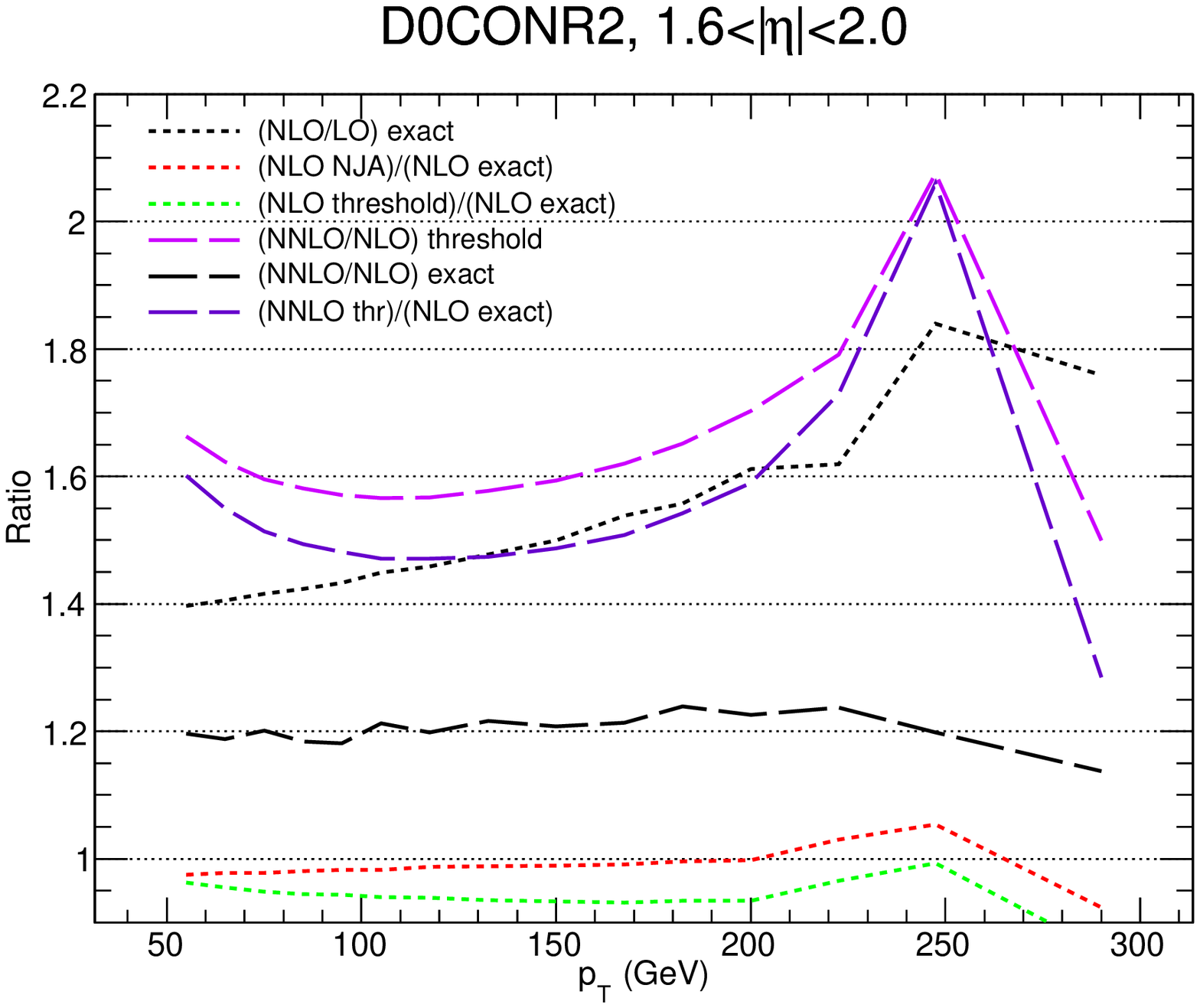}\includegraphics[scale=0.4]{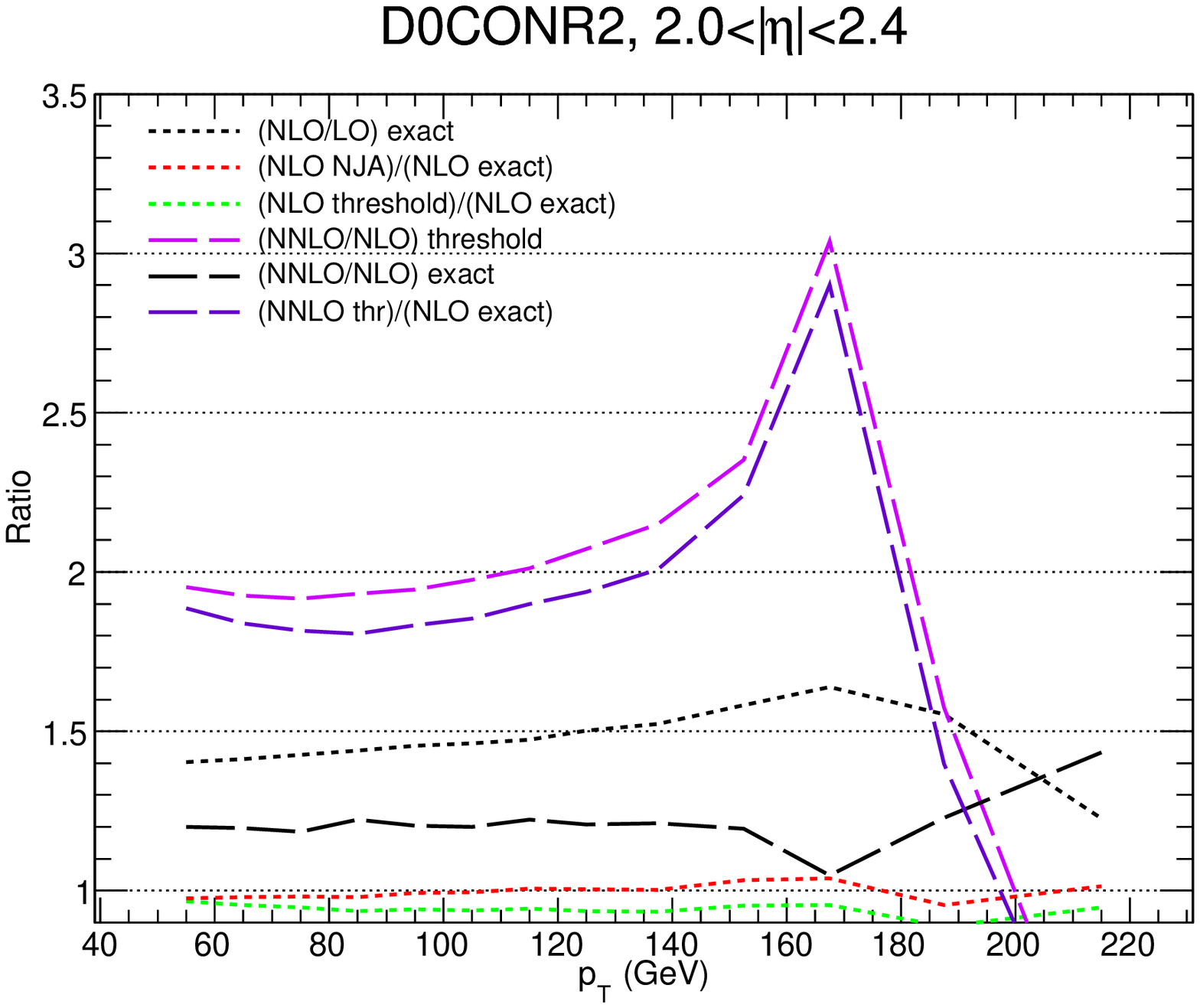}
\par\end{centering}

\caption{\label{fig:d0ratio} Ratios between exact and approximate
      predictions at the same order (LO and NLO) in perturbation theory in the gluons-only channel.
      In the same plot we present the 
      exact NLO/LO and NNLO/NLO $k$-factors (gluons-only channel) and the NNLO/NLO $k$-factors produced
      by the threshold approximation code (gluons-only channel) for the D0 $\sqrt{s}=1.96$ TeV jet binning~\cite{Abazov:2011vi}.}
\end{figure}

\subsection{D0 jets}

The results of the comparison for the D0 experimental setup are
presented in Figure~\ref{fig:d0ratio} with the respective
Tables~\ref{tab:kd01} to~\ref{tab:kd06}. As for CDF, the NLO 
prediction is in good agreement for all rapidity and $p_{T}$
bins. The NNLO predictions behave similarly to the CDF results, with
the threshold approximation code providing acceptable predictions at
least for the first three rapidity slices. Predictions for the D0
experiment have been generated using the $k_{t}$ algorithm for the jet
reconstruction instead of the MidPoint cone used for the measurement
because this algorithm is IR unsafe at NNLO.

This represents a drawback towards analysing jet data from D0 at
NNLO. At the moment D0 data has been included in PDF fits where the IR
finiteness at NLO of the MidPoint cone jet algorithm
allows the perturbative computation to be performed at this order. A possible solution to include a perturbative prediction at NNLO for the D0 data could be identifying
a relationship between the MidPoint algorithm and an IR safe cone algorithm such as SISCone. As discussed in~\cite{Salam:2007xv} relative differences between the 
two algorithms are expected to the start at ${\cal O}(\alpha_s^4)$  when we have 3 particles in a common neighbourhood plus one to balance the momentum. By producing an 
inclusive jet $p_T$ spectrum using just tree-level $2\to4$ diagrams, differences at the level of 1-2\% are observed at the Tevatron in~\cite{Salam:2007xv} between the predictions of both jet algorithms
when the SISCone algorithm is employed with cone parameters R=0.7 and $f$=0.5. A detailed study applying this prescription for the NNLO inclusive jet prediction
is beyond the scope of the current work.

\section{Exclusion criteria summary}
\label{sec:PDFfit}

In the previous sections we performed a comparison at NLO and NNLO
between the exact predictions and the approximate predictions from the
threshold resummation formalism at the same order. We performed this
comparison for all experimental hadron-hadron collider setups at the
LHC and the Tevatron from which a wealthy dataset of inclusive jet
data has been delivered.

\begin{figure}[t]
  \begin{centering}
    \includegraphics[scale=0.35]{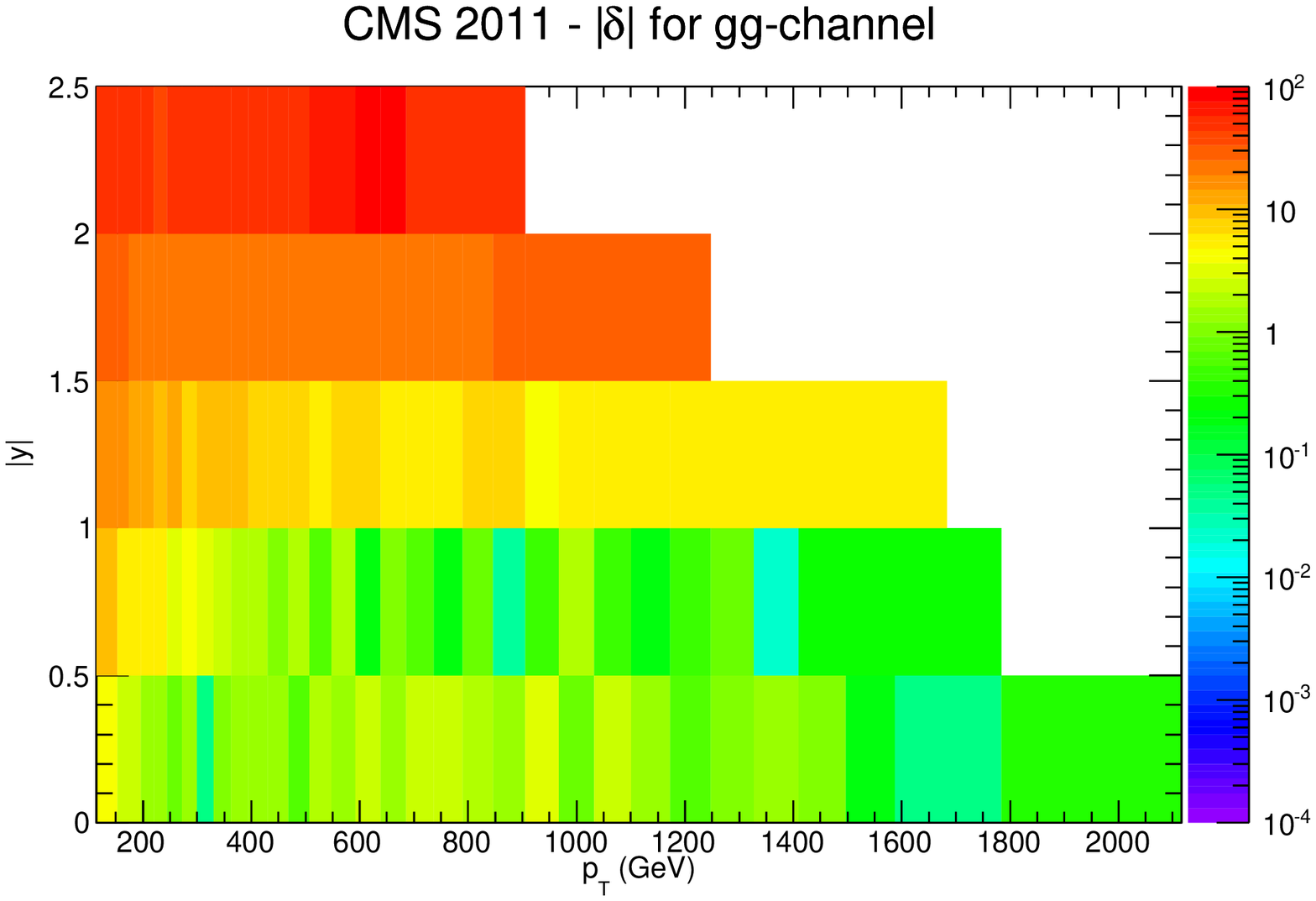}\includegraphics[scale=0.35]{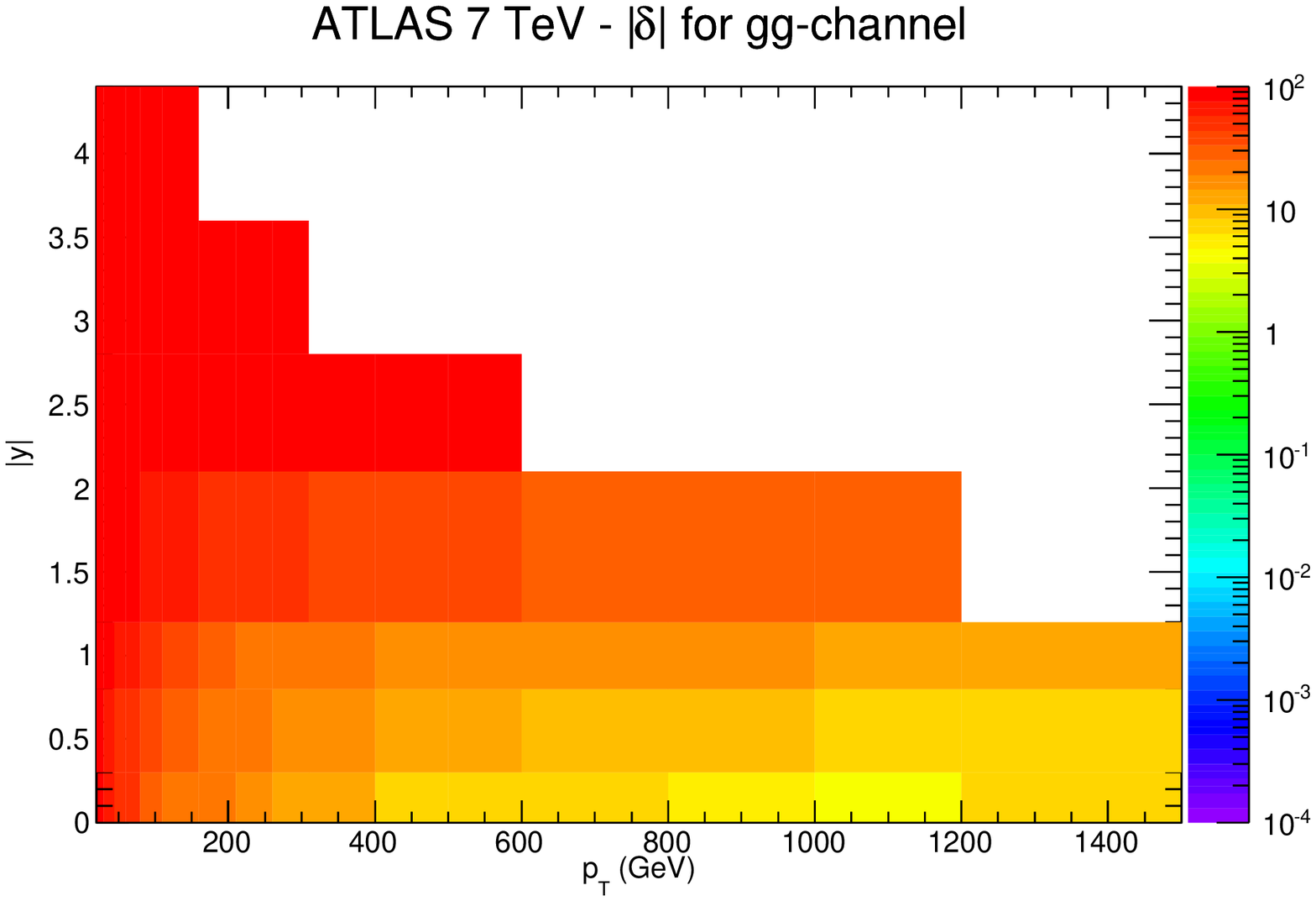}
    \par\end{centering}
  \begin{centering}
    \includegraphics[scale=0.35]{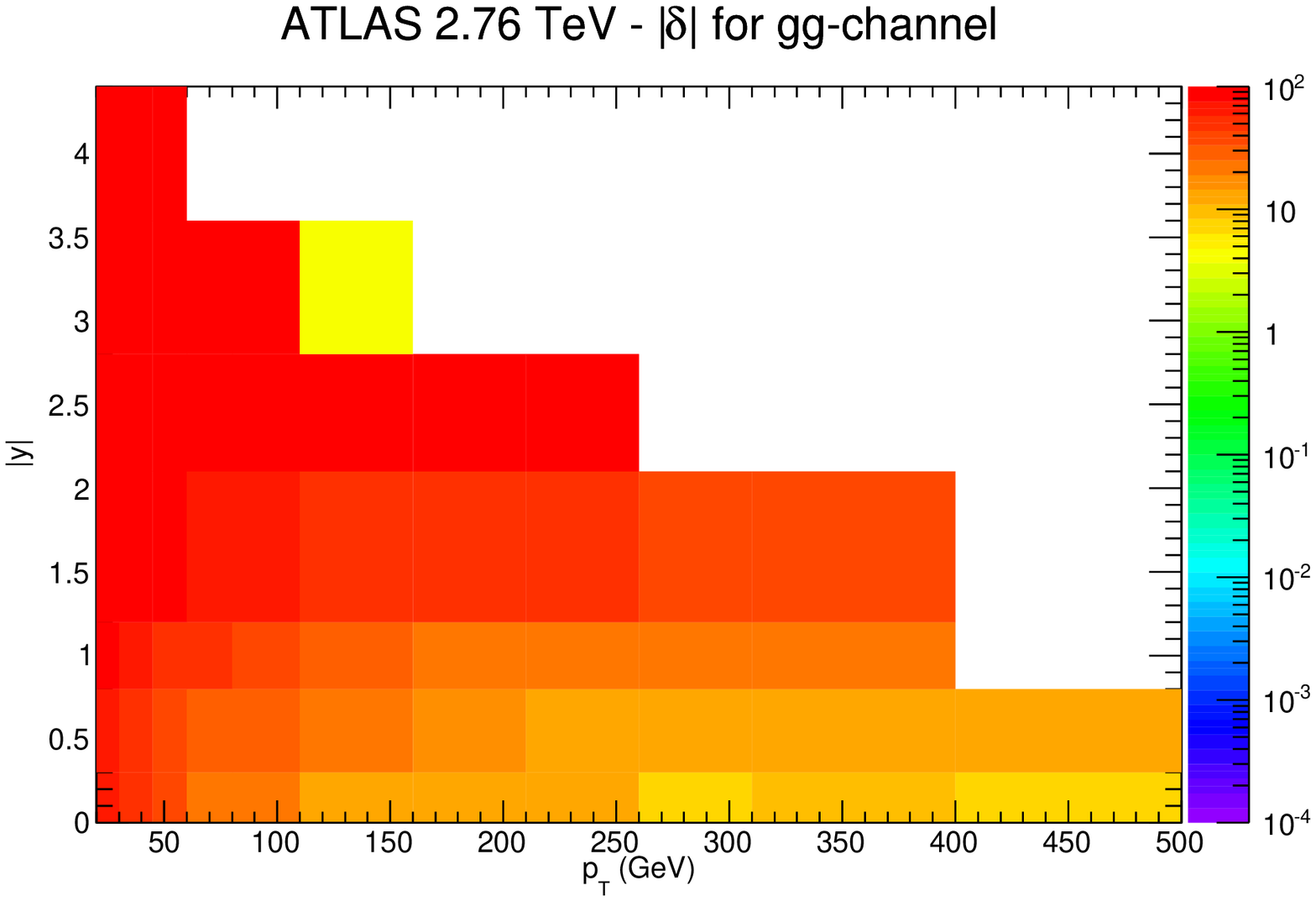}\includegraphics[scale=0.35]{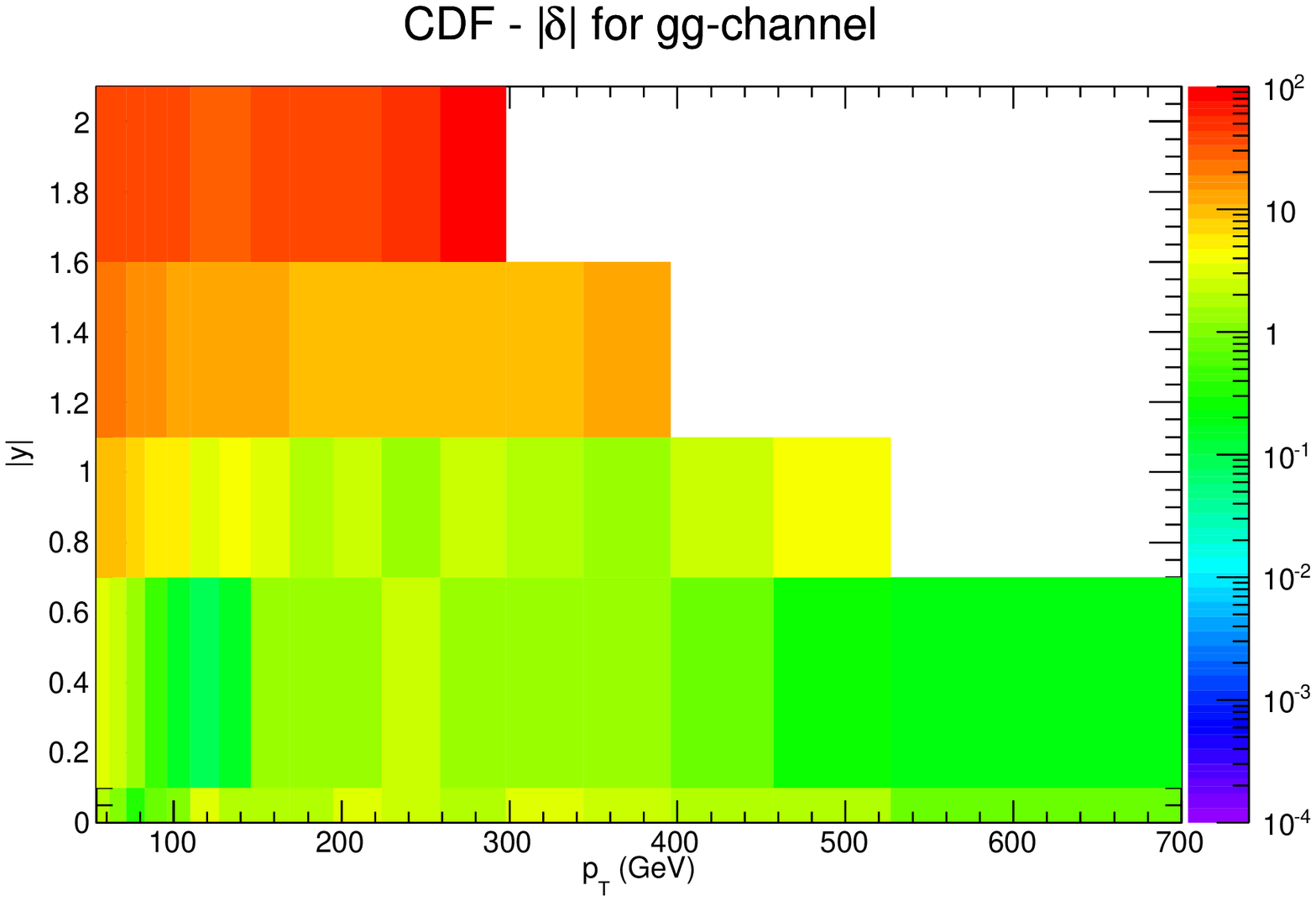}
  \par\end{centering}
\caption{
  \label{fig:exclusion} Percentage-wise relative difference $|\delta|$
  between the exact and approximate $gg$-channel NNLO $k$-factors as a
  function of $p_{T}$ and $|y|$ for CMS, ATLAS 7 TeV and 2.76 TeV and
  CDF bins. Disagreements larger than $|\delta|=100\%$ are represented
  in red.}
\end{figure}

This comparison was performed in the $gg$-channel where the exact NNLO
results were delivered first~\cite{Currie:2013dwa,Ridder:2013mf} and
therefore can be used to identify the regions where the approximate
NNLO prediction emulates the exact results. As a result of this
exercise we include in this work the full set of NLO/LO and NNLO/NLO
$k$-factors based on both approaches for the $gg$-channel and also the
approximate NNLO all-channel prediction. Using these predictions we
can identify kinematical regions where discrepancies between the two
results in the $gg$-channel are larger than $\delta$=10\%. In such
regions we conclude that the results of the approximate prediction
should not be trusted and for this reason, leads to an exclusion of
part of the full dataset of single jet inclusive cross section
measurements at hadron-hadron colliders that can be analysed. We
studied the effect of being more restrictive or relaxed with this
criteria and summarise experiment by experiment the resulting
exclusion regions of experimental data as a function of $\delta$ in
Tables~\ref{tab:PTpoints} to~\ref{tab:PTpointsCDF}.  As expected,
being more restrictive and demanding a smaller relative difference
between exact and threshold $k$-factors leads to an increased
exclusion region of data points. Using $\delta$=5\% results in
excluding more than half the data points from CMS and all data points
from ATLAS. This information is also reproduced in
Figure~\ref{fig:exclusion} where we show the relative difference
between the exact and approximate NNLO $k$-factor in the $gg$-channel
in the $(p_{T},|y|)$ plane of each experiment.

With this information at hand we aim in this section to perform
template aNNLO PDF fits of jet data including the approximate NNLO
corrections. We choose to perform four fits using a criteria of less
than 5\%, 7.5\% 10\% and 15\% relative disagreement between the exact
and the threshold $gg$-channel $k$-factors in order to exclude
approximate corrections, and assess the impact of these choices on the
quality of the fit. The kinematical regions which survive each cut are
introduced in the fit through the full channel threshold
approximation\footnote{Full channel approximate $k$-factors available
  at: \url{http://libhfill.hepforge.org/JetStudy2014}}. The study of
the effect in the fit of performing these cut variations also provides a
way for an empiric determination of the best exclusion criteria.  As
mentioned in Sect.~\ref{sec:tev} data from the D0 experiment will not
be included in this exercise.

To perform the fits we have used the latest {\tt NNPDF} fitting
technology~\cite{Ball:2012cx} to fit partial subsets of jet data
determined by the relative difference criteria $\delta$.  Moreover we
have also eliminated a few points at large rapidity bins where the
full channel $k$-factor is orders of magnitude larger that the
$gg$-channel $k$-factor and considered only datasets which after
applying the cuts contain at least 2 data points.

In Tables~\ref{tab:PTpoints} to~\ref{tab:PTpointsCDF} we present as a
function of $\delta$ the experimental $\chi^2/\textrm{dof}$ obtained
when fitting the respective subset of data points. For CMS the fits
tend to include more data points, in particular in high-$p_{T}$
regions and fairly central $|y|<1.0$ jets. In these regions, the jet
data is probing kinematics not constrained by other data and we
observe that the $\chi^2/\textrm{dof}$ has a small $|\delta|$
dependence. For CDF a large fraction of data points is included in the
fit. In this case, however, we noticed that the $\chi^2/\textrm{dof}$
improves when $\delta$ is reduced. Finally, for the ATLAS data, we
observe that by reducing $\delta$ a large fraction of data points is
excluded, and this results in large $\chi^2/\textrm{dof}$
fluctuations. In conclusion, with these results we cannot find a
precise exclusion criteria. However, we suggest a possible compromise
of $|\delta|=10\%$ which allows the inclusion of some data from all
experiments, providing a reasonable and stable $\chi^2/\textrm{dof}$
in all cases. In this way within the tolerance error chosen,
perturbative QCD corrections can be included in PDF fits and result in
a reduction of the gluon PDF uncertainty at high-$x$.

Finally we would like to point out that by computing the
$qg$~\cite{nnlojet:2014hp} and $qq$~\cite{nnlojet:2014hp} scattering
processes to exact NNLO accuracy in the fixed-order calculation along
the lines followed for $gg$~\cite{Currie:2013dwa,Ridder:2013mf} and
$q\bar{q}$~\cite{Currie:2013vh,Currie:2014upa} scattering, we avoid
the necessity of introducing a rejection criteria to exclude/include
approximate higher order predictions. Instead, the exact prediction
from the fixed-order calculation allows the use of the full dataset of
jet cross section measurements in global NNLO PDF fits of jet
data. Moreover we can test the resulting fit quality in the
description of any fully differential 2-jet observable at NNLO. We
leave this study for future work.

\begin{table}[H]
  \begin{centering}
    \begin{tabular}{l|c|c|c|l}
          \hline 
          CMS 2011 & $N_{{\rm dat}}$ & $\chi^{2}/$dof & \multicolumn{2}{c}{Exclusion regions ($y,p_{T}$)}\tabularnewline
          \hline 
          \hline 
          \multirow{2}{*}{$|\delta|<15\%$} & \multirow{2}{*}{88} & \multirow{2}{*}{1.81} & $1.0<|y|<1.5$ & $p_{T}<153$ GeV\tabularnewline
          &  &  & $|y|>1.5$ & all $p_{T}$ bins\tabularnewline
          \hline 
          \multirow{2}{*}{$|\delta|<10\%$} & \multirow{2}{*}{83} & \multirow{2}{*}{1.89} & $1.0<|y|<1.5$ & $p_{T}<272$ GeV\tabularnewline
          &  &  & $|y|>1.5$ & all $p_{T}$ bins\tabularnewline
          \hline 
          \multirow{3}{*}{$|\delta|<7.5\%$} & \multirow{3}{*}{77} & \multirow{3}{*}{1.89} & $0.5<|y|<1.0$ & $p_{T}<153$ GeV\tabularnewline
          &  &  & $1.0<|y|<1.5$ & $p_{T}<395$ GeV\tabularnewline
          &  &  & $|y|>1.5$ & all $p_{T}$ bins\tabularnewline
          \hline 
          \multirow{3}{*}{$|\delta|<5\%$} & \multirow{3}{*}{59} & \multirow{3}{*}{1.83} & $0.5<|y|<1.0$ & $p_{T}<220$ GeV\tabularnewline
          &  &  & $1.0<|y|<1.5$ & $p_{T}<737$ GeV, $p_{T}>790$ GeV \tabularnewline
          &  &  & $|y|>1.5$ & all $p_{T}$ bins\tabularnewline
          \hline 
        \end{tabular}
    
    \par\end{centering}
  
  \caption{ 
    \label{tab:PTpoints} Summary of exclusion regions in $p_T$ and rapidity $|y|$ as a
    function of the relative difference between exact and threshold
    $k$-factors for the gluon-gluon channel for the CMS 133 data points.  
    In the table we quote the $\chi^2/\textrm{dof}$ for aNNLO PDF fits performed
    with the full channel approximated $k$-factors. $N_{{\rm dat}}$ represents the number of experimental
    data points included in the fit.}
\end{table}

\begin{table}[H]
\begin{centering}
\begin{tabular}{l|c|c|c|l}
\hline 
ATLAS 2.76 TeV & $N_{{\rm dat}}$ & $\chi^{2}/$dof & \multicolumn{2}{c}{Exclusion regions ($y,p_{T}$)}\tabularnewline
\hline 
\hline 
\multirow{3}{*}{$|\delta|<15\%$} & \multirow{3}{*}{10} & \multirow{3}{*}{2.15} & $0.0<|y|<0.3$ & $p_{T}<110$ GeV\tabularnewline
 &  &  & $0.3<|y|<0.8$ & $p_{T}<210$ GeV\tabularnewline
 &  &  & $|y|>0.8$ & all $p_{T}$ bins\tabularnewline
\hline 
\multirow{2}{*}{$|\delta|<10\%$} & \multirow{2}{*}{3} & \multirow{2}{*}{0.35} & $0.0<|y|<0.3$ & $p_{T}<260$ GeV\tabularnewline
 &  &  & $|y|>0.3$ & all $p_{T}$ bins\tabularnewline
\hline 
\multirow{1}{*}{$|\delta|<7.5\%$} & \multirow{1}{*}{-} & \multirow{1}{*}{-} & all $|y|$ bins & all $p_{T}$ bins\tabularnewline
\hline 
\multirow{1}{*}{$|\delta|<5\%$} & \multirow{1}{*}{-} & \multirow{1}{*}{-} & all $|y|$ bins & all $p_{T}$ bins\tabularnewline
\hline 
\end{tabular}
\par\end{centering}

\caption{Summary of exclusion regions in $p_T$ and rapidity $|y|$ as a
    function of the relative difference between exact and threshold
    $k$-factors for the gluon-gluon channel for the ATLAS 2.76 TeV 59 data points. 
    In the table we quote the $\chi^2/\textrm{dof}$ for aNNLO PDF fits performed
    with the full channel approximated $k$-factors. $N_{{\rm dat}}$ represents the number of experimental
    data points included in the fit.}
\end{table}

\begin{table}[H]
  \begin{centering}
    \begin{tabular}{l|c|c|c|l}
          \hline 
          ATLAS 7 TeV & $N_{{\rm dat}}$ & $\chi^{2}/$dof & \multicolumn{2}{c}{Exclusion regions ($y,p_{T}$)}\tabularnewline
          \hline 
          \hline 
          \multirow{4}{*}{$|\delta|<15\%$} & \multirow{4}{*}{16} & \multirow{4}{*}{1.82} & $0.0<|y|<0.3$ & $p_{T}<260$ GeV\tabularnewline
          &  &  & $0.3<|y|<0.8$ & $p_{T}<400$ GeV\tabularnewline
          &  &  & $0.8<|y|<1.2$ & $p_{T}<1$ TeV\tabularnewline
          &  &  & $|y|>1.2$ & all $p_{T}$ bins\tabularnewline
          \hline 
          \multirow{3}{*}{$|\delta|<10\%$} & \multirow{3}{*}{9} & \multirow{3}{*}{1.58} & $0.0<|y|<0.3$ & $p_{T}<400$ GeV\tabularnewline
          &  &  & $0.3<|y|<0.8$ & $p_{T}<800$ GeV\tabularnewline
          &  &  & $|y|>0.8$ & all $p_{T}$ bins\tabularnewline
          \hline 
          \multirow{2}{*}{$|\delta|<7.5\%$} & \multirow{2}{*}{5} & \multirow{2}{*}{2.02} & $0.0<|y|<0.3$ & $p_{T}<500$ GeV\tabularnewline
          &  &  & $|y|>0.8$ & all $p_{T}$ bins\tabularnewline
          \hline 
          \multirow{2}{*}{$|\delta|<5\%$} & \multirow{2}{*}{1} & \multirow{2}{*}{-} & $0.0<|y|<0.3$ & $p_{T}<1$ TeV, $p_{T}>1.2$ TeV\tabularnewline
          &  &  & $|y|>0.3$ & all $p_{T}$ bins\tabularnewline
          \hline 
        \end{tabular}
    
    \par\end{centering}
  \caption{Summary of exclusion regions in $p_T$ and rapidity $|y|$ as a
    function of the relative difference between exact and threshold
    $k$-factors for the gluon-gluon channel for the ATLAS 7 TeV 90 data points. 
    In the table we quote the $\chi^2/\textrm{dof}$ for aNNLO PDF fits performed
    with the full channel approximated $k$-factors. $N_{{\rm dat}}$ represents the number of experimental
    data points included in the fit.}
\end{table}

\begin{table}[H]
\begin{centering}
\begin{tabular}{l|c|c|c|l}
\hline 
CDF & $N_{{\rm dat}}$ & $\chi^{2}/$dof & \multicolumn{2}{c}{Exclusion regions ($y,p_{T}$)}\tabularnewline
\hline 
\hline 
\multirow{2}{*}{$|\delta|<15\%$} & \multirow{2}{*}{60} & \multirow{2}{*}{2.32} & $1.1<|y|<1.6$ & $p_{T}<96$ GeV\tabularnewline
 &  &  & $|y|>1.6$ & all $p_{T}$ bins\tabularnewline
\hline 
\multirow{2}{*}{$|\delta|<10\%$} & \multirow{2}{*}{52} & \multirow{2}{*}{1.86} & $1.1<|y|<1.6$ & $p_{T}<224$ GeV, $p_{T}>298$ GeV\tabularnewline
 &  &  & $|y|>1.6$ & all $p_{T}$ bins\tabularnewline
\hline 
\multirow{2}{*}{$|\delta|<7.5\%$} & \multirow{2}{*}{48} & \multirow{2}{*}{1.37} & $0.7<|y|<1.1$ & $p_{T}<72$ GeV\tabularnewline
 &  &  & $|y|>1.1$ & all $p_{T}$ bins\tabularnewline
\hline 
\multirow{2}{*}{$|\delta|<5\%$} & \multirow{2}{*}{45} & \multirow{2}{*}{1.28} & $0.7<|y|<1.1$ & $p_{T}<110$ GeV\tabularnewline
 &  &  & $|y|>1.1$ & all $p_{T}$ bins\tabularnewline
\hline 
\end{tabular}
\par\end{centering}

\caption{\label{tab:PTpointsCDF} Summary of exclusion regions in $p_T$ and rapidity $|y|$ as a
    function of the relative difference between exact and threshold
    $k$-factors for the gluon-gluon channel for the CDF 76 data points.
    In the table we quote the $\chi^2/\textrm{dof}$ for aNNLO PDF fits performed
    with the full channel approximated $k$-factors. $N_{{\rm dat}}$ represents the number of experimental
    data points included in the fit.}
\end{table}

\section{Conclusion and outlook}
\label{sec:conclusion}

The purpose of this paper is to compare in the gluons-only channel predictions at NNLO for the single jet
inclusive cross section based on the exact NNLO calculation published in~\cite{Currie:2013dwa,Ridder:2013mf}
with an approximate NNLO calculation based on threshold resummation published in~\cite{deFlorian:2013qia}.
This comparison is performed using the same experimental setups employed by the Tevatron and LHC experiments in their jet analysis. 
With these results we deliver an updated description of the state of the art of the accuracy of the theoretical predictions 
for the single jet inclusive cross section in QCD at hadron colliders, and in particular revise contradictory statements
in the literature~\cite{deFlorian:2013qia}. 

We observe that when the predictions are compared using the same central scale choice for the renormalisation and factorisation
scales the disagreements are larger than previously quoted. Concerning the regions of validity of the NNLO approximation 
we conclude, based on a criteria of excluding approximate prediction which are more than 10\% off the exact prediction,
that the threshold approximation code provides predictions that are reasonably close to the exact calculation at large $p_{T}$
and central rapidity regions. We observed smaller differences between the exact calculation
and the approximate NNLO threshold calculations at the Tevatron than
at the LHC. It is important to highlight that
threshold predictions produced integrated over rapidity, as shown
in~\cite{deFlorian:2013qia}, are dominated by the central rapidity
regions and provide stable results. However, when looking at specific
rapidity bins the threshold predictions are, in some cases, far from the exact computation. This
remark is important and invites caution when using the threshold
approximation for the determination of PDFs. 

As an exercise and to test this observation we performed a PDF fit
including approximate NNLO corrections. We observed that as expected
the resulting fit quality is dependent on the criteria which is
employed to exclude/include approximate NNLO corrections. A more
conservative criteria has the effect of excluding a larger amount of
the experimental data points that go into the fit, and favours regions
where the approximate prediction gives smaller NNLO corrections in
agreement with the exact calculation.

Finally we conclude that with the current results, there is no trivial
way to determine the $p_{T}$ value for which the
threshold approximation predictions are reliable. The only possible
prescription is to check the relative difference to the exact
computation bin by bin, and admitting a tolerance which can be
correlated to the real data uncertainty. As we have shown, the regions of validity
of the threshold approximation are very dependent on the experimental setups
that we have analysed and are very likely to be different for the future high-energy Run-II of the LHC.

As a further improvement it would be interesting to repeat such study
when the NNLO exact prediction becomes available for all channels.

\acknowledgments

We thank Stefano Forte for intensive discussions and reflections about
the results presented in this document, Werner Vogelsang for providing
the threshold and NJA approximations codes and provide examples on how
to setup and run the respective codes. We also thank Juan Rojo and the
NNPDF Collaboration for discussions about PDF fits with jet data. JP
acknowledges Nigel Glover and Thomas Gehrmann for comments on the
manuscript.

JP and SC acknowledge support by an Italian PRIN2010 grant, and for SC
also by an European Investment Bank EIBURS grant, and by the European
Commission through the HiggsTools Initial Training Network
PITN-GA-2012-316704. JP thanks the Dipartimento di Fisica, Universita
di Milano-Bicocca for their kind hospitality.

\clearpage
\appendix
\section{Tables with $k$-factors for the gluon-gluon channel}

\label{sec:append}
In this Appendix we document the numerical results for the comparisons
between the NNLO threshold approximation and the NNLO exact
calculation in the gluons-only channel.  In the following tables we
show for each $p_T$ bin of each experiment in columns 2 and 3 the
experimental cross section together with its experimental uncertainty
computed as described in Sect.~\ref{sec:bench}.  Additionally we give
NNLO/NLO gluons-only $k$-factors with both NNLO and NLO results
computed in the exact calculation (column 4) and in the threshold
approximation (column 5). The percentage wise relative difference
between the two is given in column 6. For completeness we give also
the NNLO threshold $k$-factor using the NLO exact calculation in the
denominator (column 7) and using the approximate NLO threshold
calculation in the denominator (column 8). Their percentage wise
relative difference is given in column 9.

\newpage
\subsection{CMS jets}
\vspace{-0.2cm}

\begin{table}[H]
\centering{}%
\begin{tabular}{|c|cc||cc|c||cc|c|}
\hline 
\multicolumn{3}{|c||}{$|\eta|<0.5$} & \multicolumn{2}{c|}{gg $k$-factor} &  & \multicolumn{2}{c|}{gg $k$-factor, NNLO$^{{\rm thr}}/$} & \tabularnewline
\hline 
$p_{T}$ bin & $\sigma_{{\rm Exp}}$ & $\sigma_{{\rm Exp}}^{{\rm error}}$ (\%) & Exact & Thr & $\delta\%$ & NLO$^{{\rm exact}}$ & NLO$^{{\rm thr}}$ & $\delta$\%\tabularnewline
\hline 
\hline 
1 & 2.6410e+03 & 7.6605 & 1.1604 & 1.2129 & 4.52 & 1.1685 & 1.2129 & 3.80\tabularnewline
\hline 
2 & 1.2183e+03 & 7.1542 & 1.1494 & 1.1932 & 3.81 & 1.1483 & 1.1932 & 3.91\tabularnewline
\hline 
3 & 5.8988e+02 & 6.8006 & 1.1507 & 1.1771 & 2.29 & 1.1327 & 1.1771 & 3.92\tabularnewline
\hline 
4 & 2.9063e+02 & 6.4464 & 1.1393 & 1.1649 & 2.25 & 1.1214 & 1.1649 & 3.88\tabularnewline
\hline 
5 & 1.5159e+02 & 6.1977 & 1.1372 & 1.1545 & 1.52 & 1.1123 & 1.1545 & 3.79\tabularnewline
\hline 
6 & 8.1069e+01 & 6.1166 & 1.1315 & 1.1453 & 1.22 & 1.1032 & 1.1453 & 3.82\tabularnewline
\hline 
\end{tabular}{\caption{\label{tab:kcms1} Numerical results for the gluons-only exact and approximate NNLO $k$-factors as described in Sect.~\ref{sec:append}
for the CMS 2011 $\sqrt{s}=7$ TeV dataset~\cite{Chatrchyan:2012bja} in the rapidity slice $|\eta|<0.5$.}
}
\end{table}

\begin{table}[h]
\begin{centering}
\begin{tabular}{|c|cc||cc|c||cc|c|}
\hline 
\multicolumn{3}{|c||}{$0.5<|\eta|<1.0$} & \multicolumn{2}{c|}{gg $k$-factor} &  & \multicolumn{2}{c|}{gg $k$-factor, NNLO$^{{\rm thr}}/$} & \tabularnewline
\hline 
$p_{T}$ bin & $\sigma_{{\rm Exp}}$ & $\sigma_{{\rm Exp}}^{{\rm error}}$ (\%) & Exact & Thr & $\delta\%$ & NLO$^{{\rm exact}}$ & NLO$^{{\rm thr}}$ & $\delta\%$\tabularnewline
\hline 
\hline 
1 & 2.5139e+03 & 7.5804 & 1.1542 & 1.2592 & 9.10 & 1.2159 & 1.2592 & 3.56\tabularnewline
\hline 
2 & 1.1242e+03 & 7.2061 & 1.1388 & 1.2379 & 8.70 & 1.193 & 1.2379 & 3.76\tabularnewline
\hline 
3 & 5.3237e+02 & 6.7691 & 1.149 & 1.2203 & 6.21 & 1.176 & 1.2203 & 3.77\tabularnewline
\hline 
4 & 2.6596e+02 & 6.399 & 1.138 & 1.2079 & 6.14 & 1.1626 & 1.2079 & 3.90\tabularnewline
\hline 
5 & 1.3774e+02 & 6.2308 & 1.1373 & 1.1955 & 5.12 & 1.1501 & 1.1955 & 3.95\tabularnewline
\hline 
6 & 7.2171e+01 & 6.0829 & 1.131 & 1.1853 & 4.80 & 1.1406 & 1.1853 & 3.92\tabularnewline
\hline 
\end{tabular}
\par\end{centering}

\caption{\label{tab:kcms2}Numerical results for the gluons-only exact and approximate NNLO $k$-factors as described in Sect.~\ref{sec:append}
  for the CMS 2011 $\sqrt{s}=7$ TeV dataset~\cite{Chatrchyan:2012bja} in the rapidity slice $0.5<|\eta|<1.0$.}
\end{table}

\begin{table}[h]
\centering{}%
\begin{tabular}{|c|cc||cc|c||cc|c|}
\hline 
\multicolumn{3}{|c||}{$1.0<|\eta|<1.5$} & \multicolumn{2}{c|}{gg $k$-factor} &  & \multicolumn{2}{c|}{gg $k$-factor, NNLO$^{{\rm thr}}/$} & \tabularnewline
\hline 
$p_{T}$ bin & $\sigma_{{\rm Exp}}$ & $\sigma_{{\rm Exp}}^{{\rm error}}$ (\%) & Exact & Thr & $\delta\%$ & NLO$^{{\rm exact}}$ & NLO$^{{\rm thr}}$ & $\delta\%$\tabularnewline
\hline 
\hline 
1 & 2.1264e+03 & 8.5345 & 1.1464 & 1.3606 & 18.68 & 1.3193 & 1.3606 & 3.13\tabularnewline
\hline 
2 & 9.4841e+02 & 7.9455 & 1.154 & 1.3343 & 15.62 & 1.2906 & 1.3343 & 3.39\tabularnewline
\hline 
3 & 4.4306e+02 & 7.4404 & 1.1424 & 1.3102 & 14.69 & 1.2648 & 1.3102 & 3.59\tabularnewline
\hline 
4 & 2.1618e+02 & 7.0173 & 1.1386 & 1.2929 & 13.55 & 1.2443 & 1.2929 & 3.91\tabularnewline
\hline 
5 & 1.1088e+02 & 6.8155 & 1.1384 & 1.2802 & 12.46 & 1.23 & 1.2802 & 4.08\tabularnewline
\hline 
6 & 5.8338e+01 & 6.6866 & 1.1465 & 1.267 & 10.51 & 1.2159 & 1.267 & 4.20\tabularnewline
\hline 
\end{tabular}\caption{\label{tab:kcms3}Numerical results for the gluons-only exact and approximate NNLO $k$-factors as described in Sect.~\ref{sec:append}
for the CMS 2011 $\sqrt{s}=7$ TeV dataset~\cite{Chatrchyan:2012bja} in the rapidity slice $1.0<|\eta|<1.5$.}
\end{table}

\begin{table}[h]
\begin{centering}
\begin{tabular}{|c|cc||cc|c||cc|c|}
\hline 
\multicolumn{3}{|c||}{$1.5<|\eta|<2.0$} & \multicolumn{2}{c|}{gg $k$-factor} &  & \multicolumn{2}{c|}{gg $k$-factor, NNLO$^{{\rm thr}}/$} & \tabularnewline
\hline 
$p_{T}$ bin & $\sigma_{{\rm Exp}}$ & $\sigma_{{\rm Exp}}^{{\rm error}}$ (\%) & Exact & Thr & $\delta\%$ & NLO$^{{\rm exact}}$ & NLO$^{{\rm thr}}$ & $\delta\%$\tabularnewline
\hline 
\hline 
1 & 1.6383e+03 & 9.8283 & 1.1528 & 1.5131 & 31.25 & 1.4825 & 1.5131 & 2.06\tabularnewline
\hline 
2 & 7.2786e+02 & 9.4308 & 1.1491 & 1.4836 & 29.11 & 1.444 & 1.4836 & 2.74\tabularnewline
\hline 
3 & 3.3039e+02 & 9.2082 & 1.1511 & 1.4595 & 26.79 & 1.414 & 1.4595 & 3.22\tabularnewline
\hline 
4 & 1.5735e+02 & 8.903 & 1.1492 & 1.4359 & 24.95 & 1.3865 & 1.4359 & 3.56\tabularnewline
\hline 
5 & 7.8425e+01 & 8.987 & 1.1577 & 1.4236 & 22.97 & 1.3671 & 1.4236 & 4.13\tabularnewline
\hline 
6 & 3.8482e+01 & 9.0353 & 1.1385 & 1.4091 & 23.77 & 1.3517 & 1.4091 & 4.25\tabularnewline
\hline 
\end{tabular}
\par\end{centering}

\caption{\label{tab:kcms4}Numerical results for the gluons-only exact and approximate NNLO $k$-factors as described in Sect.~\ref{sec:append}
for the CMS 2011 $\sqrt{s}=7$ TeV dataset~\cite{Chatrchyan:2012bja} in the rapidity slice $1.5<|\eta|<2.0$.}
\end{table}

\begin{table}[h]
\centering{}%
\begin{tabular}{|c|cc||cc|c||cc|c|}
\hline 
\multicolumn{3}{|c||}{$2.0<|\eta|<2.5$} & \multicolumn{2}{c|}{gg $k$-factor} &  & \multicolumn{2}{c|}{gg $k$-factor, NNLO$^{{\rm thr}}/$} & \tabularnewline
\hline 
$p_{T}$ bin & $\sigma_{{\rm Exp}}$ & $\sigma_{{\rm Exp}}^{{\rm error}}$ (\%) & Exact & Thr & $\delta\%$ & NLO$^{{\rm exact}}$ & NLO$^{{\rm thr}}$ & $\delta\%$\tabularnewline
\hline 
\hline 
1 & 1.1132e+03 & 14.1934 & 1.1456 & 1.7432 & 52.16 & 1.7273 & 1.7432 & 0.92\tabularnewline
\hline 
2 & 4.7126e+02 & 14.1554 & 1.1492 & 1.7157 & 49.30 & 1.6835 & 1.7157 & 1.91\tabularnewline
\hline 
3 & 2.1037e+02 & 14.375 & 1.1575 & 1.6903 & 46.03 & 1.6466 & 1.6903 & 2.65\tabularnewline
\hline 
4 & 9.5500e+01 & 14.4644 & 1.1525 & 1.6788 & 45.67 & 1.6254 & 1.6788 & 3.29\tabularnewline
\hline 
5 & 4.2179e+01 & 14.8869 & 1.159 & 1.6734 & 44.38 & 1.608 & 1.6734 & 4.07\tabularnewline
\hline 
6 & 1.9811e+01 & 15.3918 & 1.165 & 1.6704 & 43.38 & 1.5994 & 1.6704 & 4.44\tabularnewline
\hline 
\end{tabular}\caption{\label{tab:kcms5}Numerical results for the gluons-only exact and approximate NNLO $k$-factors as described in Sect.~\ref{sec:append}
for the CMS 2011 $\sqrt{s}=7$ TeV dataset~\cite{Chatrchyan:2012bja} in the rapidity slice $2.0<|\eta|<2.5$.}
\end{table}

\clearpage

\subsection{ATLAS jets at $\sqrt{s}=7$ TeV}

\begin{table}[h]
\centering{}%
\begin{tabular}{|c|cc||cc|c||cc|c|}
\hline 
\multicolumn{3}{|c||}{$|\eta|<0.3$} & \multicolumn{2}{c|}{gg $k$-factor} &  & \multicolumn{2}{c|}{gg $k$-factor, NNLO$^{{\rm thr}}/$} & \tabularnewline
\hline 
$p_{T}$ bin & $\sigma_{{\rm Exp}}$ & $\sigma_{{\rm Exp}}^{{\rm error}}$ (\%) & Exact & Thr & $\delta\%$ & NLO$^{{\rm exact}}$ & NLO$^{{\rm thr}}$ & $\delta$\%\tabularnewline
\hline 
\hline 
1 & 4.7475e+06 & 23.3268 & 1.1405 & 2.1444 & 88.02 & 2.176 & 2.1444 & -1.45\tabularnewline
\hline 
2 & 7.2424e+05 & 18.5066 & 1.0848 & 1.8308 & 68.77 & 1.8118 & 1.8308 & 1.05\tabularnewline
\hline 
3 & 1.4949e+05 & 13.9274 & 1.0779 & 1.6208 & 50.37 & 1.5817 & 1.6208 & 2.47\tabularnewline
\hline 
4 & 3.8485e+04 & 12.0152 & 1.0383 & 1.4933 & 43.82 & 1.4502 & 1.4933 & 2.97\tabularnewline
\hline 
5 & 8.6061e+03 & 12.261 & 1.0486 & 1.3936 & 32.90 & 1.3497 & 1.3936 & 3.25\tabularnewline
\hline 
6 & 1.4949e+03 & 10.6946 & 1.0675 & 1.3075 & 22.48 & 1.2645 & 1.3075 & 3.40\tabularnewline
\hline 
\end{tabular}{\small \caption{\label{tab:katlas71}Numerical results for the gluons-only exact and approximate NNLO $k$-factors as described in Sect.~\ref{sec:append}
    for the ATLAS 2010 $\sqrt{s}=7$ TeV dataset~\cite{Aad:2010ad} in the rapidity slice $|\eta|<0.3$.}
}
\end{table}

\begin{table}[h]
\centering{}%
\begin{tabular}{|c|cc||cc|c||cc|c|}
\hline 
\multicolumn{3}{|c||}{$0.3<|\eta|<0.8$} & \multicolumn{2}{c|}{gg $k$-factor} &  & \multicolumn{2}{c|}{gg $k$-factor, NNLO$^{{\rm thr}}/$} & \tabularnewline
\hline 
$p_{T}$ bin & $\sigma_{{\rm Exp}}$ & $\sigma_{{\rm Exp}}^{{\rm error}}$ (\%) & Exact & Thr & $\delta\%$ & NLO$^{{\rm exact}}$ & NLO$^{{\rm thr}}$ & $\delta$\%\tabularnewline
\hline 
\hline 
1 & 4.7041e+06 & 23.275 & 1.0949 & 2.2051 & 101.40 & 2.2597 & 2.2051 & -2.42\tabularnewline
\hline 
2 & 7.3737e+05 & 18.9232 & 1.0858 & 1.8855 & 73.65 & 1.882 & 1.8855 & 0.19\tabularnewline
\hline 
3 & 1.5051e+05 & 14.1032 & 1.0638 & 1.6706 & 57.04 & 1.643 & 1.6706 & 1.68\tabularnewline
\hline 
4 & 3.7475e+04 & 12.0853 & 1.0535 & 1.542 & 46.37 & 1.5039 & 1.542 & 2.53\tabularnewline
\hline 
5 & 8.4343e+03 & 11.4745 & 1.0429 & 1.436 & 37.69 & 1.3944 & 1.436 & 2.98\tabularnewline
\hline 
6 & 1.4545e+03 & 11.7405 & 1.0502 & 1.3483 & 28.39 & 1.3059 & 1.3483 & 3.25\tabularnewline
\hline 
\end{tabular}{\small \caption{\label{tab:katlas72}Numerical results for the gluons-only exact and approximate NNLO $k$-factors as described in Sect.~\ref{sec:append}
for the ATLAS 2010 $\sqrt{s}=7$ TeV dataset~\cite{Aad:2010ad} in the rapidity slice $0.3<|\eta|<0.8$.}
}
\end{table}

\begin{table}[h]
\centering{}%
\begin{tabular}{|c|cc||cc|c||cc|c|}
\hline 
\multicolumn{3}{|c||}{$0.8<|\eta|<1.2$} & \multicolumn{2}{c|}{gg $k$-factor} &  & \multicolumn{2}{c|}{gg $k$-factor, NNLO$^{{\rm thr}}/$} & \tabularnewline
\hline 
$p_{T}$ bin & $\sigma_{{\rm Exp}}$ & $\sigma_{{\rm Exp}}^{{\rm error}}$ (\%) & Exact & Thr & $\delta\%$ & NLO$^{{\rm exact}}$ & NLO$^{{\rm thr}}$ & $\delta$\%\tabularnewline
\hline 
\hline 
1 & 4.0000e+06 & 27.1532 & 1.1068 & 2.3304 & 110.55 & 2.4438 & 2.3304 & -4.64\tabularnewline
\hline 
2 & 6.7959e+05 & 20.6534 & 1.1082 & 2.0116 & 81.52 & 2.0448 & 2.0116 & -1.62\tabularnewline
\hline 
3 & 1.4040e+05 & 15.5795 & 1.0721 & 1.7867 & 66.65 & 1.7808 & 1.7867 & 0.33\tabularnewline
\hline 
4 & 3.4242e+04 & 12.8432 & 1.0481 & 1.6531 & 57.72 & 1.6305 & 1.6531 & 1.39\tabularnewline
\hline 
5 & 7.6263e+03 & 12.4546 & 1.0456 & 1.5445 & 47.71 & 1.5136 & 1.5445 & 2.04\tabularnewline
\hline 
6 & 1.2929e+03 & 12.3535 & 1.0497 & 1.4457 & 37.73 & 1.4051 & 1.4457 & 2.89\tabularnewline
\hline 
7 & 2.0707e+02 & 12.6587 & 1.0332 & 1.367 & 32.31 & 1.3204 & 1.367 & 3.53\tabularnewline
\hline 
8 & 5.0100e+01 & 13.8718 & 1.0586 & 1.323 & 24.98 & 1.2765 & 1.323 & 3.64\tabularnewline
\hline 
9 & 1.5900e+01 & 14.1359 & 1.0352 & 1.2929 & 24.89 & 1.2424 & 1.2929 & 4.06\tabularnewline
\hline 
10 & 4.3800e+00 & 14.2512 & 1.0362 & 1.2711 & 22.67 & 1.2224 & 1.2711 & 3.98\tabularnewline
\hline 
11 & 9.1800e-01 & 15.507 & 1.0553 & 1.2485 & 18.31 & 1.2052 & 1.2485 & 3.59\tabularnewline
\hline 
12 & 2.1400e-01 & 16.8993 & 1.055 & 1.2346 & 17.02 & 1.1773 & 1.2346 & 4.87\tabularnewline
\hline 
13 & 4.0700e-02 & 20.2256 & 1.0547 & 1.2328 & 16.89 & 1.1807 & 1.2328 & 4.41\tabularnewline
\hline 
14 & 2.7000e-03 & 32.8892 & 1.0711 & 1.232 & 15.02 & 1.1665 & 1.232 & 5.62\tabularnewline
\hline 
15 & 4.6600e-04 & 69.3513 & 1.0877 & 1.2384 & 13.85 & 1.1774 & 1.2384 & 5.18\tabularnewline
\hline 
16 & 1.0200e-04 & 113.3008 & 1.1179 & 1.2575 & 12.49 & 1.21 & 1.2575 & 3.93\tabularnewline
\hline 
\end{tabular}{\small \caption{\label{tab:katlas73}Numerical results for the gluons-only exact and approximate NNLO $k$-factors as described in Sect.~\ref{sec:append}
for the ATLAS 2010 $\sqrt{s}=7$ TeV dataset~\cite{Aad:2010ad} in the rapidity slice $0.8<|\eta|<1.2$.}
}
\end{table}

\begin{table}[h]
\centering{}%
\begin{tabular}{|c|cc||cc|c||cc|c|}
\hline 
\multicolumn{3}{|c||}{$1.2<|\eta|<2.1$} & \multicolumn{2}{c|}{gg $k$-factor} &  & \multicolumn{2}{c|}{gg $k$-factor, NNLO$^{{\rm thr}}/$} & \tabularnewline
\hline 
$p_{T}$ bin & $\sigma_{{\rm Exp}}$ & $\sigma_{{\rm Exp}}^{{\rm error}}$ (\%) & Exact & Thr & $\delta\%$ & NLO$^{{\rm exact}}$ & NLO$^{{\rm thr}}$ & $\delta$\%\tabularnewline
\hline 
\hline 
1 & 3.8144e+06 & 28.3609 & 1.0859 & 2.6535 & 144.36 & 2.9287 & 2.6535 & -9.40\tabularnewline
\hline 
2 & 5.7938e+05 & 23.1683 & 1.079 & 2.3145 & 114.50 & 2.4513 & 2.3145 & -5.58\tabularnewline
\hline 
3 & 1.1122e+05 & 17.3887 & 1.0492 & 2.0782 & 98.07 & 2.1363 & 2.0782 & -2.72\tabularnewline
\hline 
4 & 2.7245e+04 & 14.9045 & 1.0528 & 1.9272 & 83.05 & 1.9512 & 1.9272 & -1.23\tabularnewline
\hline 
5 & 5.9592e+03 & 13.5442 & 1.0351 & 1.8033 & 74.22 & 1.7962 & 1.8033 & 0.40\tabularnewline
\hline 
6 & 9.6768e+02 & 12.2931 & 1.036 & 1.688 & 62.93 & 1.6567 & 1.688 & 1.89\tabularnewline
\hline 
\end{tabular}{\small \caption{\label{tab:katlas74}Numerical results for the gluons-only exact and approximate NNLO $k$-factors as described in Sect.~\ref{sec:append}
for the ATLAS 2010 $\sqrt{s}=7$ TeV dataset~\cite{Aad:2010ad} in the rapidity slice $1.2<|\eta|<2.1$.}
}
\end{table}

\begin{table}[h]
\centering{}%
\begin{tabular}{|c|cc||cc|c||cc|c|}
\hline 
\multicolumn{3}{|c||}{$2.1<|\eta|<2.8$} & \multicolumn{2}{c|}{gg $k$-factor} &  & \multicolumn{2}{c|}{gg $k$-factor, NNLO$^{{\rm thr}}/$} & \tabularnewline
\hline 
$p_{T}$ bin & $\sigma_{{\rm Exp}}$ & $\sigma_{{\rm Exp}}^{{\rm error}}$ (\%) & Exact & Thr & $\delta\%$ & NLO$^{{\rm exact}}$ & NLO$^{{\rm thr}}$ & $\delta$\%\tabularnewline
\hline 
\hline 
1 & 2.8750e+06 & 35.9715 & 1.0832 & 3.2222 & 197.47 & 3.8805 & 3.2222 & -16.96\tabularnewline
\hline 
2 & 4.2396e+05 & 30.3757 & 1.0753 & 2.8748 & 167.35 & 3.2745 & 2.8748 & -12.21\tabularnewline
\hline 
3 & 8.2474e+04 & 20.5697 & 1.0547 & 2.625 & 148.89 & 2.8555 & 2.625 & -8.07\tabularnewline
\hline 
4 & 1.7629e+04 & 16.4221 & 1.0525 & 2.463 & 134.01 & 2.6046 & 2.463 & -5.44\tabularnewline
\hline 
5 & 3.5102e+03 & 13.8806 & 1.0482 & 2.3323 & 122.51 & 2.4063 & 2.3323 & -3.08\tabularnewline
\hline 
6 & 5.0816e+02 & 14.279 & 1.0436 & 2.2307 & 113.75 & 2.2373 & 2.2307 & -0.29\tabularnewline
\hline 
7 & 5.9899e+01 & 16.5848 & 1.0417 & 2.1677 & 108.09 & 2.1112 & 2.1677 & 2.68\tabularnewline
\hline 
8 & 1.0101e+01 & 19.9705 & 1.0456 & 2.1603 & 106.61 & 2.055 & 2.1603 & 5.12\tabularnewline
\hline 
9 & 1.8687e+00 & 21.4024 & 1.046 & 2.1806 & 108.47 & 2.0411 & 2.1806 & 6.83\tabularnewline
\hline 
10 & 2.7879e-01 & 27.9207 & 1.0446 & 2.2199 & 112.51 & 2.0491 & 2.2199 & 8.34\tabularnewline
\hline 
11 & 1.6600e-02 & 37.8399 & 1.0667 & 2.3796 & 123.08 & 2.1761 & 2.3796 & 9.35\tabularnewline
\hline 
12 & 3.1300e-04 & 96.5422 & 1.2447 & 2.7499 & 120.93 & 2.769 & 2.7499 & -0.69\tabularnewline
\hline 
\hline 
\multicolumn{2}{|c|}{Average} & 29.3\% & 7.1\% & 146.7\% & 130.3\% & 153.8\% & 146.7\% & -1.2\%\tabularnewline
\hline 
\end{tabular}{\small \caption{\label{tab:katlas75}Numerical results for the gluons-only exact and approximate NNLO $k$-factors as described in Sect.~\ref{sec:append}
for the ATLAS 2010 $\sqrt{s}=7$ TeV dataset~\cite{Aad:2010ad} in the rapidity slice $2.1<|\eta|<2.8$.}
}
\end{table}

\begin{table}[h]
\centering{}%
\begin{tabular}{|c|cc||cc|c||cc|c|}
\hline 
\multicolumn{3}{|c||}{$2.8<|\eta|<3.6$} & \multicolumn{2}{c|}{gg $k$-factor} &  & \multicolumn{2}{c|}{gg $k$-factor, NNLO$^{{\rm thr}}/$} & \tabularnewline
\hline 
$p_{T}$ bin & $\sigma_{{\rm Exp}}$ & $\sigma_{{\rm Exp}}^{{\rm error}}$ (\%) & Exact & Thr & $\delta\%$ & NLO$^{{\rm exact}}$ & NLO$^{{\rm thr}}$ & $\delta$\%\tabularnewline
\hline 
\hline 
1 & 2.0323e+06 & 44.9498 & 1.049 & 3.9296 & 274.60 & 5.1988 & 3.9296 & -24.41\tabularnewline
\hline 
2 & 2.8404e+05 & 41.0901 & 1.0693 & 3.6019 & 236.85 & 4.4167 & 3.6019 & -18.45\tabularnewline
\hline 
3 & 4.4737e+04 & 29.701 & 1.0441 & 3.3733 & 223.08 & 3.8837 & 3.3733 & -13.14\tabularnewline
\hline 
4 & 8.9271e+03 & 22.2178 & 1.0492 & 3.2522 & 209.97 & 3.6012 & 3.2522 & -9.69\tabularnewline
\hline 
5 & 1.3505e+03 & 22.0491 & 1.0557 & 3.2034 & 203.44 & 3.4039 & 3.2034 & -5.89\tabularnewline
\hline 
6 & 1.1959e+02 & 22.3605 & 1.0341 & 3.2345 & 212.78 & 3.2729 & 3.2345 & -1.17\tabularnewline
\hline 
7 & 5.6837e+00 & 33.596 & 0.9974 & 3.4406 & 244.96 & 3.2884 & 3.4406 & 4.63\tabularnewline
\hline 
8 & 2.9490e-01 & 43.4456 & 1.1062 & 3.9922 & 260.89 & 3.7523 & 3.9922 & 6.39\tabularnewline
\hline 
9 & 9.2626e-03 & 93.156 & 1.1287 & 5.659 & 401.37 & 4.8168 & 5.659 & 17.48\tabularnewline
\hline 
\hline 
\multicolumn{2}{|c|}{Average} & 39.2\% & 5.9\% & 274.3\% & 251.9\% & 295.9\% & 274.3\% & -4.9\%\tabularnewline
\hline 
\end{tabular}{\small \caption{\label{tab:katlas76}Numerical results for the gluons-only exact and approximate NNLO $k$-factors as described in Sect.~\ref{sec:append}
for the ATLAS 2010 $\sqrt{s}=7$ TeV dataset~\cite{Aad:2010ad} in the rapidity slice $2.8<|\eta|<3.6$.}
}
\end{table}

\begin{table}[h]
\centering{}%
\begin{tabular}{|c|cc||cc|c||cc|c|}
\hline 
\multicolumn{3}{|c||}{$3.6<|\eta|<4.4$} & \multicolumn{2}{c|}{gg $k$-factor} &  & \multicolumn{2}{c|}{gg $k$-factor, NNLO$^{{\rm thr}}/$} & \tabularnewline
\hline 
$p_{T}$ bin & $\sigma_{{\rm Exp}}$ & $\sigma_{{\rm Exp}}^{{\rm error}}$ (\%) & Exact & Thr & $\delta\%$ & NLO$^{{\rm exact}}$ & NLO$^{{\rm thr}}$ & $\delta$\%\tabularnewline
\hline 
\hline 
1 & 9.1000e+05 & 64.4667 & 1.035 & 4.9707 & 380.26 & 7.3006 & 4.9707 & -31.91\tabularnewline
\hline 
2 & 9.4945e+04 & 67.1388 & 1.0528 & 4.802 & 356.12 & 6.401 & 4.802 & -24.98\tabularnewline
\hline 
3 & 9.7033e+03 & 50.5717 & 1.03 & 4.8858 & 374.35 & 5.9346 & 4.8858 & -17.67\tabularnewline
\hline 
4 & 1.0000e+03 & 40.8999 & 1.0079 & 5.1638 & 412.33 & 5.8942 & 5.1638 & -12.39\tabularnewline
\hline 
5 & 5.8478e+01 & 38.009 & 1.0187 & 5.8867 & 477.86 & 6.2008 & 5.8867 & -5.07\tabularnewline
\hline 
6 & 5.8913e-01 & 52.7968 & 0.9229 & 9.4063 & 919.21 & 8.9487 & 9.4063 & 5.11\tabularnewline
\hline 
\hline 
\multicolumn{2}{|c|}{Average} & 52.3\% & 1.1\% & 485.2\% & 486.7\% & 578\% & 485.3\% & -14.5\%\tabularnewline
\hline 
\end{tabular}{\small \caption{\label{tab:katlas77}Numerical results for the gluons-only exact and approximate NNLO $k$-factors as described in Sect.~\ref{sec:append}
for the ATLAS 2010 $\sqrt{s}=7$ TeV dataset~\cite{Aad:2010ad} in the rapidity slice $3.6<|\eta|<4.4$.}
}
\end{table}

\clearpage

\subsection{ATLAS jets at $\sqrt{s}=2.76$ TeV}

\begin{table}[h]
\centering{}%
\begin{tabular}{|c|cc||cc|c||cc|c|}
\hline 
\multicolumn{3}{|c||}{$|\eta|<0.3$} & \multicolumn{2}{c|}{gg $k$-factor} &  & \multicolumn{2}{c|}{gg $k$-factor, NNLO$^{{\rm thr}}/$} & \tabularnewline
\hline 
$p_{T}$ bin & $\sigma_{{\rm Exp}}$ & $\sigma_{{\rm Exp}}^{{\rm error}}$ (\%) & Exact & Thr & $\delta\%$ & NLO$^{{\rm exact}}$ & NLO$^{{\rm thr}}$ & $\delta$\%\tabularnewline
\hline 
\hline 
1 & 1.4864e+03 & 24.2727 & 1.0721 & 1.7853 & 66.52 & 1.7372 & 1.7853 & 2.77\tabularnewline
\hline 
2 & 2.1374e+02 & 20.195 & 1.0639 & 1.5548 & 46.14 & 1.501 & 1.5548 & 3.58\tabularnewline
\hline 
3 & 3.5667e+01 & 14.3507 & 1.0484 & 1.4088 & 34.38 & 1.3558 & 1.4088 & 3.91\tabularnewline
\hline 
4 & 7.7000e+00 & 12.8376 & 1.0613 & 1.3275 & 25.08 & 1.2789 & 1.3275 & 3.80\tabularnewline
\hline 
5 & 1.4206e+00 & 13.8038 & 1.0452 & 1.2638 & 20.91 & 1.2186 & 1.2638 & 3.71\tabularnewline
\hline 
6 & 2.0276e-01 & 12.2458 & 1.0618 & 1.2108 & 14.03 & 1.1715 & 1.2108 & 3.35\tabularnewline
\hline 
7 & 2.5778e-02 & 15.7923 & 1.0393 & 1.174 & 12.96 & 1.1431 & 1.174 & 2.70\tabularnewline
\hline 
8 & 4.2737e-03 & 24.0877 & 1.0293 & 1.1584 & 12.54 & 1.1282 & 1.1584 & 2.68\tabularnewline
\hline 
9 & 1.2081e-03 & 41.7325 & 1.06 & 1.1464 & 8.15 & 1.1147 & 1.1464 & 2.84\tabularnewline
\hline 
10 & 4.3152e-04 & 48.5072 & 1.0478 & 1.1419 & 8.98 & 1.1144 & 1.1419 & 2.47\tabularnewline
\hline 
11 & 7.1101e-05 & 116.4275 & 1.0627 & 1.1474 & 7.97 & 1.1292 & 1.1474 & 1.61\tabularnewline
\hline 
\hline 
\multicolumn{2}{|c|}{Average} & 31.3\% & 5\% & 30.2\% & 23.4\% & 26.3\% & 30.2\% & 3.04\%\tabularnewline
\hline 
\end{tabular}{\small \caption{\label{tab:katlas21}Numerical results for the gluons-only exact and approximate NNLO $k$-factors as described in Sect.~\ref{sec:append}
for the ATLAS 2011 $\sqrt{s}=2.76$ TeV dataset~\cite{Aad:2013lpa} in the rapidity slice $|\eta|<0.3$.}
}
\end{table}

\begin{table}[h]
\centering{}%
\begin{tabular}{|c|cc||cc|c||cc|c|}
\hline 
\multicolumn{3}{|c||}{$0.3<|\eta|<0.8$} & \multicolumn{2}{c|}{gg $k$-factor} &  & \multicolumn{2}{c|}{gg $k$-factor, NNLO$^{{\rm thr}}/$} & \tabularnewline
\hline 
$p_{T}$ bin & $\sigma_{{\rm Exp}}$ & $\sigma_{{\rm Exp}}^{{\rm error}}$ (\%) & Exact & Thr & $\delta\%$ & NLO$^{{\rm exact}}$ & NLO$^{{\rm thr}}$ & $\delta$\%\tabularnewline
\hline 
\hline 
1 & 1.4818e+03 & 23.9425 & 1.0863 & 1.8582 & 71.06 & 1.8202 & 1.8582 & 2.09\tabularnewline
\hline 
2 & 2.0736e+02 & 19.9388 & 1.0719 & 1.6216 & 51.28 & 1.5715 & 1.6216 & 3.19\tabularnewline
\hline 
3 & 3.2989e+01 & 14.9857 & 1.0246 & 1.464 & 42.89 & 1.4122 & 1.464 & 3.67\tabularnewline
\hline 
4 & 7.3589e+00 & 12.8073 & 1.0586 & 1.3779 & 30.16 & 1.3263 & 1.3779 & 3.89\tabularnewline
\hline 
5 & 1.3990e+00 & 12.9025 & 1.0398 & 1.3104 & 26.02 & 1.2621 & 1.3104 & 3.83\tabularnewline
\hline 
6 & 1.8980e-01 & 13.5593 & 1.0415 & 1.2539 & 20.39 & 1.2101 & 1.2539 & 3.62\tabularnewline
\hline 
7 & 2.1495e-02 & 15.6309 & 1.0365 & 1.2075 & 16.50 & 1.1686 & 1.2075 & 3.33\tabularnewline
\hline 
8 & 3.3768e-03 & 23.7035 & 1.0441 & 1.189 & 13.88 & 1.153 & 1.189 & 3.12\tabularnewline
\hline 
9 & 9.8283e-04 & 35.6697 & 1.048 & 1.1799 & 12.59 & 1.1437 & 1.1799 & 3.17\tabularnewline
\hline 
10 & 3.6303e-04 & 47.7796 & 1.0478 & 1.1774 & 12.37 & 1.1431 & 1.1774 & 3.00\tabularnewline
\hline 
11 & 8.8899e-05 & 81.2158 & 1.052 & 1.1804 & 12.21 & 1.1443 & 1.1804 & 3.15\tabularnewline
\hline 
\hline 
\multicolumn{2}{|c|}{Average} & 27.5\% & 5\% & 34.7\% & 28.1\% & 30.5\% & 34.7\% & 3.3\%\tabularnewline
\hline 
\end{tabular}{\small \caption{\label{tab:katlas22}Numerical results for the gluons-only exact and approximate NNLO $k$-factors as described in Sect.~\ref{sec:append}
for the ATLAS 2011 $\sqrt{s}=2.76$ TeV dataset~\cite{Aad:2013lpa} in the rapidity slice $0.3<|\eta|<0.8$.}
}
\end{table}

\begin{table}[h]
\centering{}%
\begin{tabular}{|c|cc||cc|c||cc|c|}
\hline 
\multicolumn{3}{|c||}{$0.8<|\eta|<1.2$} & \multicolumn{2}{c|}{gg $k$-factor} &  & \multicolumn{2}{c|}{gg $k$-factor, NNLO$^{{\rm thr}}/$} & \tabularnewline
\hline 
$p_{T}$ bin & $\sigma_{{\rm Exp}}$ & $\sigma_{{\rm Exp}}^{{\rm error}}$ (\%) & Exact & Thr & $\delta\%$ & NLO$^{{\rm exact}}$ & NLO$^{{\rm thr}}$ & $\delta$\%\tabularnewline
\hline 
\hline 
1 & 1.2854e+03 & 26.9324 & 1.0757 & 2.0263 & 88.37 & 2.0208 & 2.0263 & 0.27\tabularnewline
\hline 
2 & 1.7380e+02 & 20.4538 & 1.0475 & 1.7743 & 69.38 & 1.7394 & 1.7743 & 2.01\tabularnewline
\hline 
3 & 2.8617e+01 & 16.2391 & 1.0381 & 1.6024 & 54.36 & 1.5549 & 1.6024 & 3.05\tabularnewline
\hline 
4 & 6.1844e+00 & 14.789 & 1.0356 & 1.5034 & 45.17 & 1.4511 & 1.5034 & 3.60\tabularnewline
\hline 
5 & 1.1827e+00 & 13.0982 & 1.0486 & 1.4242 & 35.82 & 1.3704 & 1.4242 & 3.93\tabularnewline
\hline 
6 & 1.5232e-01 & 14.5727 & 1.0454 & 1.3549 & 29.61 & 1.3006 & 1.3549 & 4.17\tabularnewline
\hline 
7 & 1.7313e-02 & 18.1585 & 1.0547 & 1.3066 & 23.88 & 1.2502 & 1.3066 & 4.51\tabularnewline
\hline 
8 & 2.3051e-03 & 27.3074 & 1.039 & 1.2871 & 23.88 & 1.2309 & 1.2871 & 4.57\tabularnewline
\hline 
9 & 6.5778e-04 & 46.1972 & 1.0514 & 1.2798 & 21.72 & 1.2298 & 1.2798 & 4.07\tabularnewline
\hline 
10 & 5.2980e-05 & 126.204 & 1.0551 & 1.275 & 20.84 & 1.2148 & 1.275 & 4.96\tabularnewline
\hline 
\hline 
\multicolumn{2}{|c|}{Average} & 32.4\% & 4.91\% & 48.3\% & 41.3\% & 43.6\% & 48.3\% & 3.5\%\tabularnewline
\hline 
\end{tabular}{\small \caption{\label{tab:katlas23}Numerical results for the gluons-only exact and approximate NNLO $k$-factors as described in Sect.~\ref{sec:append}
for the ATLAS 2011 $\sqrt{s}=2.76$ TeV dataset~\cite{Aad:2013lpa} in the rapidity slice $0.8<|\eta|<1.2$.}
}
\end{table}

\begin{table}[h]
\centering{}%
\begin{tabular}{|c|cc||cc|c||cc|c|}
\hline 
\multicolumn{3}{|c||}{$1.2<|\eta|<2.1$} & \multicolumn{2}{c|}{gg $k$-factor} &  & \multicolumn{2}{c|}{gg $k$-factor, NNLO$^{{\rm thr}}/$} & \tabularnewline
\hline 
$p_{T}$ bin & $\sigma_{{\rm Exp}}$ & $\sigma_{{\rm Exp}}^{{\rm error}}$ (\%) & Exact & Thr & $\delta\%$ & NLO$^{{\rm exact}}$ & NLO$^{{\rm thr}}$ & $\delta$\%\tabularnewline
\hline 
\hline 
1 & 1.0631e+03 & 31.3513 & 1.0773 & 2.4321 & 125.76 & 2.5278 & 2.4321 & -3.79\tabularnewline
\hline 
2 & 1.3804e+02 & 25.4028 & 1.0641 & 2.145 & 101.58 & 2.1559 & 2.145 & -0.51\tabularnewline
\hline 
3 & 2.1351e+01 & 17.9496 & 1.0639 & 1.9372 & 82.08 & 1.904 & 1.9372 & 1.74\tabularnewline
\hline 
4 & 4.3594e+00 & 14.1602 & 1.0473 & 1.8043 & 72.28 & 1.7536 & 1.8043 & 2.89\tabularnewline
\hline 
5 & 7.3959e-01 & 14.6909 & 1.0501 & 1.705 & 62.37 & 1.6378 & 1.705 & 4.10\tabularnewline
\hline 
6 & 8.3714e-02 & 15.7487 & 1.0419 & 1.6173 & 55.23 & 1.5365 & 1.6173 & 5.26\tabularnewline
\hline 
7 & 9.2384e-03 & 20.4022 & 1.0441 & 1.5472 & 48.19 & 1.4631 & 1.5472 & 5.75\tabularnewline
\hline 
8 & 9.1525e-04 & 32.1716 & 1.0217 & 1.5095 & 47.74 & 1.4002 & 1.5095 & 7.81\tabularnewline
\hline 
9 & 9.0354e-05 & 81.2976 & 1.0604 & 1.4933 & 40.82 & 1.4035 & 1.4933 & 6.40\tabularnewline
\hline 
10 & 2.4030e-05 & 125.8652 & 1.0772 & 1.4978 & 39.05 & 1.4001 & 1.4978 & 6.98\tabularnewline
\hline 
\hline 
\multicolumn{2}{|c|}{Average} & 37.9\% & 5.5\% & 76.9\% & 67.5\% & 71.8\% & 76.8\% & 3.6\%\tabularnewline
\hline 
\end{tabular}{\small \caption{\label{tab:katlas24}Numerical results for the gluons-only exact and approximate NNLO $k$-factors as described in Sect.~\ref{sec:append}
for the ATLAS 2011 $\sqrt{s}=2.76$ TeV dataset~\cite{Aad:2013lpa} in the rapidity slice $1.2<|\eta|<2.1$.}
}
\end{table}

\begin{table}[h]
\centering{}%
\begin{tabular}{|c|cc||cc|c||cc|c|}
\hline 
\multicolumn{3}{|c||}{$2.1<|\eta|<2.8$} & \multicolumn{2}{c|}{gg $k$-factor} &  & \multicolumn{2}{c|}{gg $k$-factor, NNLO$^{{\rm thr}}/$} & \tabularnewline
\hline 
$p_{T}$ bin & $\sigma_{{\rm Exp}}$ & $\sigma_{{\rm Exp}}^{{\rm error}}$ (\%) & Exact & Thr & $\delta\%$ & NLO$^{{\rm exact}}$ & NLO$^{{\rm thr}}$ & $\delta$\%\tabularnewline
\hline 
\hline 
1 & 7.1034e+02 & 40.3484 & 1.0701 & 3.2055 & 199.55 & 3.558 & 3.2055 & -9.91\tabularnewline
\hline 
2 & 8.4674e+01 & 32.6038 & 1.0175 & 2.8889 & 183.92 & 3.0465 & 2.8889 & -5.17\tabularnewline
\hline 
3 & 1.0968e+01 & 22.655 & 1.0937 & 2.6712 & 144.24 & 2.702 & 2.6712 & -1.14\tabularnewline
\hline 
4 & 1.8989e+00 & 18.4278 & 1.052 & 2.5724 & 144.52 & 2.5342 & 2.5724 & 1.51\tabularnewline
\hline 
5 & 2.2594e-01 & 20.7194 & 1.0344 & 2.5309 & 144.67 & 2.4325 & 2.5309 & 4.05\tabularnewline
\hline 
6 & 1.1698e-02 & 24.7426 & 1.0123 & 2.5444 & 151.35 & 2.3683 & 2.5444 & 7.44\tabularnewline
\hline 
7 & 5.4844e-04 & 51.9687 & 1.0582 & 2.7866 & 163.33 & 2.5084 & 2.7866 & 11.09\tabularnewline
\hline 
8 & 6.5250e-05 & 123.3975 & 1.0044 & 4.8469 & 382.57 & 3.8137 & 4.8469 & 27.09\tabularnewline
\hline 
\hline 
\multicolumn{2}{|c|}{Average} & 41.8\% & 4.3\% & 200\% & 189.3\% & 187.1\% & 200\% & 4.4\%\tabularnewline
\hline 
\end{tabular}{\small \caption{\label{tab:katlas25}Numerical results for the gluons-only exact and approximate NNLO $k$-factors as described in Sect.~\ref{sec:append}
for the ATLAS 2011 $\sqrt{s}=2.76$ TeV dataset~\cite{Aad:2013lpa} in the rapidity slice $2.1<|\eta|<2.8$.}
}
\end{table}

\begin{table}[h]
\centering{}%
\begin{tabular}{|c|cc||cc|c||cc|c|}
\hline 
\multicolumn{3}{|c||}{$2.8<|\eta|<3.6$} & \multicolumn{2}{c|}{gg $k$-factor} &  & \multicolumn{2}{c|}{gg $k$-factor, NNLO$^{{\rm thr}}/$} & \tabularnewline
\hline 
$p_{T}$ bin & $\sigma_{{\rm Exp}}$ & $\sigma_{{\rm Exp}}^{{\rm error}}$ (\%) & Exact & Thr & $\delta\%$ & NLO$^{{\rm exact}}$ & NLO$^{{\rm thr}}$ & $\delta$\%\tabularnewline
\hline 
\hline 
1 & 3.6430e+02 & 50.2722 & 1.0306 & 4.2642 & 313.76 & 5.0833 & 4.2642 & -16.11\tabularnewline
\hline 
2 & 3.1697e+01 & 48.5635 & 1.0433 & 4.0624 & 289.38 & 4.4793 & 4.0624 & -9.31\tabularnewline
\hline 
3 & 2.4769e+00 & 41.0702 & 1.033 & 4.0504 & 292.10 & 4.1676 & 4.0504 & -2.81\tabularnewline
\hline 
4 & 1.9495e-01 & 36.0109 & 1.0466 & 4.2394 & 305.06 & 4.144 & 4.2394 & 2.30\tabularnewline
\hline 
5 & 4.9813e-03 & 37.8164 & 1.0301 & 4.9744 & 382.90 & 4.6503 & 4.9744 & 6.97\tabularnewline
\hline 
6 & 1.6560e-04 & 92.2274 & 1.238 & 1.1865 & -4.16 & 1.0938 & 1.1865 & 8.48\tabularnewline
\hline 
\hline 
\multicolumn{2}{|c|}{Average} & 50.9\% & 7\% & 279\% & 263.2\% & 293.6\% & 279\% & -1.8\%\tabularnewline
\hline 
\end{tabular}{\small \caption{\label{tab:katlas26}Numerical results for the gluons-only exact and approximate NNLO $k$-factors as described in Sect.~\ref{sec:append}
for the ATLAS 2011 $\sqrt{s}=2.76$ TeV dataset~\cite{Aad:2013lpa} in the rapidity slice $2.8<|\eta|<3.6$.}
}
\end{table}

\begin{table}[h]
\centering{}%
\begin{tabular}{|c|cc||cc|c||cc|c|}
\hline 
\multicolumn{3}{|c||}{$3.6<|\eta|<4.4$} & \multicolumn{2}{c|}{gg $k$-factor} &  & \multicolumn{2}{c|}{gg $k$-factor, NNLO$^{{\rm thr}}/$} & \tabularnewline
\hline 
$p_{T}$ bin & $\sigma_{{\rm Exp}}$ & $\sigma_{{\rm Exp}}^{{\rm error}}$ (\%) & Exact & Thr & $\delta\%$ & NLO$^{{\rm exact}}$ & NLO$^{{\rm thr}}$ & $\delta$\%\tabularnewline
\hline 
\hline 
1 & 5.9628e+01 & 102.8591 & 1.0091 & 6.5656 & 550.64 & 8.3872 & 6.5656 & -21.72\tabularnewline
\hline 
2 & 1.7051e+00 & 153.8379 & 0.9774 & 7.567 & 674.20 & 8.59 & 7.567 & -11.91\tabularnewline
\hline 
3 & 1.1701e-02 & 148.6082 & 0.7618 & 26.0327 & 3317.26 & 16.5523 & 26.0327 & 57.28\tabularnewline
\hline 
\hline 
\multicolumn{2}{|c|}{Average} & 135.1\% & -9\% & 13.4\% & 1514\% & 1017.7\% & 1238.8\% & 7.9\%\tabularnewline
\hline 
\end{tabular}{\small \caption{\label{tab:katlas27}Numerical results for the gluons-only exact and approximate NNLO $k$-factors as described in Sect.~\ref{sec:append}
for the ATLAS 2011 $\sqrt{s}=2.76$ TeV dataset~\cite{Aad:2013lpa} in the rapidity slice $3.6<|\eta|<4.4$.}
}
\end{table}

\clearpage

\subsection{CDF jets}

\begin{table}[h]
\centering{}%
\begin{tabular}{|c|cc||cc|c||cc|c|}
\hline 
\multicolumn{3}{|c||}{$|\eta|<0.1$} & \multicolumn{2}{c|}{gg $k$-factor} &  & \multicolumn{2}{c|}{gg $k$-factor, NNLO$^{{\rm thr}}/$} & \tabularnewline
\hline 
$p_{T}$ bin & $\sigma_{{\rm Exp}}$ & $\sigma_{{\rm Exp}}^{{\rm error}}$ (\%) & Exact & Thr & $\delta\%$ & NLO$^{{\rm exact}}$ & NLO$^{{\rm thr}}$ & $\delta$\%\tabularnewline
\hline 
\hline 
1 & 1.2332e+01 & 18.3447 & 1.1842 & 1.2089 & 2.09 & 1.1569 & 1.2089 & 4.49\tabularnewline
\hline 
2 & 5.8392e+00 & 16.3363 & 1.1805 & 1.1919 & 0.97 & 1.1439 & 1.1919 & 4.20\tabularnewline
\hline 
3 & 2.5648e+00 & 15.1286 & 1.1815 & 1.1771 & -0.37 & 1.1301 & 1.1771 & 4.16\tabularnewline
\hline 
4 & 1.1293e+00 & 14.1836 & 1.1586 & 1.1679 & 0.80 & 1.1205 & 1.1679 & 4.23\tabularnewline
\hline 
5 & 4.9055e-01 & 13.6802 & 1.1719 & 1.1587 & -1.13 & 1.1155 & 1.1587 & 3.87\tabularnewline
\hline 
6 & 2.1819e-01 & 13.5437 & 1.1882 & 1.1527 & -2.99 & 1.1124 & 1.1527 & 3.62\tabularnewline
\hline 
\end{tabular}{\small \caption{\label{tab:kcdf1}Numerical results for the gluons-only exact and approximate NNLO $k$-factors as described in Sect.~\ref{sec:append}
for the CDF $\sqrt{s}=1.96$ TeV dataset~\cite{Abulencia:2007ez} in the rapidity slice $|\eta|<0.1$.}
}
\end{table}

\begin{table}[h]
\centering{}%
\begin{tabular}{|c|cc||cc|c||cc|c|}
\hline 
\multicolumn{3}{|c||}{$0.1<|\eta|<0.7$} & \multicolumn{2}{c|}{gg $k$-factor} &  & \multicolumn{2}{c|}{gg $k$-factor, NNLO$^{{\rm thr}}/$} & \tabularnewline
\hline 
$p_{T}$ bin & $\sigma_{{\rm Exp}}$ & $\sigma_{{\rm Exp}}^{{\rm error}}$ (\%) & Exact & Thr & $\delta\%$ & NLO$^{{\rm exact}}$ & NLO$^{{\rm thr}}$ & $\delta$\%\tabularnewline
\hline 
\hline 
1 & 1.1773e+01 & 17.3309 & 1.1882 & 1.2291 & 3.44 & 1.1767 & 1.2291 & 4.45\tabularnewline
\hline 
2 & 5.3141e+00 & 15.7537 & 1.1818 & 1.2111 & 2.48 & 1.161 & 1.2111 & 4.32\tabularnewline
\hline 
3 & 2.3850e+00 & 14.375 & 1.18 & 1.1974 & 1.47 & 1.1478 & 1.1974 & 4.32\tabularnewline
\hline 
4 & 1.0304e+00 & 13.4666 & 1.1811 & 1.1861 & 0.42 & 1.1371 & 1.1861 & 4.31\tabularnewline
\hline 
5 & 4.5067e-01 & 12.9087 & 1.1752 & 1.1733 & -0.16 & 1.1276 & 1.1733 & 4.05\tabularnewline
\hline 
6 & 1.9389e-01 & 12.6869 & 1.1704 & 1.1693 & -0.09 & 1.1247 & 1.1693 & 3.97\tabularnewline
\hline 
\end{tabular}{\small \caption{\label{tab:kcdf2}Numerical results for the gluons-only exact and approximate NNLO $k$-factors as described in Sect.~\ref{sec:append}
for the CDF $\sqrt{s}=1.96$ TeV dataset~\cite{Abulencia:2007ez} in the rapidity slice $0.1<|\eta|<0.7$.}
}
\end{table}

\begin{table}[h]
\centering{}%
\begin{tabular}{|c|cc||cc|c||cc|c|}
\hline 
\multicolumn{3}{|c||}{$0.7<|\eta|<1.1$} & \multicolumn{2}{c|}{gg $k$-factor} &  & \multicolumn{2}{c|}{gg $k$-factor, NNLO$^{{\rm thr}}/$} & \tabularnewline
\hline 
$p_{T}$ bin & $\sigma_{{\rm Exp}}$ & $\sigma_{{\rm Exp}}^{{\rm error}}$ (\%) & Exact & Thr & $\delta\%$ & NLO$^{{\rm exact}}$ & NLO$^{{\rm thr}}$ & $\delta$\%\tabularnewline
\hline 
\hline 
1 & 1.0490e+01 & 17.5576 & 1.1918 & 1.308 & 9.75 & 1.2514 & 1.308 & 4.52\tabularnewline
\hline 
2 & 4.7944e+00 & 16.2211 & 1.1866 & 1.2863 & 8.40 & 1.2302 & 1.2863 & 4.56\tabularnewline
\hline 
3 & 2.1471e+00 & 14.8956 & 1.1863 & 1.2656 & 6.68 & 1.2091 & 1.2656 & 4.67\tabularnewline
\hline 
4 & 9.1016e-01 & 14.2023 & 1.1731 & 1.2457 & 6.19 & 1.1906 & 1.2457 & 4.63\tabularnewline
\hline 
5 & 3.8312e-01 & 13.8172 & 1.1749 & 1.2356 & 5.17 & 1.1795 & 1.2356 & 4.76\tabularnewline
\hline 
6 & 1.6165e-01 & 13.8845 & 1.185 & 1.2254 & 3.41 & 1.1698 & 1.2254 & 4.75\tabularnewline
\hline 
\end{tabular}{\small \caption{\label{tab:kcdf3}Numerical results for the gluons-only exact and approximate NNLO $k$-factors as described in Sect.~\ref{sec:append}
for the CDF $\sqrt{s}=1.96$ TeV dataset~\cite{Abulencia:2007ez} in the rapidity slice $0.7<|\eta|<1.1$.}
}
\end{table}

\begin{table}[h]
\centering{}%
\begin{tabular}{|c|cc||cc|c||cc|c|}
\hline 
\multicolumn{3}{|c||}{$1.1<|\eta|<1.6$} & \multicolumn{2}{c|}{gg $k$-factor} &  & \multicolumn{2}{c|}{gg $k$-factor, NNLO$^{{\rm thr}}/$} & \tabularnewline
\hline 
$p_{T}$ bin & $\sigma_{{\rm Exp}}$ & $\sigma_{{\rm Exp}}^{{\rm error}}$ (\%) & Exact & Thr & $\delta\%$ & NLO$^{{\rm exact}}$ & NLO$^{{\rm thr}}$ & $\delta$\%\tabularnewline
\hline 
\hline 
1 & 9.3784e+00 & 17.4829 & 1.1948 & 1.4397 & 20.50 & 1.3791 & 1.4397 & 4.39\tabularnewline
\hline 
2 & 3.8388e+00 & 16.4669 & 1.1745 & 1.4086 & 19.93 & 1.3458 & 1.4086 & 4.67\tabularnewline
\hline 
3 & 1.6185e+00 & 15.6789 & 1.1808 & 1.3816 & 17.01 & 1.3162 & 1.3816 & 4.97\tabularnewline
\hline 
4 & 6.5171e-01 & 15.764 & 1.1735 & 1.3624 & 16.10 & 1.293 & 1.3624 & 5.37\tabularnewline
\hline 
5 & 2.7261e-01 & 14.9935 & 1.18 & 1.3457 & 14.04 & 1.2759 & 1.3457 & 5.47\tabularnewline
\hline 
6 & 1.0535e-01 & 15.4667 & 1.1844 & 1.3325 & 12.50 & 1.2643 & 1.3325 & 5.39\tabularnewline
\hline 
7 & 3.8212e-02 & 16.014 & 1.1864 & 1.3278 & 11.92 & 1.2543 & 1.3278 & 5.86\tabularnewline
\hline 
8 & 1.3067e-02 & 17.3512 & 1.188 & 1.3254 & 11.57 & 1.2507 & 1.3254 & 5.97\tabularnewline
\hline 
9 & 3.9695e-03 & 19.2511 & 1.2008 & 1.3257 & 10.40 & 1.2544 & 1.3257 & 5.68\tabularnewline
\hline 
10 & 1.0991e-03 & 22.068 & 1.2123 & 1.3346 & 10.09 & 1.263 & 1.3346 & 5.67\tabularnewline
\hline 
11 & 2.6483e-04 & 25.8205 & 1.2328 & 1.3472 & 9.28 & 1.2774 & 1.3472 & 5.46\tabularnewline
\hline 
12 & 5.1499e-05 & 30.4573 & 1.2529 & 1.372 & 9.51 & 1.317 & 1.372 & 4.18\tabularnewline
\hline 
13 & 8.2305e-06 & 37.7877 & 1.2743 & 1.4069 & 10.41 & 1.3684 & 1.4069 & 2.81\tabularnewline
\hline 
14 & 1.1920e-06 & 58.2323 & 1.3041 & 1.4916 & 14.38 & 1.4652 & 1.4916 & 1.80\tabularnewline
\hline 
\hline 
\multicolumn{2}{|c|}{Average} & 23.1\% & 21.0\% & 37.2\% & 13.4\% & 30.9\% & 37.2\% & 4.8\%\tabularnewline
\hline 
\end{tabular}{\small \caption{\label{tab:kcdf4}Numerical results for the gluons-only exact and approximate NNLO $k$-factors as described in Sect.~\ref{sec:append}
for the CDF $\sqrt{s}=1.96$ TeV dataset~\cite{Abulencia:2007ez} in the rapidity slice $0.7<|\eta|<1.1$.}
}
\end{table}

\begin{table}[h]
\centering{}%
\begin{tabular}{|c|cc||cc|c||cc|c|}
\hline 
\multicolumn{3}{|c||}{$1.6<|\eta|<2.1$} & \multicolumn{2}{c|}{gg $k$-factor} &  & \multicolumn{2}{c|}{gg $k$-factor, NNLO$^{{\rm thr}}/$} & \tabularnewline
\hline 
$p_{T}$ bin & $\sigma_{{\rm Exp}}$ & $\sigma_{{\rm Exp}}^{{\rm error}}$ (\%) & Exact & Thr & $\delta\%$ & NLO$^{{\rm exact}}$ & NLO$^{{\rm thr}}$ & $\delta$\%\tabularnewline
\hline 
\hline 
1 & 5.9423e+00 & 16.3059 & 1.1994 & 1.6725 & 39.44 & 1.6053 & 1.6725 & 4.19\tabularnewline
\hline 
2 & 2.4098e+00 & 15.1148 & 1.1845 & 1.6347 & 38.01 & 1.5606 & 1.6347 & 4.75\tabularnewline
\hline 
3 & 9.4782e-01 & 14.659 & 1.1936 & 1.6136 & 35.19 & 1.5306 & 1.6136 & 5.42\tabularnewline
\hline 
4 & 3.4801e-01 & 14.9126 & 1.1938 & 1.5986 & 33.91 & 1.5103 & 1.5986 & 5.85\tabularnewline
\hline 
5 & 1.2369e-01 & 15.9119 & 1.1909 & 1.589 & 33.43 & 1.4929 & 1.589 & 6.44\tabularnewline
\hline 
6 & 3.9807e-02 & 17.6115 & 1.2015 & 1.587 & 32.08 & 1.4921 & 1.587 & 6.36\tabularnewline
\hline 
7 & 1.1696e-02 & 20.0832 & 1.2017 & 1.5956 & 32.78 & 1.4887 & 1.5956 & 7.18\tabularnewline
\hline 
8 & 2.8706e-03 & 23.1154 & 1.2069 & 1.6111 & 33.49 & 1.5118 & 1.6111 & 6.57\tabularnewline
\hline 
9 & 5.7736e-04 & 27.0941 & 1.2306 & 1.6507 & 34.14 & 1.5464 & 1.6507 & 6.74\tabularnewline
\hline 
10 & 8.7540e-05 & 33.0822 & 1.2463 & 1.7251 & 38.42 & 1.6413 & 1.7251 & 5.11\tabularnewline
\hline 
11 & 9.2235e-06 & 42.8557 & 1.259 & 1.9175 & 52.30 & 1.832 & 1.9175 & 4.67\tabularnewline
\hline 
12 & 5.9441e-07 & 60.5582 & 1.1491 & 1.4305 & 24.49 & 1.2008 & 1.4305 & 19.13\tabularnewline
\hline 
\hline 
\multicolumn{2}{|c|}{Average} & 25.1\% & 20.5\% & 63.5\% & 35.6\% & 53.4\% & 63.5\% & 6.9\%\tabularnewline
\hline 
\end{tabular}{\small \caption{\label{tab:kcdf5}Numerical results for the gluons-only exact and approximate NNLO $k$-factors as described in Sect.~\ref{sec:append}
for the CDF $\sqrt{s}=1.96$ TeV dataset~\cite{Abulencia:2007ez} in the rapidity slice $1.6<|\eta|<2.1$.}
}
\end{table}

\clearpage

\subsection{D0 jets}

\begin{table}[h]
\centering{}%
\begin{tabular}{|c|cc||cc|c||cc|c|}
\hline 
\multicolumn{3}{|c||}{$|\eta|<0.4$} & \multicolumn{2}{c|}{gg $k$-factor} &  & \multicolumn{2}{c|}{gg $k$-factor, NNLO$^{{\rm thr}}/$} & \tabularnewline
\hline 
$p_{T}$ bin & $\sigma_{{\rm Exp}}$ & $\sigma_{{\rm Exp}}^{{\rm error}}$ (\%) & Exact & Thr & $\delta\%$ & NLO$^{{\rm exact}}$ & NLO$^{{\rm thr}}$ & $\delta$\%\tabularnewline
\hline 
\hline 
1 & 2.3282e+04 & 11.2934 & 1.1922 & 1.2223 & 2.52 & 1.1709 & 1.2223 & 4.39\tabularnewline
\hline 
2 & 8.8665e+03 & 10.2301 & 1.1835 & 1.2008 & 1.46 & 1.1499 & 1.2008 & 4.43\tabularnewline
\hline 
3 & 3.7927e+03 & 9.7392 & 1.1769 & 1.1858 & 0.76 & 1.1366 & 1.1858 & 4.33\tabularnewline
\hline 
4 & 1.7678e+03 & 9.41 & 1.1867 & 1.1751 & -0.98 & 1.1283 & 1.1751 & 4.15\tabularnewline
\hline 
5 & 8.9746e+02 & 9.2724 & 1.1765 & 1.1663 & -0.87 & 1.1209 & 1.1663 & 4.05\tabularnewline
\hline 
6 & 4.8546e+02 & 9.0215 & 1.1694 & 1.1603 & -0.78 & 1.118 & 1.1603 & 3.78\tabularnewline
\hline 
\end{tabular}{\small \caption{\label{tab:kd01}Numerical results for the gluons-only exact and approximate NNLO $k$-factors as described in Sect.~\ref{sec:append}
for the D0 $\sqrt{s}=1.96$ TeV dataset~\cite{Abazov:2011vi} in the rapidity slice $|\eta|<0.4$.}
}
\end{table}

\begin{table}[h]
\centering{}%
\begin{tabular}{|c|cc||cc|c||cc|c|}
\hline 
\multicolumn{3}{|c||}{$0.4<|\eta|<0.8$} & \multicolumn{2}{c|}{gg $k$-factor} &  & \multicolumn{2}{c|}{gg $k$-factor, NNLO$^{{\rm thr}}/$} & \tabularnewline
\hline 
$p_{T}$ bin & $\sigma_{{\rm Exp}}$ & $\sigma_{{\rm Exp}}^{{\rm error}}$ (\%) & Exact & Thr & $\delta\%$ & NLO$^{{\rm exact}}$ & NLO$^{{\rm thr}}$ & $\delta$\%\tabularnewline
\hline 
\hline 
1 & 2.1728e+04 & 11.47 & 1.1893 & 1.2639 & 6.27 & 1.2097 & 1.2639 & 4.48\tabularnewline
\hline 
2 & 8.1627e+03 & 10.5165 & 1.1796 & 1.2379 & 4.94 & 1.1857 & 1.2379 & 4.40\tabularnewline
\hline 
3 & 3.4953e+03 & 10.061 & 1.1852 & 1.2216 & 3.07 & 1.1693 & 1.2216 & 4.47\tabularnewline
\hline 
4 & 1.6199e+03 & 9.826 & 1.1805 & 1.2085 & 2.37 & 1.1571 & 1.2085 & 4.44\tabularnewline
\hline 
5 & 8.0500e+02 & 9.7325 & 1.1707 & 1.1968 & 2.23 & 1.1472 & 1.1968 & 4.32\tabularnewline
\hline 
6 & 4.3791e+02 & 9.526 & 1.1688 & 1.1876 & 1.61 & 1.141 & 1.1876 & 4.08\tabularnewline
\hline 
\end{tabular}{\small \caption{\label{tab:kd02}Numerical results for the gluons-only exact and approximate NNLO $k$-factors as described in Sect.~\ref{sec:append}
for the D0 $\sqrt{s}=1.96$ TeV dataset~\cite{Abazov:2011vi} in the rapidity slice $0.4<|\eta|<0.8$.}
}
\end{table}

\begin{table}[h]
\centering{}%
\begin{tabular}{|c|cc||cc|c||cc|c|}
\hline 
\multicolumn{3}{|c||}{$0.8<|\eta|<1.2$} & \multicolumn{2}{c|}{gg $k$-factor} &  & \multicolumn{2}{c|}{gg $k$-factor, NNLO$^{{\rm thr}}/$} & \tabularnewline
\hline 
$p_{T}$ bin & $\sigma_{{\rm Exp}}$ & $\sigma_{{\rm Exp}}^{{\rm error}}$ (\%) & Exact & Thr & $\delta\%$ & NLO$^{{\rm exact}}$ & NLO$^{{\rm thr}}$ & $\delta$\%\tabularnewline
\hline 
\hline 
1 & 1.7739e+04 & 12.0673 & 1.188 & 1.3436 & 13.10 & 1.2873 & 1.3436 & 4.37\tabularnewline
\hline 
2 & 6.6481e+03 & 11.369 & 1.198 & 1.3133 & 9.62 & 1.256 & 1.3133 & 4.56\tabularnewline
\hline 
3 & 2.8083e+03 & 11.1644 & 1.1829 & 1.2915 & 9.18 & 1.2328 & 1.2915 & 4.76\tabularnewline
\hline 
4 & 1.2952e+03 & 11.1887 & 1.1656 & 1.2741 & 9.31 & 1.2158 & 1.2741 & 4.80\tabularnewline
\hline 
5 & 6.6021e+02 & 10.2182 & 1.1784 & 1.26 & 6.92 & 1.2008 & 1.26 & 4.93\tabularnewline
\hline 
6 & 3.4070e+02 & 10.2663 & 1.1861 & 1.2537 & 5.70 & 1.1963 & 1.2537 & 4.80\tabularnewline
\hline 
\end{tabular}{\small \caption{\label{tab:kd03}Numerical results for the gluons-only exact and approximate NNLO $k$-factors as described in Sect.~\ref{sec:append}
for the D0 $\sqrt{s}=1.96$ TeV dataset~\cite{Abazov:2011vi} in the rapidity slice $0.8<|\eta|<1.2$.}
}
\end{table}

\begin{table}[h]
\centering{}%
\begin{tabular}{|c|cc||cc|c||cc|c|}
\hline 
\multicolumn{3}{|c||}{$1.2<|\eta|<1.6$} & \multicolumn{2}{c|}{gg $k$-factor} &  & \multicolumn{2}{c|}{gg $k$-factor, NNLO$^{{\rm thr}}/$} & \tabularnewline
\hline 
$p_{T}$ bin & $\sigma_{{\rm Exp}}$ & $\sigma_{{\rm Exp}}^{{\rm error}}$ (\%) & Exact & Thr & $\delta\%$ & NLO$^{{\rm exact}}$ & NLO$^{{\rm thr}}$ & $\delta$\%\tabularnewline
\hline 
\hline 
1 & 1.5249e+04 & 13.2561 & 1.1923 & 1.4726 & 23.51 & 1.4114 & 1.4726 & 4.34\tabularnewline
\hline 
2 & 5.2668e+03 & 12.6595 & 1.1838 & 1.4355 & 21.26 & 1.3718 & 1.4355 & 4.64\tabularnewline
\hline 
3 & 2.1975e+03 & 12.5006 & 1.1882 & 1.4087 & 18.56 & 1.3407 & 1.4087 & 5.07\tabularnewline
\hline 
4 & 9.6320e+02 & 11.5469 & 1.1926 & 1.3883 & 16.41 & 1.3191 & 1.3883 & 5.25\tabularnewline
\hline 
5 & 4.6424e+02 & 11.3927 & 1.1795 & 1.3728 & 16.39 & 1.299 & 1.3728 & 5.68\tabularnewline
\hline 
6 & 2.3508e+02 & 11.4617 & 1.1722 & 1.3617 & 16.17 & 1.2904 & 1.3617 & 5.53\tabularnewline
\hline 
\end{tabular}{\small \caption{\label{tab:kd04}Numerical results for the gluons-only exact and approximate NNLO $k$-factors as described in Sect.~\ref{sec:append}
for the D0 $\sqrt{s}=1.96$ TeV dataset~\cite{Abazov:2011vi} in the rapidity slice $1.2<|\eta|<1.6$.}
}
\end{table}

\begin{table}[h]
\centering{}%
\begin{tabular}{|c|cc||cc|c||cc|c|}
\hline 
\multicolumn{3}{|c||}{$1.6<|\eta|<2.0$} & \multicolumn{2}{c|}{gg $k$-factor} &  & \multicolumn{2}{c|}{gg $k$-factor, NNLO$^{{\rm thr}}/$} & \tabularnewline
\hline 
$p_{T}$ bin & $\sigma_{{\rm Exp}}$ & $\sigma_{{\rm Exp}}^{{\rm error}}$ (\%) & Exact & Thr & $\delta\%$ & NLO$^{{\rm exact}}$ & NLO$^{{\rm thr}}$ & $\delta$\%\tabularnewline
\hline 
\hline 
1 & 1.1684e+04 & 13.2001 & 1.1967 & 1.6629 & 38.96 & 1.6007 & 1.6629 & 3.89\tabularnewline
\hline 
2 & 4.0542e+03 & 12.7548 & 1.1875 & 1.6233 & 36.70 & 1.5502 & 1.6233 & 4.72\tabularnewline
\hline 
3 & 1.5651e+03 & 12.432 & 1.2011 & 1.5954 & 32.83 & 1.5135 & 1.5954 & 5.41\tabularnewline
\hline 
4 & 6.6951e+02 & 12.5395 & 1.1844 & 1.5807 & 33.46 & 1.4936 & 1.5807 & 5.83\tabularnewline
\hline 
5 & 2.9506e+02 & 12.6582 & 1.1809 & 1.5703 & 32.97 & 1.4813 & 1.5703 & 6.01\tabularnewline
\hline 
6 & 1.3622e+02 & 12.9305 & 1.2123 & 1.5655 & 29.13 & 1.4714 & 1.5655 & 6.40\tabularnewline
\hline 
\end{tabular}{\small \caption{\label{tab:kd05}Numerical results for the gluons-only exact and approximate NNLO $k$-factors as described in Sect.~\ref{sec:append}
for the D0 $\sqrt{s}=1.96$ TeV dataset~\cite{Abazov:2011vi} in the rapidity slice $1.6<|\eta|<2.0$.}
}
\end{table}

\begin{table}[h]
\centering{}%
\begin{tabular}{|c|cc||cc|c||cc|c|}
\hline 
\multicolumn{3}{|c||}{$2.0<|\eta|<2.4$} & \multicolumn{2}{c|}{gg $k$-factor} &  & \multicolumn{2}{c|}{gg $k$-factor, NNLO$^{{\rm thr}}/$} & \tabularnewline
\hline 
$p_{T}$ bin & $\sigma_{{\rm Exp}}$ & $\sigma_{{\rm Exp}}^{{\rm error}}$ (\%) & Exact & Thr & $\delta\%$ & NLO$^{{\rm exact}}$ & NLO$^{{\rm thr}}$ & $\delta$\%\tabularnewline
\hline 
\hline 
1 & 7.2148e+03 & 14.7391 & 1.2004 & 1.9522 & 62.63 & 1.8857 & 1.9522 & 3.53\tabularnewline
\hline 
2 & 2.2811e+03 & 14.7294 & 1.196 & 1.9255 & 60.99 & 1.8391 & 1.9255 & 4.70\tabularnewline
\hline 
3 & 7.4168e+02 & 14.8434 & 1.1859 & 1.9162 & 61.58 & 1.8166 & 1.9162 & 5.48\tabularnewline
\hline 
4 & 2.7132e+02 & 15.442 & 1.2224 & 1.9309 & 57.96 & 1.8063 & 1.9309 & 6.90\tabularnewline
\hline 
5 & 9.9553e+01 & 16.0892 & 1.2034 & 1.9457 & 61.68 & 1.8335 & 1.9457 & 6.12\tabularnewline
\hline 
6 & 3.8789e+01 & 16.9895 & 1.1997 & 1.9761 & 64.72 & 1.853 & 1.9761 & 6.64\tabularnewline
\hline 
7 & 1.4726e+01 & 18.0928 & 1.222 & 2.011 & 64.57 & 1.8989 & 2.011 & 5.90\tabularnewline
\hline 
8 & 5.7305e+00 & 19.3456 & 1.208 & 2.0719 & 71.51 & 1.9374 & 2.0719 & 6.94\tabularnewline
\hline 
9 & 1.8183e+00 & 21.2101 & 1.2115 & 2.1508 & 77.53 & 2.0074 & 2.1508 & 7.14\tabularnewline
\hline 
10 & 3.8345e-01 & 24.4537 & 1.1939 & 2.352 & 97.00 & 2.2418 & 2.352 & 4.92\tabularnewline
\hline 
11 & 8.1686e-02 & 28.3113 & 1.0476 & 3.0379 & 189.99 & 2.9002 & 3.0379 & 4.75\tabularnewline
\hline 
12 & 1.0941e-02 & 35.1504 & 1.2291 & 1.5741 & 28.07 & 1.3978 & 1.5741 & 12.61\tabularnewline
\hline 
13 & 3.6139e-04 & 52.8355 & 1.4345 & 0.2947 & -79.46 & 0.279 & 0.2947 & 5.63\tabularnewline
\hline 
\hline 
\multicolumn{2}{|c|}{Average} & 22.5\% & 21.2\% & 93.4\% & 62.9\% & 82.3\% & 93.4\% & 6.3\%\tabularnewline
\hline 
\end{tabular}{\small \caption{\label{tab:kd06}Numerical results for the gluons-only exact and approximate NNLO $k$-factors as described in Sect.~\ref{sec:append}
for the D0 $\sqrt{s}=1.96$ TeV dataset~\cite{Abazov:2011vi} in the rapidity slice $2.0<|\eta|<2.4$.}
}
\end{table}

\clearpage

\bibliography{jets}

\end{document}